\documentclass[fleqn,usenatbib]{mnras}

\usepackage[T1]{fontenc}

\DeclareRobustCommand{\VAN}[3]{#2}
\let\VANthebibliography\thebibliography
\def\thebibliography{\DeclareRobustCommand{\VAN}[3]{##3}\VANthebibliography}

\usepackage{graphicx}	
\usepackage{amsmath}	
\usepackage{amssymb}	
\usepackage{newtxtext,newtxmath}

\newcommand{\velociraptor}{\textsc{VELOCIraptor}\,}
\newcommand{\swift}{\textsc{SWIFT}\,}

\title[Entropy cores at odds with observations]{EAGLE-like simulation models do not solve the entropy core problem in groups and clusters of galaxies}

\author[E. Altamura et al.]{Edoardo Altamura,$^{1}$\thanks{E-mail: \href{mailto:edoardo.altamura@manchester.ac.uk}{edoardo.altamura@manchester.ac.uk}} 
Scott T. Kay,$^{1}$ 
Richard G. Bower,$^{2}$ 
Matthieu Schaller,$^{3, 4}$ 
Yannick M. Bah{\'e},$^{4}$ 
\newauthor{Joop Schaye,$^{4}$
Josh Borrow$^{5, 2}$ and 
Imogen Towler$^{1}$}
\\
$^{1}$Jodrell Bank Centre for Astrophysics, Department of Physics and Astronomy, The University of Manchester, Oxford Road, Manchester M13 9PL, UK\\
$^{2}$Institute for Computational Cosmology, Department of Physics, University of Durham, South Road, Durham, DH1 3LE, UK\\
$^{3}$Lorentz Institute for Theoretical Physics, Leiden University, PO box 9506, 2300 RA Leiden, the Netherlands\\
$^{4}$Leiden Observatory, Leiden University, PO Box 9513, NL-2300 RA Leiden, the Netherlands\\
$^{5}$Department of Physics, Kavli Institute for Astrophysics and Space Research, Massachusetts Institute of Technology, Cambridge, MA 02139, USA
}

\date{Accepted 2023 February 06. Received 2023 January 26; in original form 2022 November 02}

\pubyear{2022}

\begin{document}
\label{firstpage}
\pagerange{\pageref{firstpage}--\pageref{lastpage}}
\maketitle

\begin{abstract}
Recent high-resolution cosmological hydrodynamic simulations run with a variety of codes systematically predict large amounts of entropy in the intra-cluster medium at low redshift, leading to flat entropy profiles and a suppressed cool-core population. This prediction is at odds with X-ray observations of groups and clusters. We use a new implementation of the EAGLE galaxy formation model to investigate the sensitivity of the central entropy and the shape of the profiles to changes in the sub-grid model applied to a suite of zoom-in cosmological simulations of a group of mass $M_{500} = 8.8 \times 10^{12}~{\rm M}_\odot$  and a cluster of mass $2.9 \times 10^{14}~{\rm M}_\odot$. Using our reference model, calibrated to match the stellar mass function of field galaxies, we confirm that our simulated groups and clusters contain hot gas with too high entropy in their cores. Additional simulations run without artificial conduction, metal cooling or AGN feedback produce lower entropy levels but still fail to reproduce observed profiles. Conversely, the two objects run without supernova feedback show a significant entropy increase which can be attributed to excessive cooling and star formation. Varying the AGN heating temperature does not greatly affect the profile shape, but only the overall normalisation. Finally, we compared runs with four AGN heating schemes and obtained similar profiles, with the exception of bipolar AGN heating, which produces a higher and more uniform entropy distribution. Our study leaves open the question of whether the entropy core problem in simulations, and particularly the lack of power-law cool-core profiles, arise from incorrect physical assumptions, missing physical processes, or insufficient numerical resolution.

\end{abstract}




\section{Introduction}
Located at the centre of massive galaxies, Super-Massive Black Holes (SMBHs) are believed to be the main cause for the self-regulation of the baryon content and the thermodynamics of the local Inter-Galactic Medium \citep[IGM, e.g.][]{2013ARA&A..51..511K, 2014ARA&A..52..589H}. The SMBH can capture cold and dense material from its environment and can generate outflows of hot plasma into the IGM. These two processes, known as \textit{feeding} and \textit{feedback} respectively, can be detected in X-ray, optical and microwave/sub-mm bands and usually affect the IGM from scales of the order of 10 kpc to 1 Mpc \citep[e.g.][]{2012ARA&A..50..455F}. Active Galactic Nuclei (AGN) are excellent examples of high-redshift galaxies showing cooling flows, relativistic jets and shock-heated plasma, resulting from a complex interaction between SMBH feeding and feedback processes \citep[see e.g.][]{agn_review_eckert_2021}. These events have an impact on the thermodynamic state of the hot IGM in groups and clusters, altering quantities such as the gas density, temperature, metallicity, and also regulating the star formation (SF) and the growth of the SMBHs.

A useful quantity to probe the thermal state of the IGM is the thermodynamic entropy $K$, which is obtained by combining the temperature $T$ and electron number density $n_e$ of the ionised gas, as $ K=k_{\rm B}T/n_e^{2/3}$, where $k_{\rm B}$ is the Boltzmann constant \citep[see][for a derivation from classical thermodynamics with a monoatomic gas]{entropy_intro_bower_1997}. Entropy profiles, obtained from measured temperature and density profiles, are sensitive to the thermodynamic state of the group/cluster atmosphere, which is influenced by feedback, cooling flows, star formation and gravitational processes. They are, in other words, conceptually simple tools which can facilitate the study of complex and interdependent processes taking place in the IGM. Probing the distribution of entropy in groups and clusters can potentially shed new light on highly debated topics in galaxy formation, such as the role of the baryon cycle across cosmic time and the link between AGN activity and galaxy quenching \citep[see][ for a review and Kay \& Pratt, in preparation]{agn_review_eckert_2021}.

X-ray observations provide measurements of the temperature and density distribution of the hot gas. Using temperature and density profiles from ROSAT observations, \cite{1996ApJ...473..692D} showed that metallicity also has a significant impact on the entropy profiles. By applying similar analysis techniques, \cite{2009ApJS..182...12C} found that, on average, galaxy clusters have self-similar entropy profiles near the virial radius, while showing great variability in the inner regions. Using a sample of 31 nearby clusters (the REXCESS sample), \cite{entropy_profiles_pratt2010} showed that the central entropy $K_0$, the entropy excess above a power-law \citep{2005ApJ...630L..13D}, and the logarithmic slope of the profile at $0.075\, r_{500}$ follow bimodal distributions\footnote{Following the spherical overdensity formalism, we define $r_{500}$ as the radius (from the gravitational potential minimum) at which the internal mean density exceeds the critical density by a factor of 500. Other properties such as the mass $M_{500}$ are computed from particles located within $r_{500}$ from the centre of the halo.}. Clusters with a low value of $K_0\approx3~\mathrm{keV~cm^2}$ also show a steep slope, while objects with high $K_0\approx75~\mathrm{keV~cm^2}$ have shallower profiles in the core. Since the gas can reach low entropy by reducing its temperature and increasing its density, the clusters with power-law entropy profiles (and steep logarithmic slopes) are classified as cool-core (CC), while the ones with shallow profiles in the inner region and large deviation from a power-law are defined as non-cool-core (NCC).



Naturally, the gas found in cooling flows is denser and cooler than the IGM in quasi-hydrostatic equilibrium and must have low entropy. In the absence of non-gravitational processes, such as radiative cooling, star formation and feedback, we expect low-entropy gas to sink to the inner region and high-entropy gas to rise buoyantly and expand towards the outskirts of the systems. These conditions were recreated in the early \textit{non-radiative} hydrodynamics simulations of galaxy clusters and produced self-similar power-law entropy profiles, showing low entropy in the cluster cores and rising towards the virial radius \citep{2001ApJ...546...63T, vkb_2005}. {In particular, Lagrangian methods, such as smoothed-particle hydrodynamics (SPH), tended to produce power-law-like entropy profiles and small cores, while Eulerian (grid-based) codes produced solutions with larger entropy cores (see e.g. the \texttt{Bryan-SAMR} code in Fig. 18 of \citealt{1999ApJ...525..554F}, and Fig. 10 of  \citealt{2016MNRAS.459.2973S}).} A study by \cite{2009MNRAS.395..180M} attributed the lack of isentropic cores in non-radiative SPH simulations of clusters to the absence of gas mixing, which is a consequence of hydrodynamic instabilities \citep[see also][]{2007MNRAS.380..963A}. More recent formulations of SPH used in simulations include an artificial conduction term to reproduce the instabilities and capture shock-heating through artificial viscosity. The importance of thermal conduction and entropy mixing is confirmed in the nIFTy comparison project, where 12 codes were used to simulate a $10^{15}$ M$_\odot$ cluster, producing a core with constant entropy, high temperature and low density \citep{2016MNRAS.457.4063S}. 

In non-radiative simulations, entropy mixing can produce large isentropic cores by suppressing cooling flows, however, studies using radiative cooling and pre-heated gas found that cooling flows can also remove low-entropy gas from the centre of halos \citep[e.g.][]{2000MNRAS.317.1029P, 2001MNRAS.326.1228B, 2002MNRAS.336..527M, 2005MNRAS.361..233B}. They demonstrated that the densest gas in the centre of clusters can radiate away energy through free-free and line emission and attain very short cooling times. The formation of cooling flows can remove this {cold and} low-entropy material from the {hot,} X-ray emitting gas phase. {Higher-entropy gas is then allowed} to sink to the centre of the system, {establishing} a high entropy level. Pre-heating was introduced in these simulations to regulate the formation of strong cooling flows which could otherwise lead to unrealistically high stellar masses {and cold gas masses}. An important drawback of this method is the production of a large entropy excess in the cores of groups and clusters, which cannot be reconciled with the observations.

Motivated by observations, modern simulations of groups and clusters dropped the pre-heating approach and, instead, explicitly model SN-driven winds and AGN outflows (feedback) to regulate the cooling flows and produce realistic galaxy populations and cluster scaling relations \citep[e.g.][]{2010MNRAS.406..822M, 2014MNRAS.442.2304H, 2014MNRAS.445.1774P, 2015ApJ...813L..17R, 2017MNRAS.465.2936M, 2017MNRAS.465..213B, ceagle.barnes.2017, 2018MNRAS.479.5385H, 2018MNRAS.481.1809B, 2020MNRAS.495.2930L, 2020MNRAS.498.3061R, 2021MNRAS.504.3922C}. As mentioned, the shape of the entropy profiles is an excellent metric for assessing the impact of stellar and AGN feedback and radiative cooling (free-free and metal lines) on the IGM. A number of these full-physics simulations show good agreement with X-ray observations. For instance, the BAHAMAS and MACSIS samples \citep{2017MNRAS.465.2936M, 2017MNRAS.465..213B} produced entropy profiles at $z=0$ and $z=1$ in agreement with observations from \cite{entropy_profiles_pratt2010, 2015MNRAS.447.3044G} and \cite{2014ApJ...794...67M}. BAHAMAS and MACSIS used a purely kinetic SN feedback scheme, based on the prescription by \cite{2008MNRAS.387.1431D}, and a purely thermal AGN scheme following \cite{2009MNRAS.398...53B}. These two simulation schemes, with gas mass resolution $\sim 10^9$ M$_\odot$, consistently produce power-law entropy profile at low redshift. However, this behaviour drastically changes at higher resolution, with particle mass of $\sim 10^6$ M$_\odot$. The C-EAGLE simulations \citep{ceagle.barnes.2017, 2017MNRAS.470.4186B}, based on the EAGLE model of \cite{eagle.schaye.2015}, provide clear examples of the entropy-core problem. The clusters, comparable in mass to the REXCESS sample from \cite{entropy_profiles_pratt2010}, show median entropy profiles which are up to 4 times larger than REXCESS at $0.01\, r_{500}$. This discrepancy can also be identified in the temperature and density profiles, one being higher and the other lower than the observations (refer to the above definition of entropy). The ROMULUS-C simulation \citep{2019MNRAS.483.3336T} breaks this trend by producing power-law entropy profiles with small entropy cores at high resolution ($\sim 10^5$ M$_\odot$) and at low redshifts. They model {one single galaxy cluster} with metal cooling for low-temperature ($T<10^4$ K) gas {\citep[see Section 2.1 of][for further details]{2022MNRAS.515...22J}}. However, their run does not include radiative cooling from metals lines (i.e. gas with $T>10^4$ K), which is known to play a pivotal role in the shape of the thermodynamic profiles and the evolution of the IGM \citep[e.g.][]{2011MNRAS.412.1965M}. Despite potentially being able to form in any group or cluster, cooling flows can easily be disrupted by non-radiative (e.g. merging events and mixing) and non-gravitational processes (e.g. cooling, SN and AGN feedback), hence breaking the self-similarity of the entropy profiles \citep{vkb_2005, 2007MNRAS.376..497M, 2008MNRAS.391.1163P}.

The thermodynamic state of the hot IGM in the centres of groups and clusters is extremely sensitive to AGN feedback due to its proximity to the central SMBH. The energy output from feedback events can increase the temperature of the gas and decrease its density by generating outflows, with the overall effect of disrupting the CC state of the object {in most simulation set-ups. Instead, studies of the ROMULUS-C cluster system \citep{2021MNRAS.504.3922C} and the Rhapsody-G data \citep{2017MNRAS.470..166H} found that mergers can disrupt cool cores.} Simulations of isolated clusters, where the hot gas starts from a hydrostatic equilibrium state, show a widely variable entropy level in their inner region due to AGN activity \citep[e.g.][]{2019A&A...631A..60B, 2022MNRAS.tmp.1955N}. {Sub-grid prescriptions based on kinetic jets and cold accretion models have been shown to preserve the cool-core properties of idealised cluster simulations down to low redshift \citep{2015ApJ...811...73L, 2015ApJ...811..108P, 2016ApJ...829...90Y, 2023MNRAS.518.4622E}.} The simulations from \cite{2022MNRAS.tmp.1955N, 2022MNRAS.516.3750H} show how galaxy clusters in quasi-hydrostatic equilibrium can also preserve a cool core in the presence of anisotropic outflows and buoyantly rising bubbles produced by AGN feedback{, with the exception of the low-mass group ($M_{200} = 10^{13}$ M$_\odot$) in \citeauthor{2022MNRAS.tmp.1955N} (\citeyear{2022MNRAS.tmp.1955N}, Section 4.5 and Fig. 10), where the cool-core is disrupted and the central entropy rises above the 50 keV cm$^2$ level irreversibly. We note, however, that their models do not include cosmological accretion and merging events, which can destabilise pre-existing cool-cores \citep[e.g.][]{2021MNRAS.504.3922C}.} In turn, the bubbles can produce weak shocks which gently transport high-entropy gas outside the cluster core, as initially suggested by \cite{2001ApJ...554..261C} and \cite{2003MNRAS.344L..43F}. {Recent works modelling kinetic feedback jets with high resolution found that weak shocks can indeed heat the cluster atmosphere isotropically via sound-wave propagation \citep{2016ApJ...829...90Y, 2017ApJ...847..106L, 2019MNRAS.483.2465M, 2021MNRAS.506..488B, 2022MNRAS.516.3750H}.}

According to the estimates of \cite{entropy_profiles_pratt2010}, CC clusters account for about one third of the population, although it is worth highlighting that the bimodality in the central entropy and logarithmic slope could be a statistical fluctuation stemming from the limited sample size and complex selection criteria. Recent numerical simulations modelling hydrodynamics, AGN feedback and other sub-grid processes, on the other hand, produce very few to no CC clusters (C-EAGLE, \cite{ceagle.barnes.2017}; SIMBA, \cite{2019MNRAS.486.2827D}; TNG, \cite{2018MNRAS.473.4077P, 2018MNRAS.481.1809B}; FABLE, \cite{2018MNRAS.479.5385H}). The entropy in these simulated galaxy clusters does not decrease towards the centre, but it flattens to a constant level and gives rise to large entropy cores \citep[see also][for a review]{2021Univ....7..209O}. {The Rhapsody-G suite of cosmological zoom-simulations of clusters stand out as a sample which reproduce the observed CC-NCC dichotomy, although it should be noted their X-ray luminosity being too high \citep{2017MNRAS.470..166H}.} In this work, we investigate several processes which may give rise to the entropy cores in simulations, ultimately inhibiting the formation of CC clusters at low redshift. Our aim is to identify the causes of excess entropy in simulations to explain why galaxy formation models like EAGLE cannot reproduce the CC cluster population found by X-ray observations.

Since its initial formulation, the EAGLE model has played a pivotal role in interpreting observations and providing theoretical predictions. Yet, it fails to produce CC clusters at $z=0$ and power-law-like entropy profiles as measured by \cite{entropy_profiles_sun2009} and \cite{entropy_profiles_pratt2010}. The causes are unclear and may be related to the modelling of the hydrodynamics, the feedback implementation, the resolution or missing physical processes which are not included in the EAGLE model. In this work, we investigate the effects of thermal conduction, metal cooling, SN and AGN feedback to identify how they affect the entropy budget and inhibit the formation of cool cores in groups and clusters at EAGLE resolution and eight times lower mass resolution. We study the sensitivity of the group/cluster properties and the entropy profiles to changes in the model and we rule out possible causes for the lack of CC clusters in our simulation sample.

This work is organised as follows. In Section \ref{sec:simulation_methods} we introduce the sample of simulated objects, the initial conditions, as well as the simulation and analysis methods. There, we also describe the SWIFT-EAGLE model used in our investigation \citep[see also][]{bahe_2021_bh_repositioning, 2022MNRAS.tmp.1955N}, highlighting the key updates from the original EAGLE Ref simulations \citep{eagle.schaye.2015}; we then lay out the changes made to our reference model to explore different physical processes. Section \ref{sec:reference_model_results} highlights the entropy core problem by comparing the results from the simulated objects at $z=0$ against observed entropy profiles from groups and clusters of comparable mass. Section \ref{sec:results_model_variations} presents and discusses the main results for two objects in the simulated sample, run with different sub-grid models in order to investigate the effects on the system-wide properties and the 3D radial profiles. Finally, we discuss our results in Section \ref{sec:discussion} and present concluding remarks in Section \ref{sec:conclusion}.

Throughout this work, we assume the Planck 2018 cosmology, given by 
$\Omega_{\rm m}=0.3111$, 
$\Omega_{\rm b}=0.04897$, 
$\Omega_\Lambda=0.6889$, 
$h=0.6766$, 
$\sigma_8=0.8102$, 
$z_{\rm reion}=7.82$, 
$T_{\rm CMB}=2.7255$ K \citep{planck.2018.cosmology}.



\section{Simulations and numerical methods}
\label{sec:simulation_methods}
AGN feedback is known to affect low- and high-mass groups differently \citep[e.g.][]{2010MNRAS.406..822M, 2022MNRAS.tmp.1955N}. The energy output from the SMBH can lead to catastrophic outflows in the IGM of small groups, while large objects might just experience a re-distribution of the gas within their virial radius. In order to study the effects of AGN feedback in these regimes, we built an \textit{extended} sample of 27 objects ranging from group-sized to cluster-sized masses. In order to study the effect of sub-grid model changes in a more controlled and detailed setting, we further selected a \textit{reduced} sample of one low- and one high-mass object, referred to as \textit{group} and \textit{cluster} respectively in the following sections.

\subsection{Sample selection}
\label{sec:simulation_methods:sample_selection}

\begin{table*}
    \centering
    \caption{Mass resolution and softening lengths for different types of simulations. The mass of dark matter (DM) particles is expressed by $m_{\rm DM}$ and the initial mass of gas particles is $m_{\rm gas}$. $\epsilon_{\rm DM,c}$  and $\epsilon_{\rm gas,c}$ indicate the comoving Plummer-equivalent gravitational softening length for DM particles and the gas respectively, $\epsilon_{\rm DM,p}$ and $\epsilon_{\rm gas,p}$ indicate the physical maximum Plummer-equivalent gravitational softening length for DM particles and the gas respectively.}
    \begin{tabular}{llcccccc}
    \hline
    Set-up      & Simulation type   & $m_{\rm DM}$           & $m_{\rm gas}$ & $\epsilon_{\rm DM,c}$ & $\epsilon_{\rm gas,c}$ & $\epsilon_{\rm DM,p}$ & $\epsilon_{\rm gas,p}$ \\ 
                &                   & (M$_\odot$)            & (M$_\odot$)   & (comoving kpc)        & (comoving kpc)         & (physical kpc)        & (physical kpc)         \\ \hline
    Parent box  & DM-only           & $5.95 \times 10^{9}$   & --            & 26.6                  & --                     & 10.4                  & --                     \\
    Zoom (low-res)  & DM-only       & $9.32 \times 10^{7}$   & --            & 6.66                  & --                     & 3.80                  & --                     \\
    Zoom (high-res) & DM-only       & $1.17 \times 10^{7}$   & --            & 3.33                  & --                     & 1.90                  & --                     \\
    Zoom (low-res)  & DM+hydro  & $7.85 \times 10^{7}$   & $1.47 \times 10^{7}$ & 6.66           & 3.80                   & 2.96                  & 1.69                   \\
    Zoom (high-res) & DM+hydro  & $9.82 \times 10^{6}$   & $1.83 \times 10^{6}$ & 3.33           & 1.90                   & 1.48                  & 0.854                  \\
    \hline
    \label{tab:resolutions}
    \end{tabular}
\end{table*}

The objects studied in this paper were selected from a $(300~\textrm{Mpc})^3$ volume (referred to as the \textit{parent} simulation), run using the \swift hydrodynamic code \citep{2016pasc.conf....2S, schaller_2018_swift} in gravity-only mode and $564^3$ DM particles with mass $m_{\rm DM} = 5.9 \times 10^{9}$ M$_\odot$ (see Table \ref{tab:resolutions}). The \velociraptor structure-finder code \citep{2019PASA...36...21E} was then used to identify DM halos in the snapshots and build a catalogue of  halos at $z=0$. We then selected the halos within the mass range $12.9 \leq \log_{10}(M_{500}/{\rm M}_\odot) \leq 14.5$ which formed in relatively isolated regions of the volume. The isolated objects are defined such that no field halos larger than 10\% of their mass can be found within a 10 $r_{500}$ radius from their centre of potential. We then split the $10^{12.9 - 14.5}~{\rm M}_\odot$ mass interval into three mass bins equally spaced in log-scale. From each bin, 9 isolated halos were randomly selected, giving an overall population of 27 objects, which we denote as the \textit{extended} halo catalogue. Spanning a wide range in masses, this sample is particularly well-suited for studying scaling relations and outlining the entropy-core problem by comparing simulated and observed entropy profile in groups and clusters (see section \ref{sec:reference_model_results:entropy_profiles}).

From the extended halo catalogue, we then focused on two objects, one of high mass and one of low mass, for studying the entropy distribution and other quantities with varying sub-grid models. The high-mass object, which we identify as \textit{cluster}, is the largest DM halo in the extended sample and has a $z=0$ mass of $M_{500} = 2.92 \times 10^{14}~{\rm M}_\odot$. The low-mass object, called \textit{group} in later sections, has $M_{500} = 8.83 \times 10^{12}~{\rm M}_\odot$ at $z=0$ and was randomly selected from the halos in the lowest mass bin. This particular subset of the extended catalogue forms the \textit{reduced} catalogue.

\begin{figure*}
	\includegraphics[width=2\columnwidth]{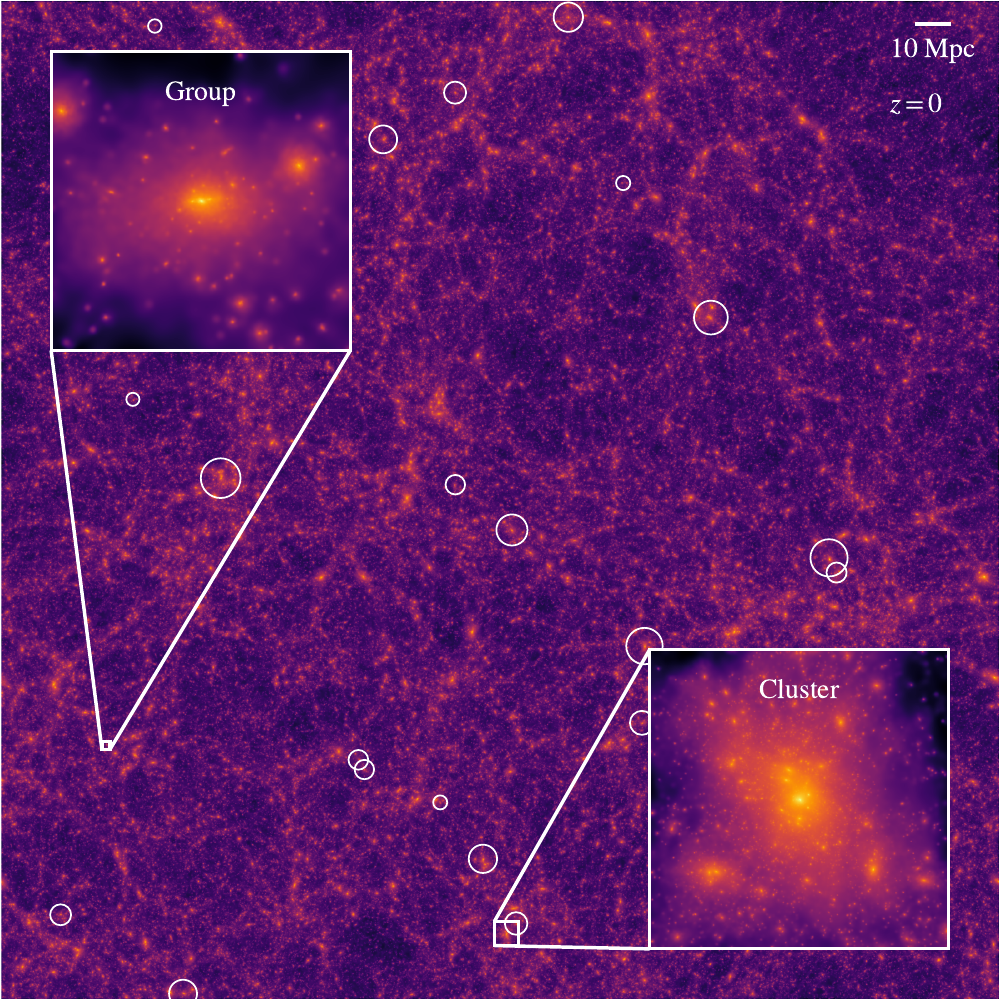}
    \caption{Projected DM mass map of the 300 Mpc volume at $z=0$, where the two insets show the objects in the reduced catalogue (group and cluster, as indicated) re-simulated in DM-only mode to $z=0$. The white circles mark the position of the remaining 25 objects in the extended catalogue; some are covered by the insets. The white circles represent the (projected) spherical 3D apertures, of radius $6\, r_{500}$, used to mask the high-resolution region in the zoom set-up. The size of the group and cluster maps is $12\, r_{500}$ along the diagonal. The spatial scale of the parent box is indicated in the top-right corner of the image.}
    \label{fig:dmo_box_with_halos}
\end{figure*}

Fig. \ref{fig:dmo_box_with_halos} shows a projected DM mass image of the parent volume, with the location of the objects in the extended catalogue (white circles) and the re-simulated objects in the reduced catalogue (square insets, zooming onto the group and the cluster). 

\subsection{Zoom initial conditions set-up}
\label{sec:simulation_methods:initial_conditions}

The process of simulating the objects in the extended catalogue is based on the zoom simulation technique \citep{1993ApJ...412..455K, 1997MNRAS.286..865T} and was developed in three stages: producing the parent simulation,  constructing the initial conditions with variable particle resolution, and re-simulating each halo at higher resolution, including hydrodynamics and sub-grid physics.
The initial conditions for both the parent simulation and the zoom set-ups were built from a set of DM particles arranged in a glass-like structure. Each particle was then displaced and assigned a velocity vector determined from first (zooms) and second (parent) order Lagrangian perturbation theory to $z=127$. The displacement and velocity fields are computed using the method of \cite{2010MNRAS.403.1859J} implemented in \textsc{Panphasia} \citep{2013MNRAS.434.2094J}, an open-source code which can generate realisations of primordial Gaussian random fields in a multi-scale setting\footnote{The phase descriptor containing the unique seed of the white noise field used for the parent simulation and the zoom simulations is $\rm [Panph1,L18,(74412,22732,260484),S3,CH1799108544,EAGLE-XL\_L0300\_VOL1]$}.

The objects in the extended sample were then re-simulated individually using the zoom-in technique. For each selected halo in the parent box at $z=0$, a spherical high-resolution region was centred on the DM halo's potential minimum and extended to $6~r_{500}$. The initial conditions, generated with the \textsc{IC\_Gen} tool and \textsc{Panphasia} \citep{2013MNRAS.434.2094J}, were produced at two resolutions. The high resolution was chosen to match that of the EAGLE L100N1504 volume \citep{eagle.schaye.2015}. In the DM-only configuration at high resolution, we used particles with mass $m_{\rm DM}=1.17 \times 10^{7}~{\rm M}_\odot$, while full-physics simulations have $m_{\rm DM}=9.82 \times 10^{6}~{\rm M}_\odot$ and initial gas particle mass $m_{\rm gas}=1.83 \times 10^{6}~{\rm M}_\odot$. An additional set of simulations was produced at a mass resolution 8 times lower than EAGLE, corresponding to (initial) particle masses that are 8 times larger. Table \ref{tab:resolutions} summarises the masses of DM and gas particles at both resolutions, in the case of DM-only simulations and full-physics runs. We also report the comoving and physical Plummer-equivalent gravitational softening length for DM and gas particles.

\subsection{The SWIFT-EAGLE subgrid model}
\label{sec:simulation_methods:ref_model}
In this study, we describe an updated version of the EAGLE sub-grid model of \cite{eagle.schaye.2015}, which we denote as SWIFT-EAGLE. This model, introduced in \cite{bahe_2021_bh_repositioning} and implemented in \swift, uses an equation of state based on a pressure floor to model the interstellar medium (\citealt{2008MNRAS.383.1210S}, using a polytrope index $\gamma=4/3$ as in Section 2.1 of \citealt{bahe_2021_bh_repositioning}), a {Kennicutt-Schmidt} star-formation law \citep{1959ApJ...129..243S, 1998ApJ...498..541K} and a metallicity-dependent star-formation number density threshold \citep{2004ApJ...609..667S, 2015MNRAS.450.1937C}. Stellar mass loss and chemical enrichment is implemented using the prescription of \cite{2009MNRAS.399..574W}. In this section, we outline the main changes that have been applied to the simulation technique since the original EAGLE simulations. Similarly to \cite{eagle.schaye.2015}, we define a fiducial sub-grid model, which we denote Ref, followed by additional ones, designed to investigate the effects of specific processes in the population of simulated clusters.

For this work, we use the SPHENIX Smoothed Particle Hydrodynamics (SPH) scheme described in \cite{sphenix_borrow2022} and configured with a quartic (M5) spline kernel. Unlike the SPH implementation ANARCHY, used in EAGLE, which uses the pressure–entropy formalism \citep[see][for a description]{eagle.schaye.2015, 2015MNRAS.454.2277S}, SPHENIX uses the traditional density-energy method, which proved to give more accurate results for the particle-particle forces and energy conservation than pressure-based schemes when coupled to astrophysical sub-grid models \citep{2021MNRAS.505.2316B}. In addition, SPHENIX makes use of an adaptive artificial viscosity (triggered in shocks using the Balsara switch, \citealt{1995JCoPh.121..357B} and \citealt{2010MNRAS.408..669C}) and artificial conduction, which is also present in ANARCHY, albeit with a different implementation. The artificial conduction is computed particle-wise and capped by the maximum artificial conduction coefficient $\alpha_{D, \rm max}\in [0, 1]$, which can be set between one, to fully enable conduction, and zero, to disable conduction completely \citep{2008JCoPh.22710040P}.

Our version of the EAGLE model includes the cooling tables presented by \cite{2020MNRAS.497.4857P}. This cooling model includes a modified version of the redshift-dependent ($z\in [0, 9]$) UV/X-ray background from \cite{2020MNRAS.493.1614F}, self-shielding for the cool interstellar medium (ISM) down to 10 K, dust, an interstellar radiation field and the effect of cosmic rays via the Kennicutt-Schmidt relation \citep{1998ApJ...498..541K}. When used with \swift, the radiative cooling for each particle is computed element-wise by interpolating the tables over redshift, temperature and hydrogen number density.

Starting from $z=19$, groups above a friends-of-friends (FoF) mass of $10^{10}$ M$_\odot$  are periodically identified and BHs with subgrid mass $M_{\rm sub} = 10^4$ M$_\odot$ are seeded within them, at the point where the local gas density is highest. Once seeded, the BHs can grow by either accreting gas or by merging with other BHs. The gas accretion rate can be estimated using the spherically symmetric Bondi-Hoyle-Lyttleton model \citep{1939PCPS...35..405H, 1944MNRAS.104..273B}, to which is applied an Eddington-limiter. Based on this prescription, the mass-accretion rate is given by
\begin{equation}
    \label{eq:bh_accretion}
    \dot{m}_{\rm BH} = \min \left[\alpha \cdot \frac{4 \pi G^2~m_{\rm BH}^2~\rho_{\rm gas}}{\left(c_{\rm s} + v_{\rm gas}^2\right)^{3/2}},~\dot{m}_{\rm Edd} \right],
\end{equation}
where $m_{\rm BH}$ is the BH subgrid mass, $c_{\rm s}$ is the gas sound speed, $v_{\rm gas}$ the bulk velocity of the gas in the kernel relative to the BH, and $\rho_{\rm gas}$ the gas density evaluated at the BH position. $\dot{m}_{\rm Edd}$ is the \cite{1926ics..book.....E} rate, computed as
\begin{equation}
\label{eq:bh_eddington}
    \dot{m}_{\rm Edd} = \frac{4 \pi G m_{\rm P}}{\epsilon_{\rm r}~c~\sigma_{\rm T}}\cdot m_{\rm BH} \approx 2.218~{\rm M_\odot~yr^{-1}}\cdot\left( \frac{m_{\rm BH}}{10^8~{\rm M_\odot}} \right),
\end{equation}
for a radiative efficiency $\epsilon_{\rm r} = 0.1$ \citep{1973A&A....24..337S}, where $c$ is the speed of light, $m_{\rm P}$ the mass of the proton and $\sigma_{\rm T}$ the Thomson cross-section \citep[see Section 2.2.4 in][for further details]{bahe_2021_bh_repositioning}. In Eq. \ref{eq:bh_accretion}, the definition of the boost factor $\alpha$ follows the \cite{2009MNRAS.398...53B} model:
\begin{equation}
    \alpha = \max\left[ \left( \frac{n_{\rm H}}{n_{\rm H}^\star} \right)^\beta,~1 \right],
\end{equation}
where $n_{\rm H}$  is the gas number density and $n_{\rm H}^\star=0.1~\mathrm{cm}^{-3}$ is a reference density marking the threshold for the onset of the cold phase in the ISM \citep{2004ApJ...609..667S}. As in EAGLE \citep{eagle.schaye.2015, 2015MNRAS.450.1937C}, we set $\alpha=1$ and the free parameter $\beta=0$ in all models. Similarly to the AGN feedback in the EAGLE-like model in \cite{2022MNRAS.tmp.1955N}, we do not include the \cite{2015MNRAS.454.1038R} angular momentum limiter. We note that our runs have $\approx$18 and 146 times higher gas {particle} mass; $\approx$3 and 6 times larger gravitational softening {than the simulations in \cite{2022MNRAS.tmp.1955N}} (see Table \ref{tab:resolutions}).

Over one timestep of length $\Delta t$, the mass involved in BH accretion is $\Delta m=\dot{m}_{\rm BH}~\Delta t$. Of this amount, a fraction $\epsilon_{\rm r}$ is converted into energy, while a fraction $(1-\epsilon_{\rm r})$ is added to the subgrid mass of the BH. To conserve the mass-energy budget in the simulations, the SMBHs can accrete a mass $\Delta m$ by "nibbling" mass from their neighbours. In this scenario, the gas particles in the kernel of the SMBH lose a small amount mass, weighted by their density contribution, to the BH accretion process \citep[refer to Eq. 4 in][]{bahe_2021_bh_repositioning}. 

{Unlike the original EAGLE model, where AGN feedback is stochastic, the AGN feedback in our Ref model uses a deterministic approach with an energy reservoir \citep{2009MNRAS.398...53B}.} Thermal AGN feedback is implemented by raising the temperature of one gas particle by $\Delta T_{\rm AGN}$, which is a (free) parameter fixed to $10^{8.5}$ K for the reference model. For every timestep, the subgrid reservoir of each SMBH is incremented by an energy $\epsilon_{\rm f}\epsilon_{\rm r} \Delta m c^2$, where $\epsilon_{\rm f}$ is the gas coupling efficiency, expressing the fraction of mass-energy radiated via BH accretion which contributes to AGN feedback. All our simulations use $\epsilon_{\rm f}=0.1$. A feedback event occurs when the reservoir contains enough energy to heat the designated gas neighbour by a temperature $\Delta T_{\rm AGN}$ \citep{2009MNRAS.398...53B, bahe_2021_bh_repositioning}. In the SWIFT-EAGLE Ref model, the target gas particle involved in thermal AGN feedback is chosen to be the closest to the SMBH (i.e. minimum-distance scheme, see Section \ref{sec:simulation_methods:model_variations}), however, we also investigate the effect of different feedback distribution methods in Section \ref{sec:results_model_variations}. Our AGN feedback implementation is purely thermal in all our simulations. 

Throughout our simulations, SMBHs are repositioned at every timestep according to the \texttt{Default} method of \cite{bahe_2021_bh_repositioning}, which aims to compensate for the inconsistencies of (unresolved) dynamical friction of SMBHs by artificially displacing them towards the minimum of the local gravitational potential. The repositioning algorithm has the effect of improving mass-accretion profiles of the SMBHs through time, as well as a self-consistent energy output from AGN feedback, compared to runs without BH repositioning.

The chemical enrichment from stellar populations follows \cite{eagle.schaye.2015} and models the effect of stellar winds from AGB and massive stars, core-collapse supernovae (SNII) and type Ia supernovae (SNIa) based on stellar ages and metallicity \citep{2009MNRAS.399..574W}. In our model, massive stars of mass $M>8$ M$_\odot$ release $10^{51}$ erg at the end of their evolution. This energy is made available for thermal SN feedback and it is modulated using a function of density and metallicity \citep{2015MNRAS.450.1937C} with a SNII energy fraction in the interval $[0.5, 5]$, as in \cite{bahe_2021_bh_repositioning}. The SN feedback implementation is stochastic and purely thermal with a fixed SN heating temperature $\Delta T_{\rm SN}=10^{7.5}$ K \citep{2012MNRAS.426..140D}. For both the SN and AGN feedback, we choose the \textit{minimum-distance} scheme as the default particle-selection method (Section \ref{sec:simulation_methods:model_variations}).

\subsection{Variations on the reference model}
\label{sec:simulation_methods:model_variations}

\begin{figure}
	\includegraphics[width=\columnwidth]{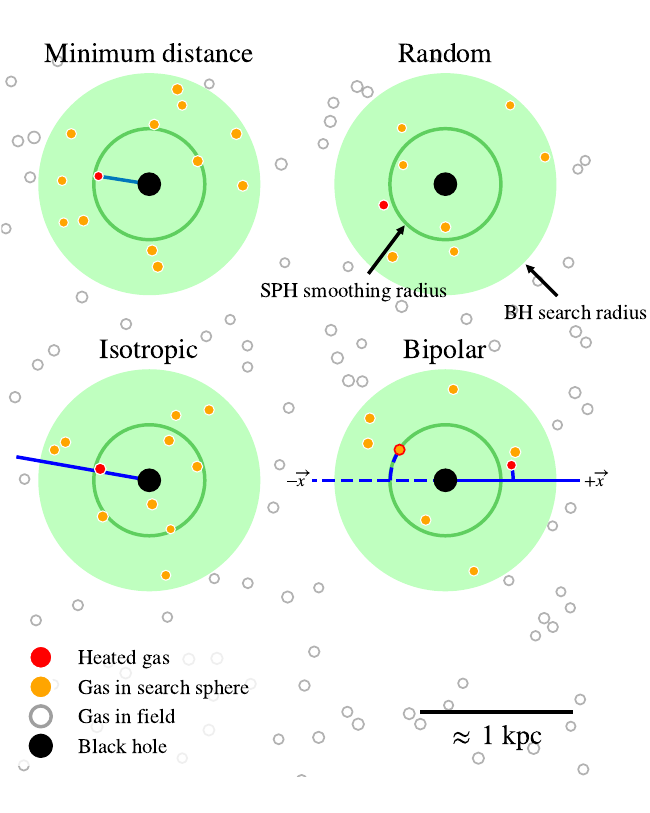}
    \caption{Schematic illustration of the rule used in the AGN feedback schemes for selecting the gas particle to heat (simplified to 2D). From the top-left to bottom-right, we illustrate the Minimum distance, Random, Isotropic and Bipolar schemes, as labelled in the figure. BHs are represented by black circles; their SPH smoothing radius is indicated by {empty} dark-green circles and the search radius by {filled} light-green circles{, around the smoothing length circle}. The gas particles are represented by empty grey circles if not found within the search radius of a BH, or filled orange circles otherwise. The gas particle selected by the heating mechanism is highlighted in red. For the Minimum distance method, we show the line connecting the BH with its nearest gas neighbour. For the Isotropic and Bipolar methods, we show the rays as solid blue lines cast from the BHs and the arc formed with the selected particle. For the Bipolar case, the orientation of the ray is randomly chosen between the $\pm \protect\overrightarrow{x}$ directions; for the positive case, we use solid lines for the ray and the arc, while we use dashed lines for the negative case. In the bottom right corner, we indicate the order of magnitude of the scale of the image.}
    \label{fig:heating_schemes}
\end{figure}

In addition to the minimum-distance method used in the Ref model summarised above, we implement three additional rules for selecting the neighbouring gas particle to be heated in SN and AGN feedback: Random, Isotropic and Bipolar. {To limit the number of scheme combinations used for SN and AGN feedback}, the energy-injection mechanism for SN feedback is the same as the one adopted for AGN feedback in all our models. {Any change in energy distribution scheme is applied to the SN and the AGN feedback in the same way.} We explain the particle selection rule using the diagram in Fig. \ref{fig:heating_schemes} \citep[see also][]{2022MNRAS.514..249C}. To produce this illustration, we generated a random distribution of gas particles (grey circles) and highlighted in orange those which are located within the search radius of the SMBH and are candidates in an AGN heating event. In all schemes, we define the SMBH \textit{search radius} to be approximately twice\footnote{For the quartic (M5) SPH kernel, we choose the search radius to be $2.018932\approx 2$ times the SPH smoothing length of the SMBH \citep[see][]{2010MNRAS.408..669C}.} its SPH smoothing length and we use it to evaluate the local gas properties. In Fig. \ref{fig:heating_schemes}, the search radius is represented by the dark-green circle around the BH. From the top-left to the bottom-right, we explain the particle selection methods as follows:
\begin{enumerate}
    \item \textit{Minimum distance}. The particle closest to the SMBH is selected and its temperature is raised by $\Delta T_{\rm AGN}$. We conventionally adopt this method in our reference model.
    
    \item \textit{Random}. One particle within the SMBH search radius is selected at random and its temperature is raised by $\Delta T_{\rm AGN}$. We note that this algorithm, also used in the original EAGLE model, results in mass-weighted sampling \cite{2022MNRAS.514..249C}.
    
    \item \textit{Isotropic}. A ray is cast from the BH in a random direction\footnote{The probability density distribution of the polar angles $(\theta, \phi)$ of the ray (centred on the BH) are uniform in $\cos\theta\in \mathopen[-1, 1\mathclose[$ and $\phi\in \mathopen[0, 2\pi\mathclose]$. In the limit of a large number of neighbours within the BH search radius and a large number of AGN feedback events, the energy is then injected isotropically around the BH.}. The gas particle that minimises the arc-length to this ray is selected and its temperature is raised by $\Delta T_{\rm AGN}$. To compute the arc-length between ray-particle pairs, we use the haversine formula reported in \cite{2022MNRAS.514..249C}. 
    
    \item \textit{Bipolar}. Same as Isotropic, but the ray is always cast along the $x$-axis of the simulation box, randomly flipping between the positive and negative directions. To decide on the sign of the $\protect\overrightarrow{x}$ direction, we draw a random float in the $\mathopen[0, 1\mathclose]$ interval and cast the ray in the $-\protect\overrightarrow{x}$ direction if the number is within $\mathopen[0, 0.5\mathclose]$ or choose $+\protect\overrightarrow{x}$ otherwise.
\end{enumerate}

 In this work, we also test the effects of the choice of $\Delta T_{\rm AGN}$ on the halo properties and the radial profiles. In addition to Ref, we introduce a model with stronger AGN feedback $\Delta T_{\rm AGN}=10^{9}$ K (AGNdT9) and one with weaker AGN feedback $\Delta T_{\rm AGN}=10^{8}$ K (AGNdT8). For models varying the heating scheme, namely Random, Isotropic and Bipolar, we adopt the fiducial $\Delta T_{\rm AGN}=10^{8.5}$ K. Crucially, different values of $\Delta T_{\rm AGN}$ do not alter the cumulative energy injected over the total simulation time. As a result, AGN feedback events are expected to occur more frequently in weaker-AGN models (e.g. AGNdT8), while stronger-AGN (e.g. AGNdT9) models lead to more "explosive" and sporadic gas-heating episodes. 

We note that the choice of a preferred AGN feedback scheme should not only be motivated by physical arguments. While the effective spatial resolution (Plummer-equivalent softening) of the simulations presented in this work is of order 1 kpc, the physical processes that govern the accretion of the SMBH in galaxies and, consequently, the exact dynamics of the relativistic jets and coupling with the local IGM occurs on sub-parsec scales, which are unresolved in our simulations. The aim of our research is to identify \textit{assumptions} on the energy-injection mechanism which can be implemented in a SMBH-based sub-grid model and lead to realistic halo properties and, crucially, are able to reproduce the observed entropy profiles.

In addition to the sub-grid models presented above, we also consider models without AGN or SN feedback, without artificial thermal conduction and without radiative cooling from metals. To switch off AGN feedback, we simply do not seed BHs at high redshift. SN feedback, instead, is switched off by imposing a zero heating temperature. Crucially, this method does not alter the metal enrichment of the gas, which is important for radiative cooling. In the SPH module, the artificial conduction is controlled by the $\alpha_{D, \mathrm{max}}$ parameter, which enters the equations of motion as a diffusive term \citep{sphenix_borrow2022}; in runs with no artificial conduction, we explicitly set $\alpha_{D, \mathrm{max}} = 0$, while we leave the default $\alpha_{D, \mathrm{max}} = 1$ for runs with artificial conduction.

In runs without metal cooling, we restrict the calculation of the cooling rates to hydrogen and helium and discard C, N, O, Ne, Mg, Si, S, Ca and Fe.

A summary of the runs used in this study is shown in Table \ref{tab:models}.

\subsection{Cluster properties}
\label{sec:analysis_methods}
The entropy is calculated by obtaining the number density of free electrons considering the chemical abundances from the chemical elements tracked by the sub-grid and hydrodynamics code. Assuming fully ionised gas, we define the total free-electron fraction for each hot ($T>10^5$ K) gas particle as
\begin{equation}
    X_{\rm e} \equiv \frac{n_{\rm e}}{n_{\rm H}} = \frac{m_{\rm H}}{f_{\rm H}}~\sum_\epsilon Z_\epsilon \frac{f_\epsilon}{m_\epsilon}
\end{equation}
and the ion fraction $X_i$
\begin{equation}
    X_{\rm i} \equiv \frac{n_{\rm i}}{n_{\rm H}} = \frac{m_{\rm H}}{f_{\rm H}}~\sum_\epsilon \frac{f_\epsilon}{m_\epsilon},
\end{equation}
which can be combined to compute the electron number density
\begin{equation}
    n_{\rm e} = \frac{X_{\rm e}}{X_{\rm e} + X_{\rm i}} \frac{\rho_g}{\mu~m_{\rm H}} = \rho_g ~ \sum_\epsilon Z_\epsilon \frac{f_\epsilon}{m_\epsilon},
\end{equation}
where $f_\epsilon$ is the gas particle mass fraction for chemical element $\epsilon = \left\{{\rm H, He, C, N, O, Ne, Mg, Si, S, Ca, Fe}\right\}$, $m_\epsilon$ the associated atomic mass and $Z_\epsilon$ the atomic number. $f_{\rm H}$ and $m_{\rm H}$ are the hydrogen mass fraction and atomic mass respectively. {Based on gas density-temperature phase-space plots for our groups and clusters, we found that the temperature cut at $T_{\rm cut}=10^5$ K is suitable for selecting the hot ionised gas and filtering out the colder, star-forming gas on the equation of state. The gas on the pressure floor and above the star formation density is neglected when calculating hot gas properties.} Finally, $\rho_g$ is the SPH density of the gas particle and $\mu$ its mean atomic weight, computed by summing the species contributions as follows:
\begin{equation}
    \mu = \left[ m_{\rm H} \sum_\epsilon \frac{f_\epsilon}{m_\epsilon} \cdot (Z_\epsilon + 1) \right]^{-1}.
\end{equation}
For our results, we present entropy, mass-weighted temperature and density profiles of the simulated groups and clusters. To compute the radial profiles, we consider 50 spherical shells centred on the centre of potential with a logarithmically increasing radius, spanning from $(0.01-2.5) \times r_{500}$. We then sum the contributions of the particles in each radial bin. 
For the $i^{\rm th}$ shell, we compute the density profile as
\begin{equation}
    \rho_{g,i} = \frac{\sum_j m_j}{V_i},
\end{equation}
where the sum is the total gas mass in shell $i$, divided by the volume $V_i$ of the shell. We also define the mass-weighted temperature $T_{{\rm MW}, i}$ as
\begin{equation}
    T_{{\rm MW}, i} = \frac{\sum_j m_j T_j}{\sum_j m_j}.
\end{equation}
The entropy profiles are then computed via the mass-weighted temperature profiles and the density profiles, as described in \cite{vikhlinin_2006_profile_slope}:
\begin{equation}
    K(r)=\frac{\mathrm{k_B}T(r)}{n_{\rm e}(r)^{2/3}}.
\end{equation}
The thermodynamic profiles are normalised to their self-similar values, appropriate for an atmosphere in hydrostatic equilibrium. We scale the density profiles to the critical density of the Universe
\begin{equation}
    \rho_{\rm crit}(z) = E^2(z) \frac{3 H_0^2}{8 \pi G},
\end{equation}
where $E^2(z)\equiv H^2(z) / H_0^2 =\Omega_{\rm m}(1+z)^3 + \Omega_\Lambda$. To select X-ray emitting gas, we only consider gas particles above a temperature of $10^5$ K. We normalise the temperature profiles to the characteristic temperature at $r_{500}$,
\begin{equation}
\label{eq:t500}
    k_\mathrm{B}T_{500}=\frac{G \bar{\mu} M_{500} m_{\rm H}}{2r_{500}},
\end{equation}
where $\bar{\mu} = 0.5954$ is the mean atomic weight for an ionized gas with primordial ($X = 0.76$, $Z = 0$) composition. Using the characteristic temperature, we also define the characteristic entropy:
\begin{equation}
    K_{500}=\frac{k_\mathrm{B}T_{500}}{\left(500 f_{\rm bary} \rho_{\rm crit}~/(\overline{\mu}_{\rm e} m_{\rm H})\right)^{2/3}},
\end{equation}
where $\overline{\mu}_{\rm e} = 1.14$ is the mean atomic weight per free electron and $f_{\rm bary}=0.157$ is the universal baryon fraction obtained by \cite{planck.2018.cosmology}.

\section{Reference model results}
\label{sec:reference_model_results}

As a first step, we introduce results for the extended catalogue at $z=0$, generated using the reference model only. We begin by illustrating the differences between the entropy profiles measured observationally and those from our simulations (Section \ref{sec:reference_model_results:entropy_profiles}), followed by a comparison between simulated and observed hot-gas fractions and star fractions (Section \ref{sec:reference_model_results:fractions}). The entropy profiles, gas and star fractions are derived from \textit{true} quantities, and do not attempt to include the hydrostatic mass bias.

\subsection{Entropy profiles}
\label{sec:reference_model_results:entropy_profiles}

\begin{figure*}
	\includegraphics[width=2\columnwidth]{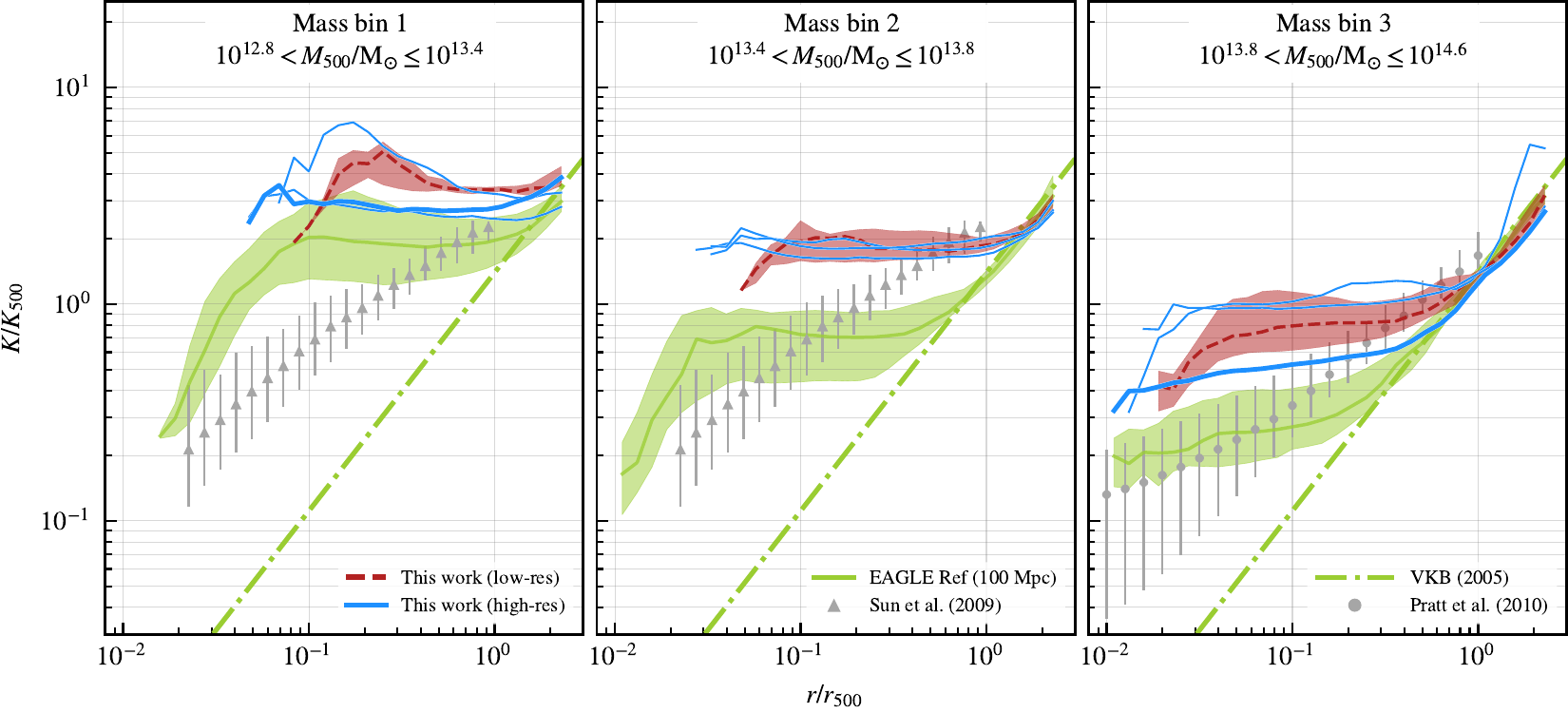}
    \caption{Comparison between the extended catalogue simulated with the Ref model (dark-red lines), the entropy profiles measured by \protect\cite{entropy_profiles_sun2009} and \protect\cite{entropy_profiles_pratt2010} using X-ray data (grey markers) and entropy profiles from the EAGLE 100 Mpc volume run with the Ref model \protect\citep[][green solid lines]{eagle.schaye.2015}. The objects in the extended sample are grouped by $M_{500}$ at $z=0$ in 3 bins, with mass ranges indicated by the label in each panel. For the objects in the extended sample, dashed red lines refer to the low-resolution simulations (combined to produce a median profile and the 16$^{\rm th}$ - 84$^{\rm th}$ percentile bands) and solid blue lines to high-resolution (plotted individually). The thicker blue lines highlight the high-res group and the high-res cluster (reduced sample) in the first and third mass bins respectively. The data from EAGLE and the observations are also reported with the percentile bands similarly to the extended sample. At small radii, the profiles are truncated where the number of particles in a 3D spherical shell falls below 50; this criterion is applied to the extended sample and the EAGLE objects. The entropy baseline profile from non-radiative simulations \protect\citep{vkb_2005} is displayed in all panels as a dot-dashed green line.}
    \label{fig:median_entropy_profiles_extended_sample}
\end{figure*}

In Fig. \ref{fig:median_entropy_profiles_extended_sample}, we compare the entropy profiles of the simulated objects at $z=0$ with observational results from \cite{entropy_profiles_sun2009} and \cite{entropy_profiles_pratt2010} over a wide range in $M_{500}$, spanning between $8.83 \times 10^{12}$ M$_\odot$ and $2.92 \times 10^{14}$ M$_\odot$. We gathered the groups and clusters in the extended sample in three mass bins, containing 7, 9 and 11 objects run with low resolution and 3 objects each run with high resolution. For the extended sample at low resolution, we computed the median entropy profile and the associated 16$^{\rm th}$ and 84$^{\rm th}$ percentile bands in each mass bin (dark red dashed lines). The entropy profiles for the high-res simulations are shown individually in blue; the group and the cluster in the reduced sample are displayed as thicker lines in the first and third mass bin respectively. In the first two bins, we show the median entropy profile from the \cite{entropy_profiles_sun2009} sample (spectroscopic\footnote{Unlike the true quantities, computed from the unprocessed simulation data, \textit{spectroscopic} quantities are estimated assuming hydrostatic equilibrium and modelling X-ray emission. Spectroscopic masses estimated from simulations are often compared to masses measured from X-ray observations. We refer to \cite{2021MNRAS.506.2533B} and references therein for further details.} $M_{500} = 1.37\times10^{13}-1.37\times10^{14}$ M$_\odot$, $0.012 <z<0.12$); the error bars span between the 10$^{\rm th}$ and 90$^{\rm th}$ percentile level. For the third mass bin, we indicate the median entropy profiles from \cite{entropy_profiles_pratt2010}, computed using only the objects with estimated $M_{500}$ in the mass range of the simulations; the error bars span between the 16$^{\rm th}$ and 84$^{\rm th}$ percentile. {The \cite{entropy_profiles_pratt2010} data set shown throughout this work includes all REXCESS clusters in the mass range, regardless of their CC/NCC classification or morphology.} The left panel of Fig. \ref{fig:median_entropy_profiles_extended_sample} focuses on group-sized objects, comparable in mass with those in the sample of \cite{entropy_profiles_sun2009}. The right panel illustrates the simulated profiles of the clusters, together with the median profiles of the objects included in the study by \cite{entropy_profiles_pratt2010}. The median profile from the objects in the intermediate mass is only compared to the results from \cite{entropy_profiles_sun2009}, since their sample included more objects within this mass range than \cite{entropy_profiles_pratt2010}. We also indicate the entropy baseline predicted by \citeauthor{vkb_2005} (\citeyear{vkb_2005}, VKB) for non-radiative simulations. We also compare our entropy profiles of the extended sample with those from the EAGLE 100 Mpc simulation run with the Ref model \citep{eagle.schaye.2015, 2015MNRAS.450.1937C, 2016A&C....15...72M}. For each panel in Fig. \ref{fig:median_entropy_profiles_extended_sample}, the green lines indicate the median entropy profiles from the groups and clusters in the EAGLE volume, binned by $M_{500}$ similarly to the extended sample. The entropy profiles from the EAGLE volume were produced from the $z=0$ snapshot; the EAGLE data and those from the extended sample were reduced using the same analysis pipeline. In both data sets, we truncate the inner region of the entropy profiles at the radius where the number of particles in the shell falls below 50.

The radial profiles for the extended sample show excess entropy compared to observations across the three mass bins. The discrepancy is smallest around $r_{500}$ and increases towards the inner region of the groups and clusters. Besides an entropy excess in the core, the objects in the first and second mass bins show a flat entropy distribution, in contrast with the power-law-like profiles observed by \cite{entropy_profiles_sun2009}. For the group, we also highlight a local peak in the entropy at $\approx 0.2~r_{500}$ in the first mass bin, which is likely produced by feedback processes coupling to the diffuse gas within these groups of galaxies. The low-mass clusters in the third bin also show excess entropy in the core, compared to the \cite{entropy_profiles_pratt2010} sample, e.g. at $0.1~r_{500}$, the low-res simulations predict a dimensionless entropy of 0.8, while the observed median is 0.35. Similarly to the lower-mass bins, the shape of the simulated cluster profiles does not match the observations. Interestingly, the radial profiles from the groups and clusters in the EAGLE 100 Mpc volume have a lower entropy level than the extended sample, despite using the same feedback heating temperature. {The difference in central entropy may be due to the use of different hydrodynamic solvers (SPHENIX in our models and ANARCHY in EAGLE Ref), which can slightly alter the properties of the IGM \citep{2015MNRAS.454.2277S}, such as the entropy}. However, similarly to the extended sample, the EAGLE median profiles also show a flat-entropy core, which disagrees with the observational data. These results from Fig. \ref{fig:median_entropy_profiles_extended_sample} illustrate the effects of the entropy-core problem in simulations of groups and clusters of galaxies, highlighting the excess in entropy in the central IGM and an unusually flat entropy distribution across radii \citep[see also][for a review]{2021Univ....7..209O}. {We also found great similarity between the entropy profiles at $z=0$ and at $z=1$, suggesting that the entropy plateau was likely established early in the evolution of these groups and clusters.}

\subsection{Hot gas and star fractions}
\label{sec:reference_model_results:fractions}

\begin{figure*}
	\includegraphics[width=2\columnwidth]{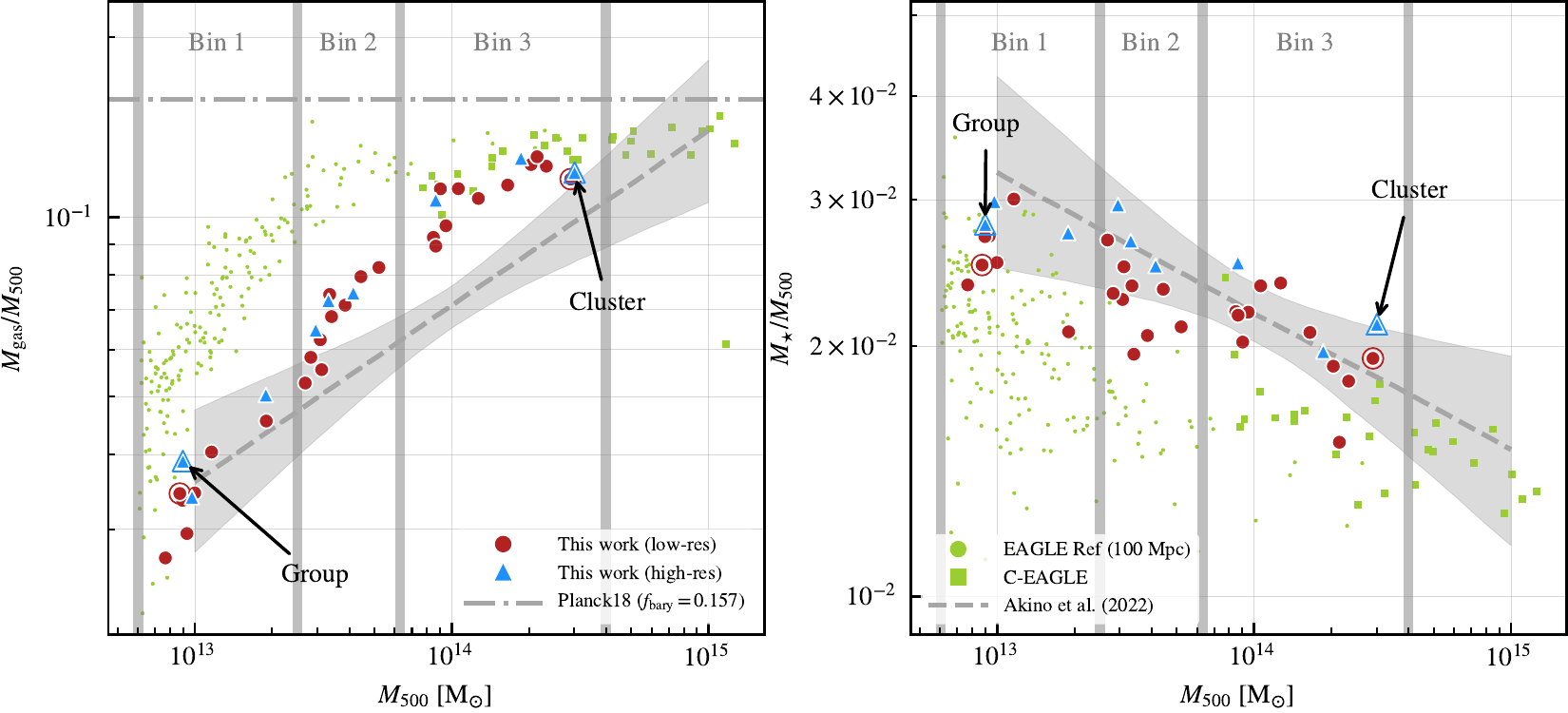}
    \caption{Scaling relations showing the gas fraction $M_{\rm gas}/M_{500}$ (left) and stellar mass fraction $M_{\star}/M_{500}$ (right) as a function of $M_{500}$ at $z=0$. In both panels, the dark-red circles and blue triangles represent the (extended) sample of objects presented in this work at low and high-resolution respectively. Double-edged markers highlight the two objects in the reduced sample (see annotations). In the left panel, the grey horizontal dash-dotted line indicates the universal baryon fraction $f_{\rm bary}=\Omega_{\rm b}/\Omega_{\rm m}=0.157$ derived from \protect\cite{planck.2018.cosmology}. The green markers are the objects in the EAGLE Ref 100 Mpc volume (dots) from \protect\cite{eagle.schaye.2015}, and the C-EAGLE objects (squares) from \protect\cite{ceagle.barnes.2017, 2017MNRAS.470.4186B}. In both panels, the dashed grey lines are the fitted scaling relations obtained by \protect\cite{xxl.baryons.akino2022} for the gas and stellar mass fractions, with the 16$^{\rm th}$ and 84$^{\rm th}$ percentile levels from the errors in the fit parameters, shown by the grey area. We emphasise that the grey area does not represent the intrinsic scatter, which instead extends for about 1 dex along the $x$-axis and 0.2 dex along the $y$-axis. In the right panel, the observations include the group and cluster sample from \protect\cite{xxl.baryons.akino2022}. The three mass bins specified in Fig. \ref{fig:median_entropy_profiles_extended_sample} are also indicated.}
    \label{fig:scaling_relations_extended_sample}
\end{figure*}

For the extended sample, we also compute the hot gas fraction $f_{\rm gas} = M_{\rm gas}/M_{500}$, where $M_{\rm gas}$ defines the mass of the gas with temperature $T > 10^5$ K within $r_{500}$. Similarly, we define the star fraction as $f_\star = m_\star/M_{500}$, where $m_\star$ is the stellar mass within $r_{500}$. These results are used to construct the $f_{\rm gas}$-$M_{500}$ and $f_\star$-$M_{500}$ scaling relations shown in Fig. \ref{fig:scaling_relations_extended_sample}. 

The $f_{\rm gas}$-$M_{500}$ relation for the extended sample (dark-red circles for low-resolution and blue triangles for high-resolution) is compared to the results from the HSC-XXL weak lensing survey \cite[][136 objects with $0\leq z \leq 1$]{xxl.baryons.akino2022}. The weak lensing method of the HSC-XXL survey introduces a lower mass bias relative to X-ray measurements, motivating the use of true masses for our analysis \citep[e.g.][]{2012MNRAS.421.1073B}. Our sample of simulated objects shows good convergence at the two resolutions. When compared to the scaling relation from weak lensing, our sample shows larger $f_{\rm gas}$ for $M_{500}\sim 10^{14}$ M$_\odot$ clusters (we note, however, that the scaling relation from \cite{xxl.baryons.akino2022} was obtained by averaging a {binned} sample of 136 objects{; without binning, the individual data points from their sample} would otherwise show a much larger scatter at the $10^{14}$ M$_\odot$ level). From group-sized objects towards large clusters, the hot gas fraction approaches the universal baryon fraction measured by \cite{planck.2018.cosmology}, $f_{\rm bary}=\Omega_{\rm m}/\Omega_{\rm b}=0.157$. This is a well-known consequence of feedback processes being more efficient at expelling hot gas from low-mass groups than in large clusters, which can retain most of the baryons throughout their formation history \citep{2010MNRAS.406..822M}. 

In this comparison, we also include the groups and low-mass clusters from the EAGLE Ref 100 Mpc simulation \cite{eagle.schaye.2015} and the C-EAGLE clusters \citep{ceagle.barnes.2017}, both displayed as green markers. For groups ($10^{13}<M_{500}/{\rm M_\odot}<10^{14}$) the EAGLE Ref model produces systematically higher gas fractions than our reference model; this discrepancy is largest in the second mass bin, where our runs predict $f_{\rm gas}\approx0.07$, while EAGLE produces values around 0.12. The cause of this effect is unclear, as multiple changes have been applied to the simulation code (gravity and hydrodynamic solver) and the subgrid model \citep[see Appendix B of][for a comparison between the star-formation rate history in EAGLE and SWIFT-EAGLE]{bahe_2021_bh_repositioning}. Despite the gas fractions still being systematically higher than the observations, the $M_{500}$-$f_{\rm gas}$ relation produced with our objects is closer to the \cite{xxl.baryons.akino2022} data than EAGLE for groups. In the third mass bin, our results agree with those from the EAGLE and C-EAGLE samples, with the former showing a slightly lower $f_{\rm gas}$ than the latter. While the reference model used in this work uses the same AGN heating temperature as the EAGLE Ref model, the C-EAGLE sample was run with a higher value (equivalent to the AGNdT9 model described in \citealt{eagle.schaye.2015}) than EAGLE Ref, introduced to expel more hot gas from the clusters.
The high-resolution simulations, indicated with blue triangles in the $M_{500}$-$f_{\rm gas}$ relation, show good numerical convergence with the low-res objects. The group in the reduced sample aligns with the parametrised relation from the HSC-XXL survey, while the cluster is slightly more gas rich than the observed median relation ($f_{\rm gas}=0.1$ at $M_{500}=2.9\times10^{14}~{\rm M_\odot}$).

The star fractions are also compared to the observational results by \cite{xxl.baryons.akino2022}. In this case, the extended sample is in excellent agreement with the parametrised scaling relation from the HSC-XXL survey. The high-resolution simulations systematically yield a slightly higher stellar mass than their low-resolution version. 
%
In summary, the original EAGLE model produces a poor representation of gas and stellar properties, while showing a closer agreement between the entropy profiles and the observations. Our version of the same model yields instead more realistic gas and stellar properties, but much larger entropy cores. We stress that our SWIFT-EAGLE model has not been fully calibrated. {The stellar mass function, galaxy sizes and black hole-stellar mass relations were reasonably matched to observations, however this agreement was not tested for a simulated cosmological volumes as large as our parent simulation (300 Mpc).} Moreover, this model does not reproduce the original EAGLE galaxy properties presented in \cite{eagle.schaye.2015} and \cite{2015MNRAS.450.1937C}. We refer to \cite{2022JOSS....7.4240K} for an illustration of the sub-grid calibration methodology and to Borrow and EAGLE-XL Collaboration, in preparation, for a detailed discussion of the galaxy properties with the SWIFT-EAGLE model.

\begin{table*}
\centering
\caption{List of the models used for simulating the objects in the reduced catalogue. In this work, each model is identified with a label summarising its key features. These include the distribution scheme used in the AGN heating mode, the temperature increase $\Delta T_{\rm AGN}$ of the gas particles when heated by an AGN outburst, whether the SNe or the AGNs are enabled in \swift, whether the metals are included in the radiative cooling calculation, and the $\alpha_{D,\mathrm{max}}$ parameter, which is set to 1 if the SPH scheme uses artificial thermal conduction and to 0 otherwise.}
\begin{tabular}{llccccc}
\hline
Model label            & AGN distribution mode & $\log_{10} (\Delta T_{\rm AGN}/\mathrm{K})$ & SN active & AGN active & Metal cooling enabled & $\alpha_{D,\mathrm{max}}$ \\ \hline
Ref                      & Minimum distance & 8.5     & Yes         & Yes          & Yes                     & 1          \\
No-conduction            & Minimum distance & 8.5     & Yes         & Yes          & Yes                     & 0          \\
No-metals                & Minimum distance & 8.5     & Yes         & Yes          & No                      & 1          \\
No-SN                    & Minimum distance & 8.5     & No          & Yes          & Yes                     & 1          \\
No-AGN                   & Minimum distance & 8.5     & Yes         & No           & Yes                     & 1          \\
AGNdT8                   & Minimum distance & 8.0     & Yes         & Yes          & Yes                     & 1          \\
AGNdT9                   & Minimum distance & 9.0     & Yes         & Yes          & Yes                     & 1          \\
Random                   & Random           & 8.5     & Yes         & Yes          & Yes                     & 1          \\
Isotropic                & Isotropic        & 8.5     & Yes         & Yes          & Yes                     & 1          \\
Bipolar                  & Bipolar          & 8.5     & Yes         & Yes          & Yes                     & 1          \\
\hline
\label{tab:models}
\end{tabular}
\end{table*}

\section{Sub-grid model variations}
\label{sec:results_model_variations}

\begin{figure*}
	\includegraphics[width=2\columnwidth]{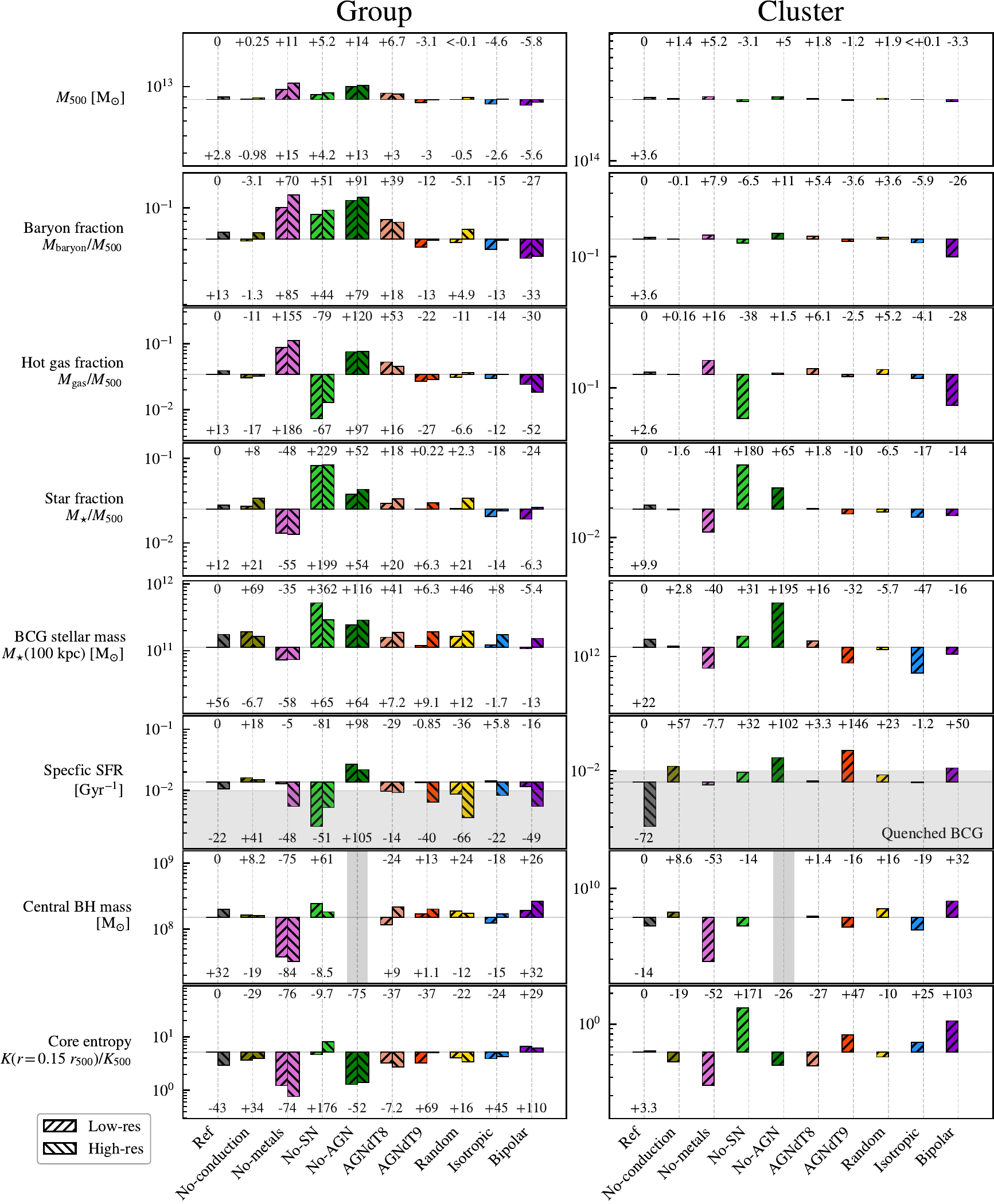}
    \caption{Summary of the $z=0$ properties of the group (left panels) and the cluster (right panels) in the reduced catalogue. For each panel, the quantity of interest is represented on the vertical axis, while the horizontal axis spans the models listed in Table \ref{tab:models}. For both objects, the results at low resolution (1/8$^{\rm th}$ EAGLE) and high resolution (EAGLE) are represented by hatched bars with forward and backward slashes respectively. The layout highlights the relative differences compared to the Ref model at low resolution, with the relative change, in per cent, for the low and high resolution annotated at the top and the bottom of each panel respectively. For the No-AGN model, the central BH mass is not defined, as indicated by the excluded grey area.}
    \label{fig:z0properties}
\end{figure*}

Given these results, we now investigate the sensitivity of the hot gas distribution to changes in the sub-grid model. The sub-grid parameters which are changed between different models are summarised in Table \ref{tab:models}. In Fig. \ref{fig:z0properties}, we show the variation of $M_{500}$, $f_{\rm gas}$ and $f_\star$, the stellar mass of the BCG within a 100 kpc spherical aperture $M_\star(\rm 100 kpc)$, the specific SFR\footnote{We average the SFR over 1 Gyr using the birth scale factor of stars in a 100 kpc spherical aperture.} of the BCG within 100 kpc and the mass of the central BH, $M_{\rm BH}$, all computed at $z=0$ for the reduced sample, i.e. 1 group and 1 cluster. In addition, we also report the entropy in the core, obtained from interpolating the radial profiles and normalised to $K_{500}$. The sub-grid models are distributed along the horizontal axis and listed at the bottom of the figure. Each row of plots focuses on one of the quantities mentioned above; the panels on the left show the properties of the group, whilst those on the right refer to the cluster. In order to highlight the relative differences in the quantities with changing sub-grid model and mass resolution, we fixed the low-resolution Ref model at the centre of the vertical plot range, where we show a grey solid line to guide the eye. The variation in the quantities for each model and resolution can be seen by inspecting the height of the bars from the horizontal grey line. The bars are grouped in pairs, with the one on the left-hand side representing the low resolution run and the one on the right-hand side the high-resolution run. At the top (bottom) of each panel, we report the percentile change of the quantities for the low (high) resolution simulations, relative to the low (high)-res Ref runs. The No-AGN simulations do not seed BHs in the centres of galaxies, and therefore the central BH mass is not defined for this model, as represented by the crossed out region in the bottom two panels.

We complement our discussion by presenting the 3-dimensional mass-weighted radial profiles of the density, mass-weighted temperature and entropy profiles for the hot ICM. These are included in Figs. \ref{fig:profiles_noconduction} - \ref{fig:profiles_schemes}. For each figure, the top three panels show the entropy, temperature and density profiles (from left to right) of the group, while the bottom panels show those for the cluster, in the same order. The sub-grid variations use the same colour scheme as in Fig. \ref{fig:z0properties}. The profiles for low- and high-resolution runs are illustrated using solid and dashed lines respectively. In all panels, the horizontal axis indicates the radius scaled by $r_{500}$ and the vertical axes show the dimensionless entropy $K/K_{500}$, the scaled temperature $T/T_{500}$, and the normalised density $\rho/\rho_{\rm crit} (r/r_{500})^2$. In all profiles, we use the same axis ranges in order to facilitate a visual comparison between the different figures.

In the rest of this section, we describe the overall variation in the halo mass across different models and resolutions, before detailing the results from models with varying AGN heating temperature (Ref, AGNdT8 and AGNdT9), AGN energy-distribution scheme (Ref, Random and Isotropic), the effects of directional heating (Isotropic vs Bipolar), artificial conduction (Ref vs No-conduction), the contribution from individual feedback processes (Ref, No-AGN and No-SN) and the effect of metal cooling (Ref vs No-metals).

For most models, we do not find significant differences in $M_{500}$, with the exception of the group run with AGNdT8, No-AGN, No-SN and No-metals due to their larger baryon fractions. The variation in sub-grid model is, in fact, expected to have minimal influence on the halo mass at $z=0$ (although effects are still significant for precision cluster cosmology, e.g. \citealt{2021MNRAS.505..593D}), since the gravitational potential is dominated by dark matter, particularly at $r_{500}$. This claim finds confirmation in Fig. \ref{fig:z0properties} for both the group and the cluster, with the latter exhibiting even smaller differences in $M_{500}$ due to its deeper potential well. We note that the same effect also applies to the quantities that are directly linked to hydrodynamics, such as the hot gas mass and hot gas fraction. Most models produce objects with well-converged $M_{500}$ at both resolutions. We find this not to be true for the No-metals model, which generated a 4\% discrepancy in the group's mass at different resolutions. We discuss the effect of changes in the models on the other properties in the sections below.

\subsection{Effect of artificial conduction}
\label{sec:results_model_variations:conduction}

Switching off the artificial conduction has small effects on the hot gas fractions. The change in star fraction compared to Ref is $\sim$1\% for the cluster, however, we report a 8\% increase in the group at low resolution, and a 21\% increase at high resolution. Overall, the baryon fractions of both objects run with Ref and No-conduction are very similar.

When switching off artificial conduction, we report larger differences in the BCG stellar mass, which is 69\% higher in the group at low resolution. The same object run without conduction at high resolution, however, forms a BCG 7\% less massive than with Ref. The mass of the central BH is not strongly impacted by the change in $\alpha_{D, {\rm max}}$ in either of the objects in the reduced sample. The same is true for the sSFR in the group's BCG, while the cluster's BCG shows a 18\% increase in sSFR, which makes it non-quenched at $z=0$.

\begin{figure*}
	\includegraphics[width=2\columnwidth]{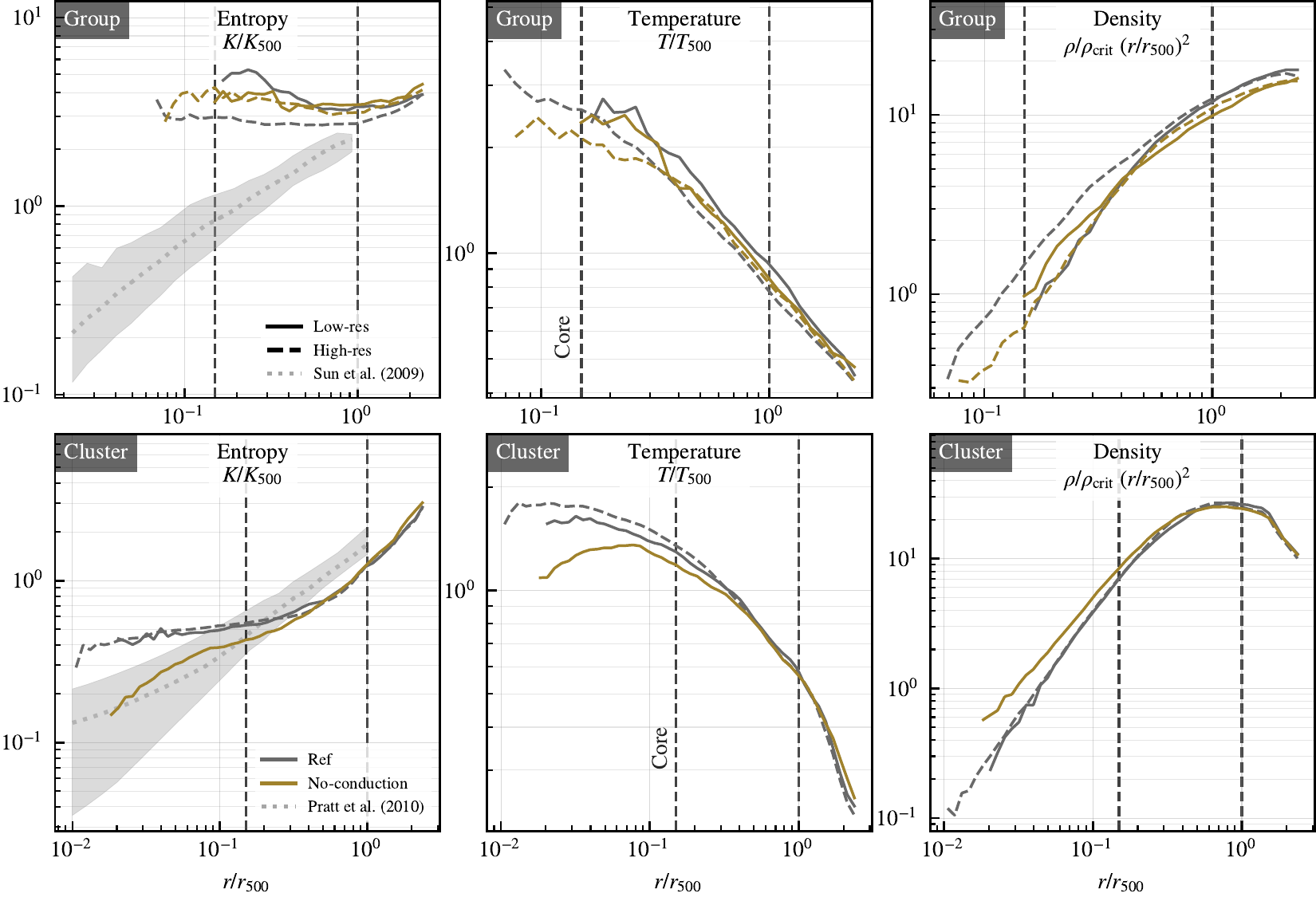}
    \caption{Radial profiles at $z=0$ for models with (Ref) and without (No-conduction) artificial conduction. \textit{Top row:} results from the group at both resolutions; \textit{bottom row:} results from the cluster at low resolution (and high resolution for the reference model). The panels on the left show the entropy profiles, normalised to $K_{500}$, the self-similar entropy scale. The central panels show the temperature profiles, scaled by $T_{500}$, and the right panels show the hot gas density profile in dimensionless units. We use 25 log-spaced bins from $0.01~r_{500}$ out to $2.5~r_{500}$. For the inner part, the profiles are truncated where the number of hot gas particles in the spherical shell drops below 50. The vertical dashed lines indicate $0.15~r_{500}$ (which we use to define the core as in \citealt{entropy_profiles_pratt2010}) and $r_{500}$.}
    \label{fig:profiles_noconduction}
\end{figure*}

As shown in Fig. \ref{fig:profiles_noconduction}, artificial conduction does not impact the shape of the group's entropy profile, which remains flat at both resolutions. For the cluster, switching off artificial conduction lowers the entropy inside the core radius compared to the reference model. Artificial conduction causes more gas to mix and prevents low-entropy gas from sinking towards the centre of the halo. In the No-conduction model, instead, low-entropy gas does not mix with high-entropy gas and can collapse into denser structures, as can be seen in the cluster's density profile.

The No-conduction model also produces cooler cores in both the group and the cluster than in the Ref model. This result is compatible with the $\alpha_{D, {\rm max}}=0$ model allowing the inner IGM to collapse further and to form denser cores. By increasing the density, the gas loses more energy due to radiative cooling, which can be observed in Fig. \ref{fig:profiles_noconduction}.

{\cite{2015ApJ...813L..17R} also produced simulations with and without artificial conduction to probe its effect on the entropy profiles of their clusters.} The cluster run without artificial conduction does not form a cool core by $z=0$, unlike what \cite{2015ApJ...813L..17R} found for their "cluster-2" object with similar mass. Their simulations, however, are run at about 10 times lower resolution than EAGLE and use the SPH scheme by \cite{2016MNRAS.455.2110B}. These differences could have an impact on the mixing of high- and low-entropy gas. We emphasise that these SPH schemes require artificial conduction to correctly reproduce hydrodynamic instabilities and the No-conduction model gives an unphysical representation of the simulated objects \citep[e.g.][]{sphenix_borrow2022}. This comparison is nevertheless useful to show that artificial conduction cannot dictate the formation of a power-law-like entropy profile.

\subsection{Effect of metal cooling}

{When metal line cooling is included in the simulation model, the cooling rate of the gas increases, and so does the amount of material that cools down and exits the low-density phase, characteristic of the intra-cluster medium. Moreover, this process favours the formation of a multi-phase IGM, because it removes some of the gas from the hot phase ($T > 10^7$ K, near the virial temperature), and allows that gas to cool down to the warm and cool gas phases ($T\sim 10^5$ K and below, \citealt{2007ApJS..168..213G, 2013MNRAS.434.1043O}). The regulation of the amount of gas in the cold phase available to form stars, and potentially the cold gas fuelling SMBHs, is therefore tightly linked to the cooling of the metals.}

Without metal cooling, heavy elements radiate energy away on longer cooling timescales and fewer gas particles become cool and dense enough to form stars \citep[e.g.][]{2011MNRAS.412.1965M}. Correspondingly, Fig. \ref{fig:z0properties} confirms this claim by showing a lower $r_{500}$ star fraction (and BCG mass), as well as more hot gas in the No-metals runs. No-metals also produces a larger baryon fraction, which can be interpreted as the net effect of the reduced amount of stars and the consequent abundance of non-star-forming gas \citep{2021Univ....7..209O}. The baryon fraction is therefore a reliable metric for assessing the impact of feedback \citep[e.g.][]{2010MNRAS.406..822M}, which appears to be weaker in the No-metals runs. This effect could be due to the suppressed star formation at high redshift, which leads to weaker stellar winds and, more importantly, less SN feedback. Compared to Ref, the central BH also grows to a smaller mass and the overall AGN feedback is consequently weakened. These effects have a stronger impact on the group than on the cluster, since the former has a shallower gravitational potential and therefore is more susceptible to the amount of energy injected into the IGM.

\begin{figure*}
	\includegraphics[width=2\columnwidth]{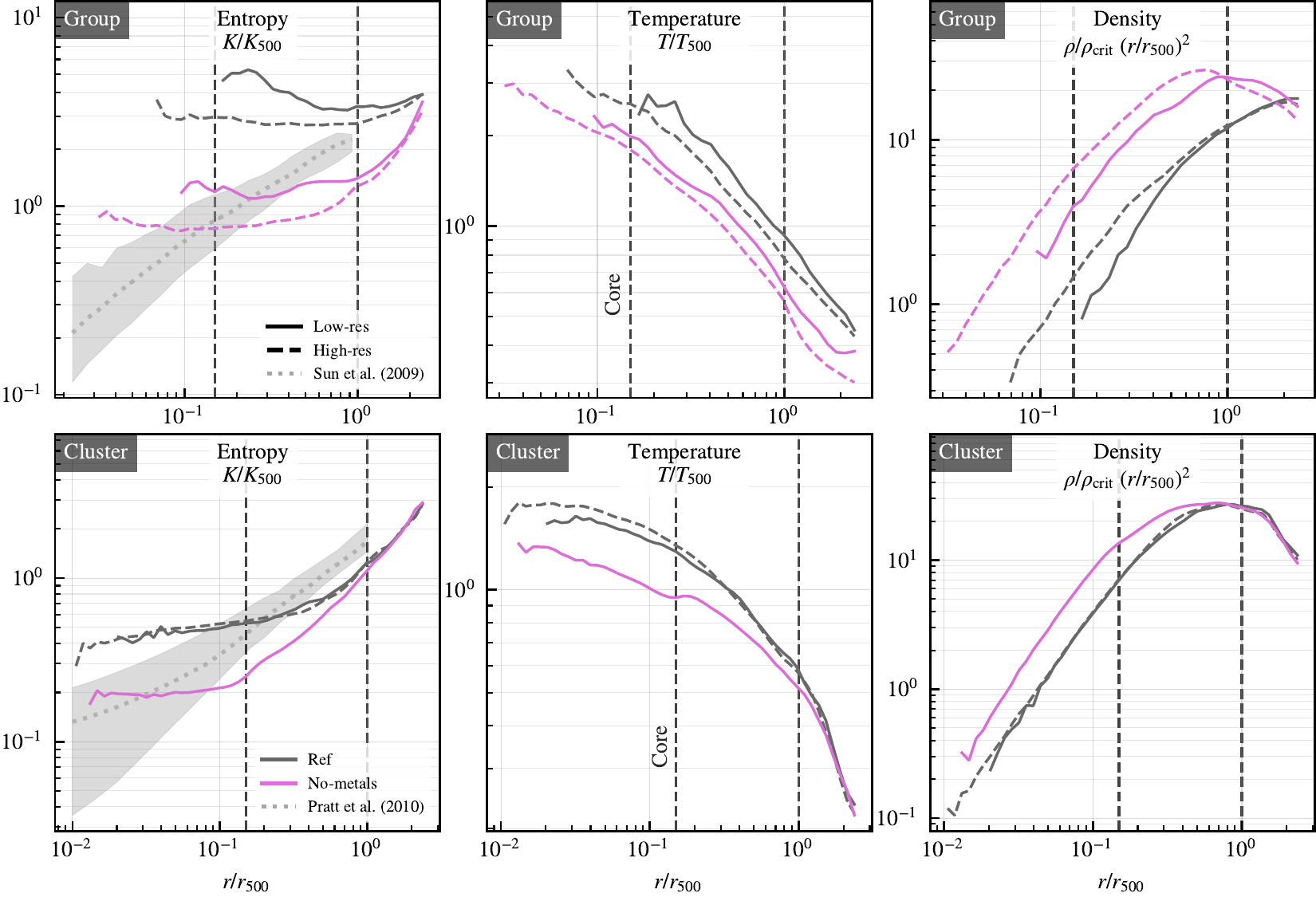}
    \caption{As in Fig. \ref{fig:profiles_noconduction}, but comparing models with and without metal cooling.}
    \label{fig:profiles_nometals}
\end{figure*}

The absence of metal-line cooling also has important consequences for the profiles, as we show in Fig. \ref{fig:profiles_nometals}. The reduced effect of thermal feedback and the suppressed star formation allow the formation of cooler and denser cores, which in turn lead to lower entropy. While the entropy level in the core is reduced, the central plateau persists and prevents the formation of a power-law entropy profile in the central region. Simulated at a mass resolution of $2.1 \times 10^5$ M$_\odot$ (i.e. $\sim 10$ times higher resolution than our high-res set-up) without metal cooling, the ROMULUS-C cluster does form a cool core and a plateau-less entropy profile \citep{2019MNRAS.483.3336T}, {and maintains it until $z=0.3$, when the onset of a merger forms an entropy core \citep{2021MNRAS.504.3922C}}. Our sub-grid set-up differs from that of ROMULUS-C in several ways, however, it seems that, without metal cooling, the cluster run with the EAGLE model still cannot form a stable cool core. {Our results also agree with those of \cite{2011MNRAS.417.1853D}, where a suite Adaptive Mesh Refinement (AMR) simulations of a galaxy cluster (with similar mass to ours) showed a smaller entropy core without metal cooling (\texttt{AGNHEATrun}) and a larger core at the $K\approx 200$ keV cm$^2$ level when metal cooling was included (\texttt{ZAGNHEATrun})}. As in the runs without artificial conduction, the No-metals model is unphysical and cannot produce the correct global properties of clusters, groups, and especially galaxies. This comparison shows that, despite the moderate reduction in the central entropy level, switching off metal cooling simply does not lead to a cool-core scenario in our simulated objects. 

\subsection{Models without feedback}

\begin{figure*}
	\includegraphics[width=2\columnwidth]{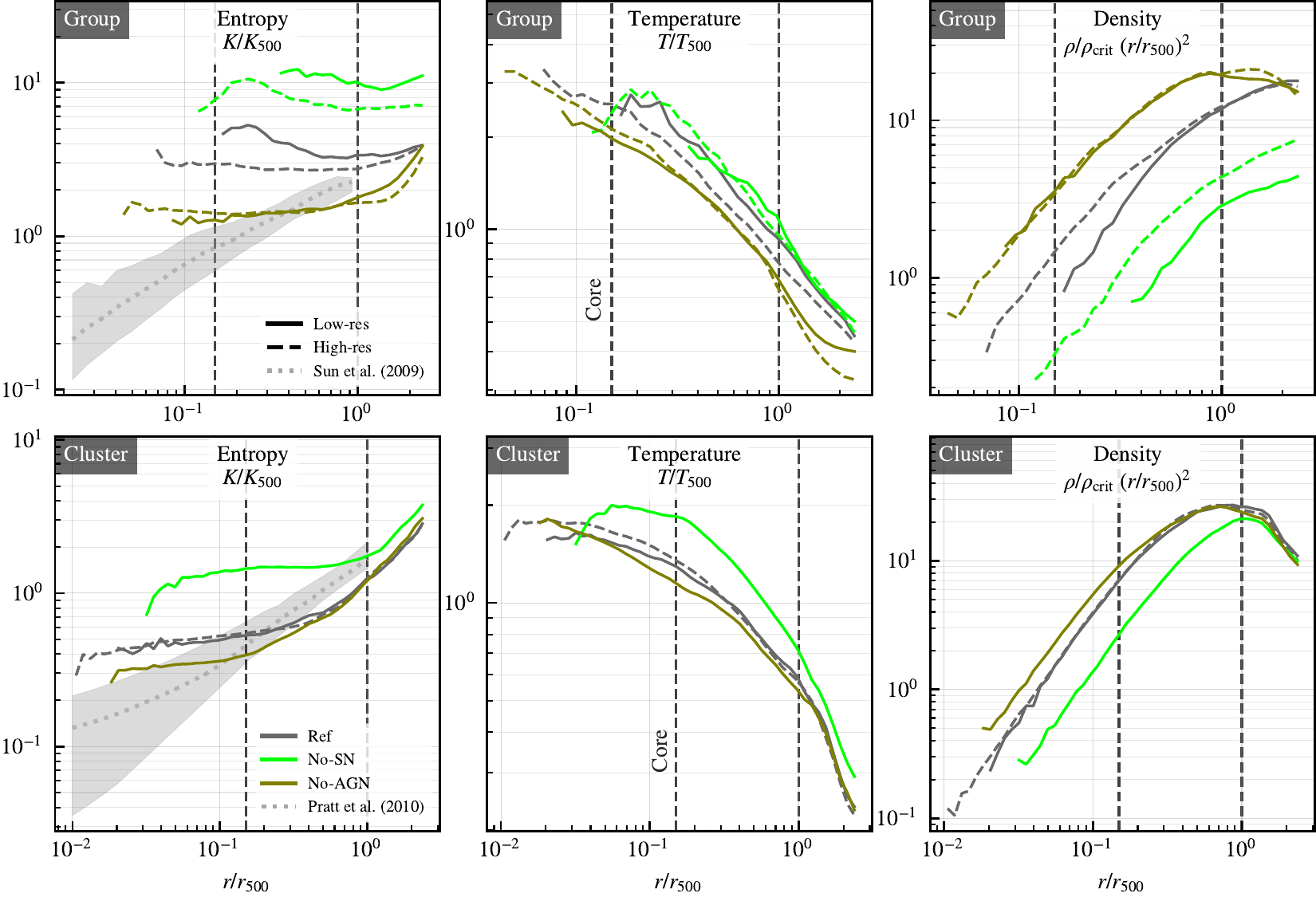}
    \caption{As Fig. \ref{fig:profiles_noconduction}, but including models with AGN or SN feedback switched off.}
    \label{fig:profiles_nofeedback}
\end{figure*}

We now compare the Ref model, which includes both SN and AGN feedback processes, with runs where they are switched off one at a time. We present these results by considering the properties in Fig. \ref{fig:z0properties} and the profiles in Fig. \ref{fig:profiles_nofeedback}. In both the No-SN and No-AGN group runs, we measure a 50-90\% larger baryon fractions than in Ref, which are qualitatively consistent with what is expected from weaker feedback. The cluster's total baryon budget is less susceptible to these changes and only shows per-cent level differences at low resolution \citep[see also][]{2017MNRAS.465...32B}. The absence of one feedback channel, however, has major consequences on the baryon cycle, i.e. on the hot gas and star fraction, as we discuss below.

All our No-SN runs show two peculiarities: a 80-70\% lower hot gas fraction, and a doubled star fraction. We begin by discussing the first effect. The missing hot gas could not have \textit{only} been expelled by AGN feedback, {since the central SMBH has grown to a similar mass and could not have produced AGN feedback outflows strong enough to justify such a large depletion of hot gas. Instead}, as the higher baryon {(and star)} fraction suggests, {the missing hot gas} must have formed a larger amount of stars than Ref. As for the second effect, SN feedback regulates star formation by preventing gas from cooling down and collapsing. When no SNe are present, the collapse of the IGM on small scales is unregulated and the objects experience enhanced star formation, which leads to large stellar masses at $z=0$ and up to three times more massive BCGs. The absence of SNe also affects the central BH. In the group, the BH grows 10-60\% more massive than in Ref. 
We did not find evidence of this mechanism in the cluster, whose BH grows slightly less in the No-SN run than in Ref. 

The entropy profiles for the No-SN runs, shown in  Fig. \ref{fig:profiles_nofeedback}, are higher than in the Ref runs. While the shape of the profiles remains flat as in Ref, the overall normalisation is affected, resulting in upward-shifted profiles. A change in entropy normalisation is, in fact, expected to take place when a large amount of cool and low-entropy gas becomes star-forming and is removed from the core \citep[see Section 4.1 of][]{2002ApJ...576..601V}, as our density profiles suggest. The remaining hot and non-star-forming gas also has a higher mass-weighted temperature. Fig. \ref{fig:z0properties} also shows that SN feedback is tightly connected with the BCG star formation history and can affect its quenching time. The specific SFR (sSFR) is used to define whether a BCG is quenched at $z=0$; using the threshold adopted by \citeauthor{2012MNRAS.424..232W} (\citeyear{2012MNRAS.424..232W}, Section 3.1), we define a BCG with sSFR $<10^{-2}$ Gyr$^{-1}$ quenched, and non-quenched otherwise. The BCG in the No-SN group shows a lower sSFR than Ref, suggesting that the intense star-burst epoch that converted large amounts of gas into stars could have taken place earlier than in Ref, resulting in a quenched BCG populated by old stars.
Motivated by the $\sim$ 1 dex increase in entropy and the $\sim$ 1 dex decrease in density profiles compared to Ref, we checked that the SWIFT code, coupled with the EAGLE model, models the star formation accurately even for such an extreme set-up. We re-ran the group at EAGLE resolution {and the cluster at 8 times lower resolution} with a reduced SN heating temperature of $10^{7}$ K to study a scenario where $\Delta T_{\rm SN} < 10^{7.5}~{\rm K}$ and cooling losses are more prevalent than in Ref. As expected, we obtained profiles with intermediate values between those in Ref and No-SN, suggesting that the star-formation processes were captured as expected. {We show the profiles for these additional runs, labelled as SNdT7, in Appendix \ref{appx:sn-heating-temperature}.}

In the No-SN cluster run, however, the central BH was found outside the BCG, likely due to the repositioning algorithm. In the rare cases where the SMBH, and not the host galaxy, dominates the local gravitational potential, the repositioning stops being effective and the SMBH is no longer shifted towards the centre of the galaxy \citep[see][for a discussion of this behaviour]{bahe_2021_bh_repositioning}. This effect is nevertheless unimportant in our analysis. In fact, the BH in the No-SN cluster appears to have reached the expected mass at the end of its rapid growth phase \citep{2018MNRAS.481.3118M}, and delivered most of its feedback \textit{before} the onset of the undesired repositioning. Since BH accretion rates become rapidly suppressed at low redshift, we estimate that the BH mass could only have been minimally affected by the off-centering and therefore does not impact our results and conclusions.

In the simulations where BHs are removed and no AGN feedback is present, the hot gas fraction of the group doubles compared to Ref (the cluster is largely unaffected), and the star fraction/BCG mass are also greater by 50-100\%. {We note that the absence of SMBHs also avoids some gas being accreted and removed from the local environment. Nevertheless, the accreted gas would be negligible compared to the total gas content and, therefore, it does not affect our analysis.} Unlike the No-SN model, a smaller amount of gas cools and becomes star-forming because the SN feedback is now able to regulate this process. On the other hand, less hot gas was expelled via thermal feedback due to the missing AGN channel (the baryon fraction is higher than in Ref), allowing more gas to remain within $r_{500}$ and, while being regulated by SN feedback, form stars and add to the stellar budget of the system, in agreement with \cite{2012MNRAS.422.2816B}. Due to SN feedback, the No-AGN runs do not show the signature of a runaway star formation as in the No-SN model, and produce cores with lower entropy (see Fig. \ref{fig:profiles_nofeedback}). The absence of AGN feedback, once again, does not lead to power-law-like entropy profiles; the entropy plateaus still persist in all No-AGN runs, although their entropy level (i.e. normalisation) is lower than in Ref. The role of AGN feedback in stopping star formation and quenching the BCG is commonplace in current studies of galaxy formation \citep[see e.g.][]{2013MNRAS.428.2885D} and the No-AGN runs illustrate a scenario where, instead, this process is absent, i.e. unphysical. In Fig. \ref{fig:z0properties}, we show that the No-AGN runs consistently produce non-quenched BCGs, where significant star formation is still ongoing at $z=0$.

Switching off one feedback channel completely (SN or AGN in turn) does not transform the core of our simulated objects from non-cool to cool. While the presence of feedback is causally linked to a change in entropy, the shape of the profile, as well as the entropy plateaus, do not appear to be set by the thermal feedback in the model we consider. We stress that the reduced sample may only contain objects whose accretion history does not admit the formation of a cool core, however, we have shown here evidence that feedback cannot be solely responsible for the formation of a power-law-like entropy profile in our simulations. 

\subsection{AGN heating temperature}
\begin{figure*}
	\includegraphics[width=2\columnwidth]{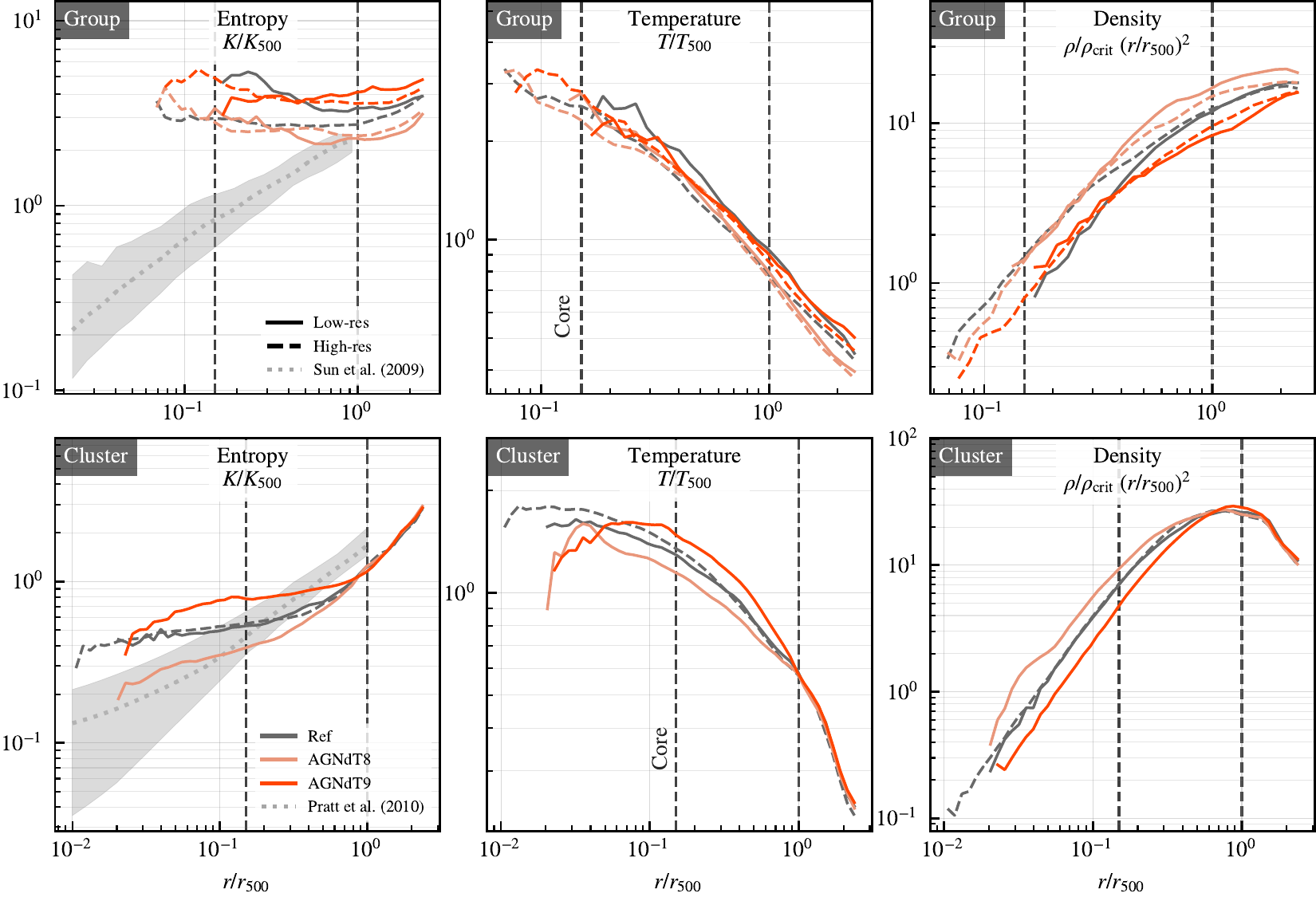}
    \caption{As in Fig. \ref{fig:profiles_noconduction}, but comparing models with varying AGN heating temperature.}
    \label{fig:profiles_agndt}
\end{figure*}

To study the effects of the AGN heating temperature on the cluster properties, we compare runs performed using EAGLE's fiducial value of $10^{8.5}$ K, with a lower AGN heating temperature, $10^{8}$ K in AGNdT8, and a higher one, $10^{9}$ K in AGNdT9. As shown in Fig. \ref{fig:z0properties}, we find that AGNdT8 produces objects with larger baryon fraction than Ref, as expected from a more continuous energy injection consisting of more numerous but gentler heating events. A lower hot gas fraction is measured in the AGNdT9 runs, which use fewer and more energetic events. Similar results were found for the (lower-resolution) Cosmo-OWLS models \citep{2014MNRAS.441.1270L}. Moreover, varying the AGN heating temperature results in the baryon fraction changing in the same way for both the group and the cluster. Our calculations show that higher AGN heating temperature leads to more baryons being expelled from the system. The hot gas fraction follows the same behaviour as the baryon fraction in both objects. The star fractions and the BCG masses indicate that the star formation is well regulated in both AGNdT8 and AGNdT9, with values broadly consistent with Ref.

 We find that none of the models considered in this section produce a quenched group at low resolution. The high-resolution simulations show a consistently lower sSFR in the group and produce a quenched BCG in the group run with AGNdT8. For the group at high resolution, AGNdT8 yield an sSFR 29\% lower than the Ref version and the AGNdT9 an 1\% lower sSFR. For the cluster, we register a 3\% higher sSFR for the AGNdT8 and a 146\% higher sSFR in the AGNdT9, in contrast with what was found by \cite{2010MNRAS.401.1670F}. Previous simulation studies have confirmed AGN activity is the main cause of the quenching of the BCG at low redshifts \citep{2010MNRAS.406..822M, 2012MNRAS.420.2859M}. While this effect can be measured for an ensemble of objects, individual ones may exhibit discrepant behaviour, such as in our case.

The mass of the BH at the centre of the BCG increases by $\approx$20\%, with the high-resolution group hosting a 30-40\% more massive BH than the low-resolution counterparts. We note that the BH mass is largely determined by its accretion history. This variation can be attributed to different accretion events occurring at different times in the formation history of the host galaxy.

As anticipated in section \ref{sec:reference_model_results:entropy_profiles}, the group shows entropy profiles that are relatively flat regardless of $\Delta T_{\rm AGN}$ (Fig. \ref{fig:profiles_agndt}), but with a mild change in normalisation, especially for AGNdT8 (a factor of $\approx 2$ lower than Ref). This effect can be explained by the lower amount of energy injected into the IGM per AGN feedback event (note that the \textit{cumulative} energy injected by AGNs throughout cosmic time remains similar). Crucially, we demonstrate that the Ref and AGNdT9 profiles are very similar and that the minimum-distance AGN model does not produce a significant entropy excess when increasing $\Delta T_{\rm AGN}$ by 0.5 dex. More appreciable differences in the entropy profiles can be seen in the cluster. All models produce consistent profiles beyond $r_{500}$, however, AGNdT8 generates lower entropy in the core of this object compared to Ref (by a factor of $\approx 2$). The AGNdT9 model, on the other hand, predicts an $\approx$ 80\% higher entropy level than Ref in the core. At small radii, the AGNdT9 profile for the cluster turns downwards, resembling the shape of a power law. 


The dependence of the hot gas fraction on the AGN heating temperature can also be inferred from the density profiles in Fig. \ref{fig:profiles_agndt}. In the two right panels, the AGNdT8 (orange) yields a density profile that is always higher than in Ref, while the AGNdT9 profile is the lowest. This result confirms that AGNs are responsible for the removal of hot gas from the centre of the objects. The group is affected by AGN feedback out to $2r_{500}$, visible as an offset between the AGNdT8, Ref and AGNdT9 models. The cluster, on the other hand, has very similar profiles around $r_{500}$ and only shows differences in the inner region due to the feedback outflows. The cooling flow in the cluster run with AGNdT8 contributes to the lowering of the entropy level in the core, by allowing high-density, low-temperature gas to sink towards the centre of the halo. 

By comparing runs with different AGN heating temperatures, we conclude that the entropy core cannot be removed by simply varying $\Delta T_{\rm AGN}$. The AGN heating temperature impacts the overall normalisation of the entropy profiles in the core, but does not modify their shape from flat to a power-law. While AGNdT8 seems to produce a steeper slope of the cluster entropy profile which approaches the slope measured by \cite{entropy_profiles_pratt2010}, we emphasise that lowering $\Delta T_{\rm AGN}$ too much reduces the amount of gas expelled from the system and produces objects with unrealistically high gas (and baryon) fractions (see the discussion in \citealt{eagle.schaye.2015}, and the motivation for using AGNdT9 in the C-EAGLE simulations in \citealt{ceagle.barnes.2017}).

\subsection{AGN {and SN} feedback distribution schemes}
\begin{figure*}
	\includegraphics[width=2\columnwidth]{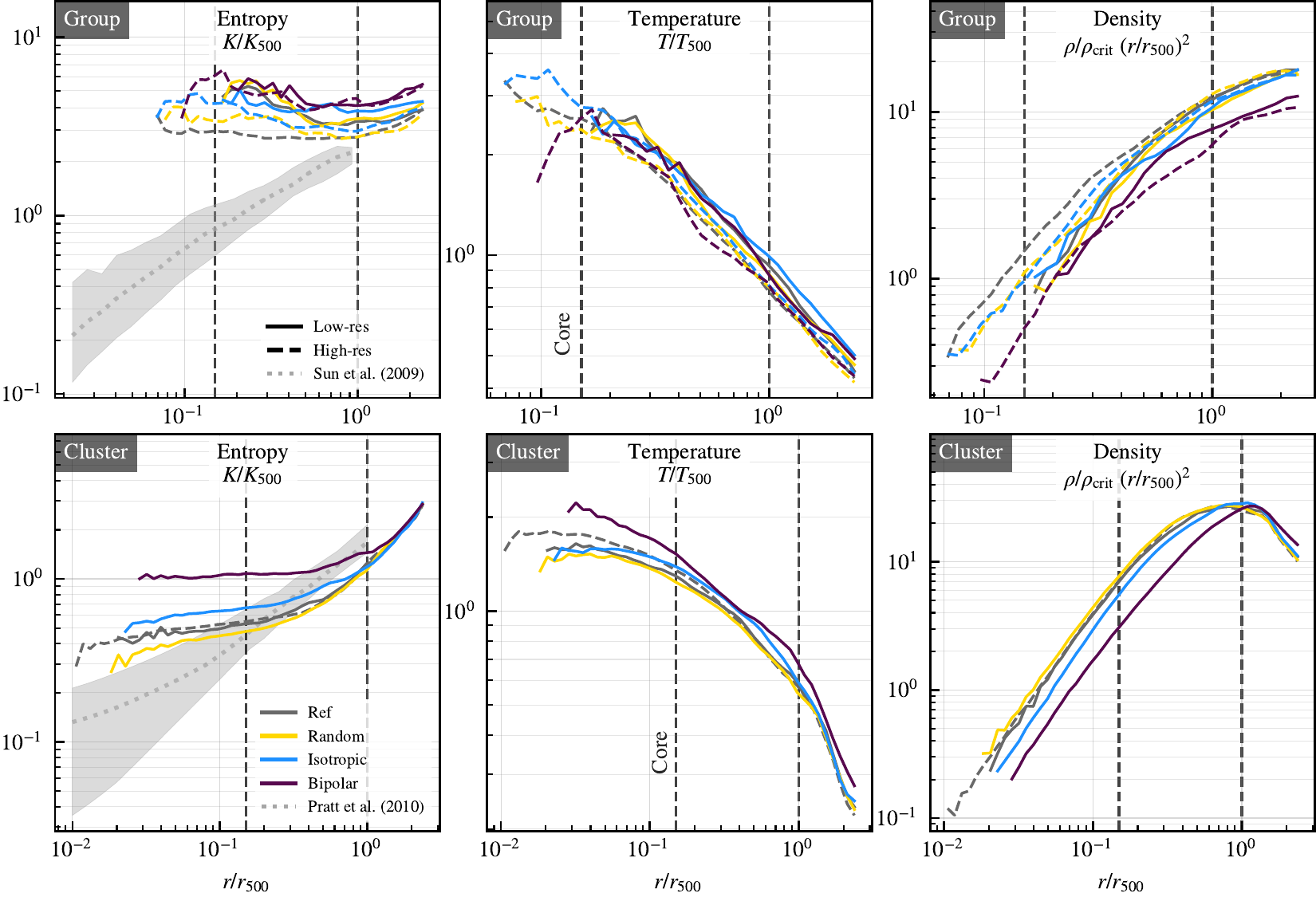}
    \caption{As in Fig. \ref{fig:profiles_noconduction}, but with varying AGN feedback distribution schemes (minimum distance, random, isotropic and bipolar).}
    \label{fig:profiles_schemes}
\end{figure*}

Finally, we compare runs with different AGN energy distribution schemes: Minimum-Distance (i.e. Ref), Isotropic, Random (used in the EAGLE Ref model) and Bipolar.
We first discuss the properties of the group and cluster with reference to Fig. \ref{fig:z0properties} focusing on the grey, yellow and blue bars, which correspond to the first three models mentioned above. Isotropic and Random, configured with fixed $\Delta T_{\rm AGN}=10^{8.5}$ K, produce objects with baryon fractions comparable to Ref and show minimal differences in the hot gas and star fractions ($\sim$ 10\%) at both resolutions. Similarly, the choice of heating scheme does not seem to impact other properties significantly. An exception involves the Isotropic cluster run, which produced a BCG 47\% less massive than Ref. Despite the slightly lower baryon fraction, this difference is unlikely to be the result of slightly more efficient AGN feedback. {The central SMBH is less massive than Ref and, consequently, it could not have produced stronger outflows. However,} SN feedback could potentially be responsible for this effect. {The thermal energy injected by SNe into the IGM is distributed to the gas on small scales. This process has two effects: raising the temperature of the gas locally and producing pressure gradients which raise the velocity dispersion. Therefore, when SN feedback is more effective, we expect that star formation is delayed, leading to a less massive BCG.}

At constant  $\Delta T_{\rm AGN}=10^{8.5}$ K, the entropy profiles shown in Fig. \ref{fig:profiles_schemes} for the group and the cluster are insensitive to the choice of the feedback distribution scheme. This similarity is evident between the Ref and the Random model, which only differ in the inner region around $0.02\, r_{500}$. This result differs from what found by \cite{2022MNRAS.514..249C}. In their study, the Random (mass-weighted) distribution scheme produced less efficient feedback than the Minimum-distance scheme and led to significant differences in the properties of the IGM. The entropy profile produced by the Isotropic model is instead consistently higher within $r_{500}$. This effect is best observed in the cluster, although it also affects the group to a smaller extent. Crucially, the overall shape of the inner entropy profiles remains flat regardless of the the energy distribution scheme.


The results shown in Fig. \ref{fig:profiles_schemes} show that the choice of AGN distribution scheme does not have a large impact on the shape or the normalisation of the profiles for both objects at both resolutions. We note that the rule for selecting the target gas particles to be heated is a purely numerical choice, without any direct physical implication. Albeit still debated, the actual heating mechanism coupling the BHs and the gas may occur on much smaller scales{, yet unresolved. Other studies have implemented feedback mechanisms which couple the energy output to the IGM via jets on \textit{resolved} scales \citep{2016ApJ...829...90Y, 2017ApJ...847..106L, 2017MNRAS.472.4707B, 2017MNRAS.470.4530W, 2019MNRAS.483.2465M, 2021MNRAS.506..488B}.}


To expand our investigation, we also explore the effect of choosing a particular fixed direction (the x-axis of the parent box) to select the particle to heat, as opposed to heating particles in random directions at every feedback event. This scheme is implemented in the Bipolar model, and can be regarded as a limiting case of a collimated outflow. From Fig. \ref{fig:z0properties}, we immediately notice that the Bipolar model has the overall effect of reducing the baryon fraction of both objects, compared to the Isotropic case. Since the star fractions are similar, the variation in total baryon fraction is dominated by the differences in hot gas fraction. The group run with bipolar feedback contains $\sim$10\% less baryons than Isotropic, and this change is greater in the cluster, which shows a 12\% reduction in baryon fraction when switching from the Isotropic to the Bipolar model.

We find that changing the directionality of the AGN feedback also has an impact on the profiles, as shown in Fig. \ref{fig:profiles_schemes}. The group is, once again, less affected, but the Bipolar cluster run produced a larger isentropic core and a flatter entropy profile compared to the other schemes. We find that a collimated AGN feedback scheme does not allow the formation of a cool core, but rather has the opposite effect of producing an even flatter entropy profile. However, our implementation of the Bipolar model is remarkably simple and different results may be produced by a more sophisticated method based on semi-analytic models \citep[e.g.][]{2022MNRAS.516.3750H}.

\section{Discussion}
\label{sec:discussion}

Large entropy cores, defined by flat entropy profiles with extended plateaus, are commonplace in recent cluster simulations run at high resolution (e.g. C-EAGLE, \citealt{2017MNRAS.465..213B}, SIMBA, \citealt{2019MNRAS.486.2827D}, TNG, \citealt{2018MNRAS.473.4077P, 2018MNRAS.481.1809B}, FABLE, \citealt{2018MNRAS.479.5385H}) . Power-law entropy profiles, typical in cool-core systems, are absent in all these examples, except for the higher-resolution Romulus-C cluster which, however, excludes the effects of metal cooling. Interestingly, these cores/plateaus appear to form irrespective of the hydrodynamics methods and feedback schemes. The study in this paper shows that this effect also persists when varying relevant parameters (e.g. feedback temperature, particle heating method) within a fixed model. In this section, we discuss alternative scenarios which could possibly explain the absence of cool-core clusters in contemporary simulations.

\begin{enumerate}
    \item \textbf{Object selection}. Cool cores in groups and clusters are rare and we have not performed \textit{ad hoc} selections of such objects in our sample of 27 objects. In particular, a power-law-like, cool-core system will only form when the cluster is moving into its relaxed phase and is no longer experiencing merger activity. We expect such objects to be less common on cluster scales given their complex dynamical youth. While we cannot rule out this explanation, we note that the general profile shape of X-ray-selected groups and clusters, such as those studied by \cite{entropy_profiles_sun2009} and \cite{entropy_profiles_pratt2010}, is considerably different from our simulated systems. However, X-ray luminosity selection can implicitly exclude objects with under-dense cores, particularly in lower-mass groups, where the high entropy is a likely result of AGN feedback, as concluded by \cite{2022A&A...663A...2C}. Furthermore, profiles from larger samples e.g. in the TNG simulation \citep{2018MNRAS.481.1809B}, are also at odds with observations, so we do not regard object selection as a likely explanation.
    
    \item \textbf{Hydrostatic mass bias}. We have not included the effects of hydrostatic mass bias explicitly in our analysis. The expected positive bias (i.e. a lower hydrostatic mass than the true mass) would not affect the shape of the profile; instead, it would shift the radial scale ($r_{500}$) and entropy scale ($K_{500}$) to smaller values for the simulations, leading to a higher normalisation in the scaled profile at fixed scaled radius. This would result in even poorer agreement with observations in the central region, as we found in the group and cluster run with the Ref model. Furthermore, we have verified that typical bias levels for our simulated objects are 10-20\% at $r_{500}$, in line with previous work. Therefore, hydrostatic bias does not strongly affect the entropy profiles in our analysis.
    
    \item \textbf{Projection effects}. Obtaining the 3D entropy profile from X-ray data is a complex procedure. A significant source of uncertainty comes from the deprojection of data (X-ray emission-measure/surface brightness and temperature profiles), based on the assumption of spherical symmetry. Projection effects are unlikely to dramatically change the shape of the entropy profile. To produce a cool-core system, we would require a radially-dependent bias that results in a decrease in the central temperature and/or increase in the central density. On the other hand, incorrectly accounting for gas at larger projected distances from the centre would likely have the opposite effect since the gas there is at higher entropy. Nevertheless, we have verified this by modelling projected data and using this to reconstruct the 3D entropy profile. 
    
    \item \textbf{X-ray weighting}. A related issue to the previous point is that X-ray observables are not mass-weighted. Firstly, since the gas density is derived from the emission measure (which depends on the square of the density), it is susceptible to clumping effects \citep[see e.g.][]{2022arXiv221101239T}. However, recent analyses take this effect into account by calculating the azimuthal median profile \citep[e.g. in X-COP,][]{2017AN....338..293E} in order to mitigate substructure effects. Secondly, the X-ray spectroscopic temperature is also well known to be biased relative to the mass-weighted value, due to the assumption of isothermality when fitting the spectrum. Spectroscopic temperatures are lower than mass-weighted temperatures \citep[e.g.][]{ceagle.barnes.2017, 2021MNRAS.506.2533B}, so the effect of this bias would reduce the entropy. {We have conducted simple tests to check the effect of X-ray weighting on both the group and the cluster run with the Ref model. We compared the mass-weighted thermodynamic profiles with the spectroscopic-like profiles, weighted by the square of the hot-gas density, and the profiles weighted by the X-ray luminosity computed using the same cooling tables used in the sub-grid gas cooling \citep{2020MNRAS.497.4857P}. We did not detect any changes in the shape of the scaled profiles and discrepancies relative to Ref were $\sim 10\%$.} Nevertheless, the X-ray weighting does not necessarily reduce the size of the entropy cores, as shown by the \textit{spectroscopic} profiles of C-EAGLE and IllustrisTNG \citep{ceagle.barnes.2017, 2018MNRAS.481.1809B}.
    
    \item \textbf{Jet feedback}. Observational evidence for AGN feedback in clusters comes from radio observations of {relativistic} jets. {Their effect of} displacing the thermal X-ray-emitting gas with bubbles of super-heated relativistic plasma is not modelled in our simulations. Such {\it jet feedback} is widely thought to be the solution to the cluster cooling flow problem, quenching star formation in the BCG, but the exact mechanisms by which the jets couple to the gas are poorly understood. Directional feedback may reduce the high entropy levels if these are mainly produced by quasi-spherical shock-heating in the current models (inspection of our simulation outputs show that such events do occur). However, it is difficult to imagine such directional heating being sustained on cosmological timescales, given that there is evidence for jet precession and that we expect precipitation (and star formation) along axes perpendicular to the jet direction \citep{2021A&A...646A..38T}. \\
    {Sub-grid models tracking the spin (magnitude and orientation) of the SMBH explicitly and dividing AGN feedback into quasar/radio modes have also shown that SMBH spins re-orient as mergers, gas accretion and SN feedback take place \citep[e.g.][]{2014MNRAS.440.2333D, 2019A&A...631A..60B}. Recent implementations of jet feedback, based on the \cite{1977MNRAS.179..433B} self-similar jet model, were successfully tuned to match the properties of the circum-galactic medium of observed in Seyfert galaxies with active SMBHs \citep{2021MNRAS.504.3619T}. Idealised simulations of galaxy clusters showed that the SMBH spins can re-orient more frequently in high-mass systems than in low-mass ones, and produce pressurised lobes which can uplift the cold, low-entropy gas and remove it from the core \citep{2022MNRAS.516.3750H}.} \\
    {The SIMBA simulations attempted to include the effects of jet feedback in a cosmological setting by using bipolar kinetic jets \citep{2019MNRAS.486.2827D}. Their implementation temporarily decouples the jetted gas particles from the hydrodynamic scheme, avoiding a directly interaction on small scales. This strategy led to realistic hot-gas and star fractions, but still produced large entropy cores \citep{2020MNRAS.498.3061R}. Finally, simulations of kinetic jets with \textit{chaotic cold gas accretion} \citep{2013MNRAS.432.3401G} in cluster simulations have been able to maintain cool-cores on long timescales, albeit only using idealised initial conditions \cite[e.g.][]{2015ApJ...811...73L, 2015ApJ...811..108P, 2016ApJ...829...90Y, 2023MNRAS.518.4622E}.} 
    
    \item \textbf{AGN heating cycle}. The shape and level of the inner region of the entropy profiles is largely dictated by the AGN activity. \cite{2022MNRAS.tmp.1955N} and \cite{2022MNRAS.516.3750H} have identified periodic behaviour in the core entropy, which rises as the AGN injects energy into the surrounding medium and decreases as the radiative cooling then takes over and prompts the formation of a cool core. Our simulations do not show such periodicity. Therefore, the high entropy and lack of cool cores at $z=0$ could not have been accidentally captured during a particularly active phase of a possible AGN heating cycle. A detailed time-evolution study of the entropy distribution will appear in a follow-up paper.

\end{enumerate}

We appear to be left with the scenario in which entropy cores are the outcome of feedback prescriptions currently being used to regulate star formation and black hole growth in the current generation of galaxy formation simulations. Interestingly, the problem is less acute at lower resolution (e.g. in \citealt{2015ApJ...813L..17R}, Dianoga Ref, \citealt{2020A&A...642A..37B}, BAHAMAS, \citealt{2017MNRAS.465.2936M} and MACSIS, \citealt{2017MNRAS.465..213B}) where good agreement with observed entropy profiles is found. Two possibilities for this difference are as follows. Firstly, it may be that more low entropy gas is being heated and ejected from proto-group and proto-cluster regions at early times in the higher-resolution simulations. This would lead to the larger characteristic entropy scale of the remaining gas, largely unaffected by the feedback. However, as discussed by \cite{ceagle.barnes.2017}, such a scenario is hard to justify when the low-redshift gas fractions are larger than observed. A second possibility is that the feedback is coupling more strongly to the gas (through shocks) as a result of improved resolution. This leads to the conclusion that we may be missing (or not resolving) important physics, allowing the energy to be deposited in a gentler way in sound waves (as suggested by X-ray observations of Perseus) or other, non-thermal mechanisms such as cosmic-ray heating and transport processes that would require us to model magnetic fields.

{A study based on the Dianoga simulations \citep{2021MNRAS.507.5703P} suggested that not only can the shock-heating mechanism resulting from AGN feedback thermalise the gas, but it can also transfer energy into populations of relativistic cosmic rays via diffusive shock acceleration (\citealt{2007ApJ...669..729K}, and the review by \citealt{2007NuPhS.165..122B}) and therefore alter the thermodynamic properties of the IGM. Furthermore, magnetic fields themselves can suppress the thermal conduction via processes of magnetic draping \citep{2006MNRAS.373...73L}, which is important in shock fronts generated by e.g. AGN outflows or mergers \citep{2022MNRAS.512.2157C}. Where thermal conduction is reduced, so is the mixing between high- and low-entropy gas; this scenario could indeed help preserving the low-entropy gas phase and cool-cores, as suggested by our No-conduction simulations (Section \ref{sec:results_model_variations:conduction}). On the other hand, magnetic fields could also favour the emergence of processes which produce the opposite effect. For instance, if AGN-produced magnetised bubbles are preserved, they can drive small-scale, often resolution-dependent, hydrodynamic instabilities. These motions can lead to a powerful turbulent cascade, which ultimately thermalises the local IGM and can compromise the integrity of a cool core \citep{2005MNRAS.363..891F, 2019ApJ...886...78B}.} Current models are also missing a cold (i.e., $T < 10^4$ K) interstellar gas phase that could also play an important role in how the central galaxy grows and how the feedback interacts with the gas, {similarly to the chaotic cold accretion mode in \cite{2023MNRAS.518.4622E}.}

\section{Conclusions}
\label{sec:conclusion}
Using a new implementation of the EAGLE galaxy formation model, we produced zoom-in simulations of 27 objects selected from a 300 Mpc parent volume. The re-simulated groups and clusters of galaxies showed reasonable agreement with the star fraction-halo mass relation from \cite{xxl.baryons.akino2022}, despite overshooting the hot gas fractions at halo masses $\sim 10^{14}$\,M$_\odot$. In spite of this shortcoming, our simulations provide hot gas and star fractions closer to the observed values than the EAGLE model in \cite{eagle.schaye.2015}. On the other hand, our simulations predict larger entropy cores than EAGLE, which still cannot be reconciled with the observations from \cite{entropy_profiles_sun2009} and \cite{entropy_profiles_pratt2010}. To investigate the sensitivity of the entropy profiles to sub-grid model changes, we selected one group and one cluster from the object sample and produced a suite of simulations with different sub-grid schemes and resolutions. We summarise the main results as follows:
\begin{enumerate}
    \item Artificial conduction does not have a large impact on the global properties of the objects; the thermodynamic profiles for the group are also generally unaltered. The cluster profiles, however, show the formation at $z=0$ of a colder, denser and anisentropic core when artificial conduction is switched off, which can be attributed to a smaller degree of mixing between gas in the high- and low-entropy phase. While yielding more power-law-like entropy profiles for clusters, the hydrodynamic scheme without artificial conduction cannot reliably reproduce instabilities and smooth particle behaviour at interfaces between fluids \citep{2008JCoPh.22710040P}.
    
    \item The simulations without metal cooling produced objects with gas fractions nearly twice as large and 50\% lower star fractions than the reference model. In the group, the lower-mass BCG does not differ significantly in its specific SFR, while the central BH is more than 70\% lighter in the runs without metal cooling. Both the group and the cluster have lower entropy overall and the cluster forms a smaller entropy core, with a power-law-like profile outside the core, in analogy to the results by \cite{2019MNRAS.483.3336T}, whose simulations also did not include metal cooling. The absence of metal cooling only affects the inner region of the cluster, but has a large impact on the group's environment out to $\approx2.5~r_{500}$.
    
    \item The runs without AGN also produced a large isentropic core, although the entropy level in the inner regions is lower than for the reference model. Turning off SN feedback, instead, produced a vast increase in entropy throughout the group and cluster atmospheres, associated with hotter and less dense cores. The absence of thermal SN feedback has a large impact on the self-regulation of the central BH, boosts the star formation in the group and cluster core.
    
    \item When increasing the AGN heating temperature from $10^8$ to $10^9$ K in steps of 0.5 dex, we observed an entropy increase at all radii in the group, which develops an overall less dense atmosphere. The cluster, on the other hand, only developed a larger entropy core, without impacting the hot gas outside $r_{500}$. High values of $\Delta T_{\rm AGN}$ lead to less frequent, but more energetic feedback events, which remove the low-entropy gas from the core.
    
    \item The minimum-distance, random and isotropic AGN feedback distribution schemes produced similar thermodynamic profiles for the group, and minimal differences can be seen in the cluster core. We conclude that, for the two objects considered in this comparative study, the entropy distribution is generally insensitive to the choice of distribution scheme for thermal AGN feedback. Our simple implementation of a fixed-direction bipolar scheme produces a marginally less dense, higher entropy IGM in the group. Larger effects are observed in the cluster core, where the entropy profile flattens to a constant level out to $0.5~r_{500}$, correlated with a higher core temperature and lower density. The bipolar feedback scheme is more effective in expelling hot gas from the system, however, it does not preserve the low-entropy gas and produces larger entropy cores.
    
\end{enumerate}
We find a close similarity between entropy profiles at $z=0$ and at $z=1$, suggesting that the formation of large, isentropic cores may occur at higher redshift, during the phase of maximum AGN activity. {The AGN favours the rise if entropy in the cores the group and the cluster. Additionally, a further investigation has shown that the shape of the entropy profiles changes significantly during and after merger episodes. This scenario is similar to that of the ROMULUS-C system \citep{2021MNRAS.504.3922C}. Indeed, the progenitor halo of our cluster appears to show a cool core at very high redshift ($z\approx 4$), but a core/plateau forms during the complex formation process that involves both mergers and AGN activity. The merger between our cluster and its neighbour produced large shock waves, which could play an important role in boosting the entropy amplification in the core.} In a follow-up paper, we provide a detailed overview of how the thermodynamic profiles and gas properties of the group and cluster evolve over time. The AGN model used in our simulations uses a simple prescription, which cannot simultaneously reproduce the gas properties and entropy profiles of groups and clusters. More sophisticated feedback mechanisms will be implemented and studied in future work.

\section*{Acknowledgements}
{The authors thank the anonymous referee for providing comments which improved the quality, clarity and depth of our work.} This work used the DiRAC@Durham facility managed by the Institute for Computational Cosmology on behalf of the STFC DiRAC HPC Facility (www.dirac.ac.uk). The equipment was funded by BEIS capital funding via STFC capital grants ST/K00042X/1, ST/P002293/1, ST/R002371/1 and ST/S002502/1, Durham University and STFC operations grant ST/R000832/1. DiRAC is part of the National e-Infrastructure. EA acknowledges the STFC studentship grant ST/T506291/1. YMB acknowledges support from NWO under Veni grant number 639.041.751. {The work received support under the Project HPC-EUROPA3 (INFRAIA-2016-1-730897), with the support of the European Commission Research Innovation Action under the H2020 Programme; in particular, EA gratefully acknowledges the support of the Sterrewacht@Leiden University and the computer resources and technical support provided by SURFsara, the Dutch national high-performance computing facility.} The research in this paper made use of the \swift open-source simulation code (\href{https://swiftsim.com/}{swiftsim.com}, \citealt{schaller_2018_swift}) version 0.9.0. In addition, this work made use of the following software packages and libraries: 
\textsc{Python} \citep{van1995python},
\textsc{Numpy} \citep{harris2020array},
\textsc{Scipy} \citep{virtanen2020scipy},
\textsc{Numba} \citep{lam2015numba},
\textsc{Matplotlib} \citep{hunter2007matplotlib, caswell2020matplotlib},
\textsc{SWIFTsimIO} \citep{Borrow2020, Borrow2021b} and
\textsc{Astropy} \citep{robitaille2013astropy, price2022astropy}.

\section*{Data Availability}

The \swift and \velociraptor structure-finding codes are public and open-source, and can be downloaded from GitHub or GitLab. The initial conditions for the halos, the snapshots and the halo catalogues can be made available upon reasonable request to the corresponding author. The code used in the analysis is publicly available on the corresponding author's GitHub repository (\href{https://github.com/edoaltamura/entropy-core-problem}{github.com/edoaltamura/entropy-core-problem}) and we include the data used to generate the figures presented throughout the document.

\bibliographystyle{mnras}
\bibliography{main} 

\begin{thebibliography}{}
\makeatletter
\relax
\def\mn@urlcharsother{\let\do\@makeother \do\$\do\&\do\#\do\^\do\_\do\%\do\~}
\def\mn@doi{\begingroup\mn@urlcharsother \@ifnextchar [ {\mn@doi@}
  {\mn@doi@[]}}
\def\mn@doi@[#1]#2{\def\@tempa{#1}\ifx\@tempa\@empty \href
  {http://dx.doi.org/#2} {doi:#2}\else \href {http://dx.doi.org/#2} {#1}\fi
  \endgroup}
\def\mn@eprint#1#2{\mn@eprint@#1:#2::\@nil}
\def\mn@eprint@arXiv#1{\href {http://arxiv.org/abs/#1} {{\tt arXiv:#1}}}
\def\mn@eprint@dblp#1{\href {http://dblp.uni-trier.de/rec/bibtex/#1.xml}
  {dblp:#1}}
\def\mn@eprint@#1:#2:#3:#4\@nil{\def\@tempa {#1}\def\@tempb {#2}\def\@tempc
  {#3}\ifx \@tempc \@empty \let \@tempc \@tempb \let \@tempb \@tempa \fi \ifx
  \@tempb \@empty \def\@tempb {arXiv}\fi \@ifundefined
  {mn@eprint@\@tempb}{\@tempb:\@tempc}{\expandafter \expandafter \csname
  mn@eprint@\@tempb\endcsname \expandafter{\@tempc}}}

\bibitem[\protect\citeauthoryear{{Agertz} et~al.,}{{Agertz}
  et~al.}{2007}]{2007MNRAS.380..963A}
{Agertz} O.,  et~al., 2007, \mn@doi [\mnras]
  {10.1111/j.1365-2966.2007.12183.x}, \href
  {https://ui.adsabs.harvard.edu/abs/2007MNRAS.380..963A} {380, 963}

\bibitem[\protect\citeauthoryear{{Akino} et~al.,}{{Akino}
  et~al.}{2022}]{xxl.baryons.akino2022}
{Akino} D.,  et~al., 2022, \mn@doi [\pasj] {10.1093/pasj/psab115}, \href
  {https://ui.adsabs.harvard.edu/abs/2022PASJ...74..175A} {74, 175}

\bibitem[\protect\citeauthoryear{{Astropy Collaboration} et~al.,}{{Astropy
  Collaboration} et~al.}{2013}]{robitaille2013astropy}
{Astropy Collaboration} et~al., 2013, \mn@doi [\aap]
  {10.1051/0004-6361/201322068}, \href
  {https://ui.adsabs.harvard.edu/abs/2013A&A...558A..33A} {558, A33}

\bibitem[\protect\citeauthoryear{{Astropy Collaboration} et~al.,}{{Astropy
  Collaboration} et~al.}{2022}]{price2022astropy}
{Astropy Collaboration} et~al., 2022, \mn@doi [\apj]
  {10.3847/1538-4357/ac7c74}, \href
  {https://ui.adsabs.harvard.edu/abs/2022ApJ...935..167A} {935, 167}

\bibitem[\protect\citeauthoryear{{Bah{\'e}}, {McCarthy}  \& {King}}{{Bah{\'e}}
  et~al.}{2012}]{2012MNRAS.421.1073B}
{Bah{\'e}} Y.~M.,  {McCarthy} I.~G.,   {King} L.~J.,  2012, \mn@doi [\mnras]
  {10.1111/j.1365-2966.2011.20364.x}, \href
  {https://ui.adsabs.harvard.edu/abs/2012MNRAS.421.1073B} {421, 1073}

\bibitem[\protect\citeauthoryear{{Bah{\'e}} et~al.,}{{Bah{\'e}}
  et~al.}{2017}]{2017MNRAS.470.4186B}
{Bah{\'e}} Y.~M.,  et~al., 2017, \mn@doi [\mnras] {10.1093/mnras/stx1403},
  \href {https://ui.adsabs.harvard.edu/abs/2017MNRAS.470.4186B} {470, 4186}

\bibitem[\protect\citeauthoryear{{Bah{\'e}} et~al.,}{{Bah{\'e}}
  et~al.}{2022}]{bahe_2021_bh_repositioning}
{Bah{\'e}} Y.~M.,  et~al., 2022, \mn@doi [\mnras] {10.1093/mnras/stac1339},
  \href {https://ui.adsabs.harvard.edu/abs/2022MNRAS.516..167B} {516, 167}

\bibitem[\protect\citeauthoryear{{Balogh}, {Pearce}, {Bower}  \&
  {Kay}}{{Balogh} et~al.}{2001}]{2001MNRAS.326.1228B}
{Balogh} M.~L.,  {Pearce} F.~R.,  {Bower} R.~G.,   {Kay} S.~T.,  2001, \mn@doi
  [\mnras] {10.1111/j.1365-2966.2001.04667.x}, \href
  {https://ui.adsabs.harvard.edu/abs/2001MNRAS.326.1228B} {326, 1228}

\bibitem[\protect\citeauthoryear{{Balsara}}{{Balsara}}{1995}]{1995JCoPh.121..357B}
{Balsara} D.~S.,  1995, \mn@doi [Journal of Computational Physics]
  {10.1016/S0021-9991(95)90221-X}, \href
  {https://ui.adsabs.harvard.edu/abs/1995JCoPh.121..357B} {121, 357}

\bibitem[\protect\citeauthoryear{{Bambic} \& {Reynolds}}{{Bambic} \&
  {Reynolds}}{2019}]{2019ApJ...886...78B}
{Bambic} C.~J.,  {Reynolds} C.~S.,  2019, \mn@doi [\apj]
  {10.3847/1538-4357/ab4daf}, \href
  {https://ui.adsabs.harvard.edu/abs/2019ApJ...886...78B} {886, 78}

\bibitem[\protect\citeauthoryear{{Barnes}, {Kay}, {Henson}, {McCarthy},
  {Schaye}  \& {Jenkins}}{{Barnes} et~al.}{2017a}]{2017MNRAS.465..213B}
{Barnes} D.~J.,  {Kay} S.~T.,  {Henson} M.~A.,  {McCarthy} I.~G.,  {Schaye} J.,
    {Jenkins} A.,  2017a, \mn@doi [\mnras] {10.1093/mnras/stw2722}, \href
  {https://ui.adsabs.harvard.edu/abs/2017MNRAS.465..213B} {465, 213}

\bibitem[\protect\citeauthoryear{{Barnes} et~al.,}{{Barnes}
  et~al.}{2017b}]{ceagle.barnes.2017}
{Barnes} D.~J.,  et~al., 2017b, \mn@doi [\mnras] {10.1093/mnras/stx1647}, \href
  {https://ui.adsabs.harvard.edu/abs/2017MNRAS.471.1088B} {471, 1088}

\bibitem[\protect\citeauthoryear{{Barnes} et~al.,}{{Barnes}
  et~al.}{2018}]{2018MNRAS.481.1809B}
{Barnes} D.~J.,  et~al., 2018, \mn@doi [\mnras] {10.1093/mnras/sty2078}, \href
  {https://ui.adsabs.harvard.edu/abs/2018MNRAS.481.1809B} {481, 1809}

\bibitem[\protect\citeauthoryear{{Barnes}, {Vogelsberger}, {Pearce}, {Pop},
  {Kannan}, {Cao}, {Kay}  \& {Hernquist}}{{Barnes}
  et~al.}{2021}]{2021MNRAS.506.2533B}
{Barnes} D.~J.,  {Vogelsberger} M.,  {Pearce} F.~A.,  {Pop} A.-R.,  {Kannan}
  R.,  {Cao} K.,  {Kay} S.~T.,   {Hernquist} L.,  2021, \mn@doi [\mnras]
  {10.1093/mnras/stab1276}, \href
  {https://ui.adsabs.harvard.edu/abs/2021MNRAS.506.2533B} {506, 2533}

\bibitem[\protect\citeauthoryear{{Bassini} et~al.,}{{Bassini}
  et~al.}{2020}]{2020A&A...642A..37B}
{Bassini} L.,  et~al., 2020, \mn@doi [\aap] {10.1051/0004-6361/202038396},
  \href {https://ui.adsabs.harvard.edu/abs/2020A&A...642A..37B} {642, A37}

\bibitem[\protect\citeauthoryear{{Beck} et~al.,}{{Beck}
  et~al.}{2016}]{2016MNRAS.455.2110B}
{Beck} A.~M.,  et~al., 2016, \mn@doi [\mnras] {10.1093/mnras/stv2443}, \href
  {https://ui.adsabs.harvard.edu/abs/2016MNRAS.455.2110B} {455, 2110}

\bibitem[\protect\citeauthoryear{{Beckmann} et~al.,}{{Beckmann}
  et~al.}{2019}]{2019A&A...631A..60B}
{Beckmann} R.~S.,  et~al., 2019, \mn@doi [\aap] {10.1051/0004-6361/201936188},
  \href {https://ui.adsabs.harvard.edu/abs/2019A&A...631A..60B} {631, A60}

\bibitem[\protect\citeauthoryear{{Blandford} \& {Znajek}}{{Blandford} \&
  {Znajek}}{1977}]{1977MNRAS.179..433B}
{Blandford} R.~D.,  {Znajek} R.~L.,  1977, \mn@doi [\mnras]
  {10.1093/mnras/179.3.433}, \href
  {https://ui.adsabs.harvard.edu/abs/1977MNRAS.179..433B} {179, 433}

\bibitem[\protect\citeauthoryear{{Blasi}}{{Blasi}}{2007}]{2007NuPhS.165..122B}
{Blasi} P.,  2007, \mn@doi [Nuclear Physics B Proceedings Supplements]
  {10.1016/j.nuclphysbps.2006.11.022}, \href
  {https://ui.adsabs.harvard.edu/abs/2007NuPhS.165..122B} {165, 122}

\bibitem[\protect\citeauthoryear{{Bondi} \& {Hoyle}}{{Bondi} \&
  {Hoyle}}{1944}]{1944MNRAS.104..273B}
{Bondi} H.,  {Hoyle} F.,  1944, \mn@doi [\mnras] {10.1093/mnras/104.5.273},
  \href {https://ui.adsabs.harvard.edu/abs/1944MNRAS.104..273B} {104, 273}

\bibitem[\protect\citeauthoryear{{Booth} \& {Schaye}}{{Booth} \&
  {Schaye}}{2009}]{2009MNRAS.398...53B}
{Booth} C.~M.,  {Schaye} J.,  2009, \mn@doi [\mnras]
  {10.1111/j.1365-2966.2009.15043.x}, \href
  {https://ui.adsabs.harvard.edu/abs/2009MNRAS.398...53B} {398, 53}

\bibitem[\protect\citeauthoryear{{Borgani}, {Finoguenov}, {Kay}, {Ponman},
  {Springel}, {Tozzi}  \& {Voit}}{{Borgani} et~al.}{2005}]{2005MNRAS.361..233B}
{Borgani} S.,  {Finoguenov} A.,  {Kay} S.~T.,  {Ponman} T.~J.,  {Springel} V.,
  {Tozzi} P.,   {Voit} G.~M.,  2005, \mn@doi [\mnras]
  {10.1111/j.1365-2966.2005.09158.x}, \href
  {https://ui.adsabs.harvard.edu/abs/2005MNRAS.361..233B} {361, 233}

\bibitem[\protect\citeauthoryear{Borrow \& Borrisov}{Borrow \&
  Borrisov}{2020}]{Borrow2020}
Borrow J.,  Borrisov A.,  2020, \mn@doi [Journal of Open Source Software]
  {10.21105/joss.02430}, 5, 2430

\bibitem[\protect\citeauthoryear{{Borrow} \& {Kelly}}{{Borrow} \&
  {Kelly}}{2021}]{Borrow2021b}
{Borrow} J.,  {Kelly} A.~J.,  2021, arXiv e-prints, \href
  {https://ui.adsabs.harvard.edu/abs/2021arXiv210605281B} {p. arXiv:2106.05281}

\bibitem[\protect\citeauthoryear{{Borrow}, {Schaller}  \& {Bower}}{{Borrow}
  et~al.}{2021}]{2021MNRAS.505.2316B}
{Borrow} J.,  {Schaller} M.,   {Bower} R.~G.,  2021, \mn@doi [\mnras]
  {10.1093/mnras/stab1423}, \href
  {https://ui.adsabs.harvard.edu/abs/2021MNRAS.505.2316B} {505, 2316}

\bibitem[\protect\citeauthoryear{{Borrow}, {Schaller}, {Bower}  \&
  {Schaye}}{{Borrow} et~al.}{2022}]{sphenix_borrow2022}
{Borrow} J.,  {Schaller} M.,  {Bower} R.~G.,   {Schaye} J.,  2022, \mn@doi
  [\mnras] {10.1093/mnras/stab3166}, \href
  {https://ui.adsabs.harvard.edu/abs/2022MNRAS.511.2367B} {511, 2367}

\bibitem[\protect\citeauthoryear{{Bourne} \& {Sijacki}}{{Bourne} \&
  {Sijacki}}{2017}]{2017MNRAS.472.4707B}
{Bourne} M.~A.,  {Sijacki} D.,  2017, \mn@doi [\mnras] {10.1093/mnras/stx2269},
  \href {https://ui.adsabs.harvard.edu/abs/2017MNRAS.472.4707B} {472, 4707}

\bibitem[\protect\citeauthoryear{{Bourne} \& {Sijacki}}{{Bourne} \&
  {Sijacki}}{2021}]{2021MNRAS.506..488B}
{Bourne} M.~A.,  {Sijacki} D.,  2021, \mn@doi [\mnras]
  {10.1093/mnras/stab1662}, \href
  {https://ui.adsabs.harvard.edu/abs/2021MNRAS.506..488B} {506, 488}

\bibitem[\protect\citeauthoryear{{Bower}}{{Bower}}{1997}]{entropy_intro_bower_1997}
{Bower} R.~G.,  1997, \mn@doi [\mnras] {10.1093/mnras/288.2.355}, \href
  {https://ui.adsabs.harvard.edu/abs/1997MNRAS.288..355B} {288, 355}

\bibitem[\protect\citeauthoryear{{Bower}, {Benson}  \& {Crain}}{{Bower}
  et~al.}{2012}]{2012MNRAS.422.2816B}
{Bower} R.~G.,  {Benson} A.~J.,   {Crain} R.~A.,  2012, \mn@doi [\mnras]
  {10.1111/j.1365-2966.2012.20516.x}, \href
  {https://ui.adsabs.harvard.edu/abs/2012MNRAS.422.2816B} {422, 2816}

\bibitem[\protect\citeauthoryear{{Bower}, {Schaye}, {Frenk}, {Theuns},
  {Schaller}, {Crain}  \& {McAlpine}}{{Bower}
  et~al.}{2017}]{2017MNRAS.465...32B}
{Bower} R.~G.,  {Schaye} J.,  {Frenk} C.~S.,  {Theuns} T.,  {Schaller} M.,
  {Crain} R.~A.,   {McAlpine} S.,  2017, \mn@doi [\mnras]
  {10.1093/mnras/stw2735}, \href
  {https://ui.adsabs.harvard.edu/abs/2017MNRAS.465...32B} {465, 32}

\bibitem[\protect\citeauthoryear{Caswell et~al.,}{Caswell
  et~al.}{2023}]{caswell2020matplotlib}
Caswell T.~A.,  et~al., 2023, matplotlib/matplotlib: REL: v3.7.0rc1,
  \mn@doi{10.5281/zenodo.7570264}, \url
  {https://doi.org/10.5281/zenodo.7570264}

\bibitem[\protect\citeauthoryear{{Cavagnolo}, {Donahue}, {Voit}  \&
  {Sun}}{{Cavagnolo} et~al.}{2009}]{2009ApJS..182...12C}
{Cavagnolo} K.~W.,  {Donahue} M.,  {Voit} G.~M.,   {Sun} M.,  2009, \mn@doi
  [\apjs] {10.1088/0067-0049/182/1/12}, \href
  {https://ui.adsabs.harvard.edu/abs/2009ApJS..182...12C} {182, 12}

\bibitem[\protect\citeauthoryear{{Chadayammuri}, {Tremmel}, {Nagai}, {Babul}
  \& {Quinn}}{{Chadayammuri} et~al.}{2021}]{2021MNRAS.504.3922C}
{Chadayammuri} U.,  {Tremmel} M.,  {Nagai} D.,  {Babul} A.,   {Quinn} T.,
  2021, \mn@doi [\mnras] {10.1093/mnras/stab1010}, \href
  {https://ui.adsabs.harvard.edu/abs/2021MNRAS.504.3922C} {504, 3922}

\bibitem[\protect\citeauthoryear{{Chadayammuri}, {ZuHone}, {Nulsen}, {Nagai}
  \& {Russell}}{{Chadayammuri} et~al.}{2022}]{2022MNRAS.512.2157C}
{Chadayammuri} U.,  {ZuHone} J.,  {Nulsen} P.,  {Nagai} D.,   {Russell} H.,
  2022, \mn@doi [\mnras] {10.1093/mnras/stac594}, \href
  {https://ui.adsabs.harvard.edu/abs/2022MNRAS.512.2157C} {512, 2157}

\bibitem[\protect\citeauthoryear{{Chaikin}, {Schaye}, {Schaller}, {Bah{\'e}},
  {Nobels}  \& {Ploeckinger}}{{Chaikin} et~al.}{2022}]{2022MNRAS.514..249C}
{Chaikin} E.,  {Schaye} J.,  {Schaller} M.,  {Bah{\'e}} Y.~M.,  {Nobels} F.
  S.~J.,   {Ploeckinger} S.,  2022, \mn@doi [\mnras] {10.1093/mnras/stac1132},
  \href {https://ui.adsabs.harvard.edu/abs/2022MNRAS.514..249C} {514, 249}

\bibitem[\protect\citeauthoryear{{Churazov}, {Br{\"u}ggen}, {Kaiser},
  {B{\"o}hringer}  \& {Forman}}{{Churazov} et~al.}{2001}]{2001ApJ...554..261C}
{Churazov} E.,  {Br{\"u}ggen} M.,  {Kaiser} C.~R.,  {B{\"o}hringer} H.,
  {Forman} W.,  2001, \mn@doi [\apj] {10.1086/321357}, \href
  {https://ui.adsabs.harvard.edu/abs/2001ApJ...554..261C} {554, 261}

\bibitem[\protect\citeauthoryear{{Crain} et~al.,}{{Crain}
  et~al.}{2015}]{2015MNRAS.450.1937C}
{Crain} R.~A.,  et~al., 2015, \mn@doi [\mnras] {10.1093/mnras/stv725}, \href
  {https://ui.adsabs.harvard.edu/abs/2015MNRAS.450.1937C} {450, 1937}

\bibitem[\protect\citeauthoryear{{Crossett} et~al.,}{{Crossett}
  et~al.}{2022}]{2022A&A...663A...2C}
{Crossett} J.~P.,  et~al., 2022, \mn@doi [\aap] {10.1051/0004-6361/202142057},
  \href {https://ui.adsabs.harvard.edu/abs/2022A&A...663A...2C} {663, A2}

\bibitem[\protect\citeauthoryear{{Cullen} \& {Dehnen}}{{Cullen} \&
  {Dehnen}}{2010}]{2010MNRAS.408..669C}
{Cullen} L.,  {Dehnen} W.,  2010, \mn@doi [\mnras]
  {10.1111/j.1365-2966.2010.17158.x}, \href
  {https://ui.adsabs.harvard.edu/abs/2010MNRAS.408..669C} {408, 669}

\bibitem[\protect\citeauthoryear{{Dalla Vecchia} \& {Schaye}}{{Dalla Vecchia}
  \& {Schaye}}{2008}]{2008MNRAS.387.1431D}
{Dalla Vecchia} C.,  {Schaye} J.,  2008, \mn@doi [\mnras]
  {10.1111/j.1365-2966.2008.13322.x}, \href
  {https://ui.adsabs.harvard.edu/abs/2008MNRAS.387.1431D} {387, 1431}

\bibitem[\protect\citeauthoryear{{Dalla Vecchia} \& {Schaye}}{{Dalla Vecchia}
  \& {Schaye}}{2012}]{2012MNRAS.426..140D}
{Dalla Vecchia} C.,  {Schaye} J.,  2012, \mn@doi [\mnras]
  {10.1111/j.1365-2966.2012.21704.x}, \href
  {https://ui.adsabs.harvard.edu/abs/2012MNRAS.426..140D} {426, 140}

\bibitem[\protect\citeauthoryear{{Dav{\'e}}, {Angl{\'e}s-Alc{\'a}zar},
  {Narayanan}, {Li}, {Rafieferantsoa}  \& {Appleby}}{{Dav{\'e}}
  et~al.}{2019}]{2019MNRAS.486.2827D}
{Dav{\'e}} R.,  {Angl{\'e}s-Alc{\'a}zar} D.,  {Narayanan} D.,  {Li} Q.,
  {Rafieferantsoa} M.~H.,   {Appleby} S.,  2019, \mn@doi [\mnras]
  {10.1093/mnras/stz937}, \href
  {https://ui.adsabs.harvard.edu/abs/2019MNRAS.486.2827D} {486, 2827}

\bibitem[\protect\citeauthoryear{{David}, {Jones}  \& {Forman}}{{David}
  et~al.}{1996}]{1996ApJ...473..692D}
{David} L.~P.,  {Jones} C.,   {Forman} W.,  1996, \mn@doi [\apj]
  {10.1086/178182}, \href
  {https://ui.adsabs.harvard.edu/abs/1996ApJ...473..692D} {473, 692}

\bibitem[\protect\citeauthoryear{{Debackere}, {Schaye}  \&
  {Hoekstra}}{{Debackere} et~al.}{2021}]{2021MNRAS.505..593D}
{Debackere} S. N.~B.,  {Schaye} J.,   {Hoekstra} H.,  2021, \mn@doi [\mnras]
  {10.1093/mnras/stab1326}, \href
  {https://ui.adsabs.harvard.edu/abs/2021MNRAS.505..593D} {505, 593}

\bibitem[\protect\citeauthoryear{{Donahue}, {Voit}, {O'Dea}, {Baum}  \&
  {Sparks}}{{Donahue} et~al.}{2005}]{2005ApJ...630L..13D}
{Donahue} M.,  {Voit} G.~M.,  {O'Dea} C.~P.,  {Baum} S.~A.,   {Sparks} W.~B.,
  2005, \mn@doi [\apjl] {10.1086/462416}, \href
  {https://ui.adsabs.harvard.edu/abs/2005ApJ...630L..13D} {630, L13}

\bibitem[\protect\citeauthoryear{{Dubois}, {Devriendt}, {Teyssier}  \&
  {Slyz}}{{Dubois} et~al.}{2011}]{2011MNRAS.417.1853D}
{Dubois} Y.,  {Devriendt} J.,  {Teyssier} R.,   {Slyz} A.,  2011, \mn@doi
  [\mnras] {10.1111/j.1365-2966.2011.19381.x}, \href
  {https://ui.adsabs.harvard.edu/abs/2011MNRAS.417.1853D} {417, 1853}

\bibitem[\protect\citeauthoryear{{Dubois}, {Pichon}, {Devriendt}, {Silk},
  {Haehnelt}, {Kimm}  \& {Slyz}}{{Dubois} et~al.}{2013}]{2013MNRAS.428.2885D}
{Dubois} Y.,  {Pichon} C.,  {Devriendt} J.,  {Silk} J.,  {Haehnelt} M.,  {Kimm}
  T.,   {Slyz} A.,  2013, \mn@doi [\mnras] {10.1093/mnras/sts224}, \href
  {https://ui.adsabs.harvard.edu/abs/2013MNRAS.428.2885D} {428, 2885}

\bibitem[\protect\citeauthoryear{{Dubois}, {Volonteri}, {Silk}, {Devriendt}  \&
  {Slyz}}{{Dubois} et~al.}{2014}]{2014MNRAS.440.2333D}
{Dubois} Y.,  {Volonteri} M.,  {Silk} J.,  {Devriendt} J.,   {Slyz} A.,  2014,
  \mn@doi [\mnras] {10.1093/mnras/stu425}, \href
  {https://ui.adsabs.harvard.edu/abs/2014MNRAS.440.2333D} {440, 2333}

\bibitem[\protect\citeauthoryear{{Eckert}, {Ettori}, {Pointecouteau},
  {Molendi}, {Paltani}  \& {Tchernin}}{{Eckert}
  et~al.}{2017}]{2017AN....338..293E}
{Eckert} D.,  {Ettori} S.,  {Pointecouteau} E.,  {Molendi} S.,  {Paltani} S.,
  {Tchernin} C.,  2017, \mn@doi [Astronomische Nachrichten]
  {10.1002/asna.201713345}, \href
  {https://ui.adsabs.harvard.edu/abs/2017AN....338..293E} {338, 293}

\bibitem[\protect\citeauthoryear{{Eckert}, {Gaspari}, {Gastaldello}, {Le Brun}
  \& {O'Sullivan}}{{Eckert} et~al.}{2021}]{agn_review_eckert_2021}
{Eckert} D.,  {Gaspari} M.,  {Gastaldello} F.,  {Le Brun} A. M.~C.,
  {O'Sullivan} E.,  2021, \mn@doi [Universe] {10.3390/universe7050142}, \href
  {https://ui.adsabs.harvard.edu/abs/2021Univ....7..142E} {7, 142}

\bibitem[\protect\citeauthoryear{{Eddington}}{{Eddington}}{1926}]{1926ics..book.....E}
{Eddington} A.~S.,  1926, {The Internal Constitution of the Stars}.
Cambridge Science Classics

\bibitem[\protect\citeauthoryear{{Ehlert}, {Weinberger}, {Pfrommer}, {Pakmor}
  \& {Springel}}{{Ehlert} et~al.}{2023}]{2023MNRAS.518.4622E}
{Ehlert} K.,  {Weinberger} R.,  {Pfrommer} C.,  {Pakmor} R.,   {Springel} V.,
  2023, \mn@doi [\mnras] {10.1093/mnras/stac2860}, \href
  {https://ui.adsabs.harvard.edu/abs/2023MNRAS.518.4622E} {518, 4622}

\bibitem[\protect\citeauthoryear{{Elahi}, {Ca{\~n}as}, {Poulton}, {Tobar},
  {Willis}, {Lagos}, {Power}  \& {Robotham}}{{Elahi}
  et~al.}{2019}]{2019PASA...36...21E}
{Elahi} P.~J.,  {Ca{\~n}as} R.,  {Poulton} R. J.~J.,  {Tobar} R.~J.,  {Willis}
  J.~S.,  {Lagos} C. d.~P.,  {Power} C.,   {Robotham} A. S.~G.,  2019, \mn@doi
  [\pasa] {10.1017/pasa.2019.12}, \href
  {https://ui.adsabs.harvard.edu/abs/2019PASA...36...21E} {36, e021}

\bibitem[\protect\citeauthoryear{{Fabian}}{{Fabian}}{2012}]{2012ARA&A..50..455F}
{Fabian} A.~C.,  2012, \mn@doi [\araa] {10.1146/annurev-astro-081811-125521},
  \href {https://ui.adsabs.harvard.edu/abs/2012ARA&A..50..455F} {50, 455}

\bibitem[\protect\citeauthoryear{{Fabian}, {Sanders}, {Allen}, {Crawford},
  {Iwasawa}, {Johnstone}, {Schmidt}  \& {Taylor}}{{Fabian}
  et~al.}{2003}]{2003MNRAS.344L..43F}
{Fabian} A.~C.,  {Sanders} J.~S.,  {Allen} S.~W.,  {Crawford} C.~S.,  {Iwasawa}
  K.,  {Johnstone} R.~M.,  {Schmidt} R.~W.,   {Taylor} G.~B.,  2003, \mn@doi
  [\mnras] {10.1046/j.1365-8711.2003.06902.x}, \href
  {https://ui.adsabs.harvard.edu/abs/2003MNRAS.344L..43F} {344, L43}

\bibitem[\protect\citeauthoryear{{Fabian}, {Reynolds}, {Taylor}  \&
  {Dunn}}{{Fabian} et~al.}{2005}]{2005MNRAS.363..891F}
{Fabian} A.~C.,  {Reynolds} C.~S.,  {Taylor} G.~B.,   {Dunn} R.~J.~H.,  2005,
  \mn@doi [\mnras] {10.1111/j.1365-2966.2005.09484.x}, \href
  {https://ui.adsabs.harvard.edu/abs/2005MNRAS.363..891F} {363, 891}

\bibitem[\protect\citeauthoryear{{Fabjan}, {Borgani}, {Tornatore}, {Saro},
  {Murante}  \& {Dolag}}{{Fabjan} et~al.}{2010}]{2010MNRAS.401.1670F}
{Fabjan} D.,  {Borgani} S.,  {Tornatore} L.,  {Saro} A.,  {Murante} G.,
  {Dolag} K.,  2010, \mn@doi [\mnras] {10.1111/j.1365-2966.2009.15794.x}, \href
  {https://ui.adsabs.harvard.edu/abs/2010MNRAS.401.1670F} {401, 1670}

\bibitem[\protect\citeauthoryear{{Faucher-Gigu{\`e}re}}{{Faucher-Gigu{\`e}re}}{2020}]{2020MNRAS.493.1614F}
{Faucher-Gigu{\`e}re} C.-A.,  2020, \mn@doi [\mnras] {10.1093/mnras/staa302},
  \href {https://ui.adsabs.harvard.edu/abs/2020MNRAS.493.1614F} {493, 1614}

\bibitem[\protect\citeauthoryear{{Frenk} et~al.,}{{Frenk}
  et~al.}{1999}]{1999ApJ...525..554F}
{Frenk} C.~S.,  et~al., 1999, \mn@doi [\apj] {10.1086/307908}, \href
  {https://ui.adsabs.harvard.edu/abs/1999ApJ...525..554F} {525, 554}

\bibitem[\protect\citeauthoryear{{Gaspari}, {Ruszkowski}  \& {Oh}}{{Gaspari}
  et~al.}{2013}]{2013MNRAS.432.3401G}
{Gaspari} M.,  {Ruszkowski} M.,   {Oh} S.~P.,  2013, \mn@doi [\mnras]
  {10.1093/mnras/stt692}, \href
  {https://ui.adsabs.harvard.edu/abs/2013MNRAS.432.3401G} {432, 3401}

\bibitem[\protect\citeauthoryear{{Giles}, {Maughan}, {Hamana}, {Miyazaki},
  {Birkinshaw}, {Ellis}  \& {Massey}}{{Giles}
  et~al.}{2015}]{2015MNRAS.447.3044G}
{Giles} P.~A.,  {Maughan} B.~J.,  {Hamana} T.,  {Miyazaki} S.,  {Birkinshaw}
  M.,  {Ellis} R.~S.,   {Massey} R.,  2015, \mn@doi [\mnras]
  {10.1093/mnras/stu2679}, \href
  {https://ui.adsabs.harvard.edu/abs/2015MNRAS.447.3044G} {447, 3044}

\bibitem[\protect\citeauthoryear{{Gnat} \& {Sternberg}}{{Gnat} \&
  {Sternberg}}{2007}]{2007ApJS..168..213G}
{Gnat} O.,  {Sternberg} A.,  2007, \mn@doi [\apjs] {10.1086/509786}, \href
  {https://ui.adsabs.harvard.edu/abs/2007ApJS..168..213G} {168, 213}

\bibitem[\protect\citeauthoryear{{Hahn}, {Martizzi}, {Wu}, {Evrard}, {Teyssier}
   \& {Wechsler}}{{Hahn} et~al.}{2017}]{2017MNRAS.470..166H}
{Hahn} O.,  {Martizzi} D.,  {Wu} H.-Y.,  {Evrard} A.~E.,  {Teyssier} R.,
  {Wechsler} R.~H.,  2017, \mn@doi [\mnras] {10.1093/mnras/stx001}, \href
  {https://ui.adsabs.harvard.edu/abs/2017MNRAS.470..166H} {470, 166}

\bibitem[\protect\citeauthoryear{Harris et~al.,}{Harris
  et~al.}{2020}]{harris2020array}
Harris C.~R.,  et~al., 2020, Nature, 585, 357

\bibitem[\protect\citeauthoryear{{Heckman} \& {Best}}{{Heckman} \&
  {Best}}{2014}]{2014ARA&A..52..589H}
{Heckman} T.~M.,  {Best} P.~N.,  2014, \mn@doi [\araa]
  {10.1146/annurev-astro-081913-035722}, \href
  {https://ui.adsabs.harvard.edu/abs/2014ARA&A..52..589H} {52, 589}

\bibitem[\protect\citeauthoryear{{Henden}, {Puchwein}, {Shen}  \&
  {Sijacki}}{{Henden} et~al.}{2018}]{2018MNRAS.479.5385H}
{Henden} N.~A.,  {Puchwein} E.,  {Shen} S.,   {Sijacki} D.,  2018, \mn@doi
  [\mnras] {10.1093/mnras/sty1780}, \href
  {https://ui.adsabs.harvard.edu/abs/2018MNRAS.479.5385H} {479, 5385}

\bibitem[\protect\citeauthoryear{{Hirschmann}, {Dolag}, {Saro}, {Bachmann},
  {Borgani}  \& {Burkert}}{{Hirschmann} et~al.}{2014}]{2014MNRAS.442.2304H}
{Hirschmann} M.,  {Dolag} K.,  {Saro} A.,  {Bachmann} L.,  {Borgani} S.,
  {Burkert} A.,  2014, \mn@doi [\mnras] {10.1093/mnras/stu1023}, \href
  {https://ui.adsabs.harvard.edu/abs/2014MNRAS.442.2304H} {442, 2304}

\bibitem[\protect\citeauthoryear{{Hoyle} \& {Lyttleton}}{{Hoyle} \&
  {Lyttleton}}{1939}]{1939PCPS...35..405H}
{Hoyle} F.,  {Lyttleton} R.~A.,  1939, \mn@doi [Proceedings of the Cambridge
  Philosophical Society] {10.1017/S0305004100021150}, \href
  {https://ui.adsabs.harvard.edu/abs/1939PCPS...35..405H} {35, 405}

\bibitem[\protect\citeauthoryear{Hunter}{Hunter}{2007}]{hunter2007matplotlib}
Hunter J.~D.,  2007, Computing in science \& engineering, 9, 90

\bibitem[\protect\citeauthoryear{{Hu{\v{s}}ko}, {Lacey}, {Schaye}, {Schaller}
  \& {Nobels}}{{Hu{\v{s}}ko} et~al.}{2022}]{2022MNRAS.516.3750H}
{Hu{\v{s}}ko} F.,  {Lacey} C.~G.,  {Schaye} J.,  {Schaller} M.,   {Nobels} F.
  S.~J.,  2022, \mn@doi [\mnras] {10.1093/mnras/stac2278}, \href
  {https://ui.adsabs.harvard.edu/abs/2022MNRAS.516.3750H} {516, 3750}

\bibitem[\protect\citeauthoryear{{Jenkins}}{{Jenkins}}{2010}]{2010MNRAS.403.1859J}
{Jenkins} A.,  2010, \mn@doi [\mnras] {10.1111/j.1365-2966.2010.16259.x}, \href
  {https://ui.adsabs.harvard.edu/abs/2010MNRAS.403.1859J} {403, 1859}

\bibitem[\protect\citeauthoryear{{Jenkins}}{{Jenkins}}{2013}]{2013MNRAS.434.2094J}
{Jenkins} A.,  2013, \mn@doi [\mnras] {10.1093/mnras/stt1154}, \href
  {https://ui.adsabs.harvard.edu/abs/2013MNRAS.434.2094J} {434, 2094}

\bibitem[\protect\citeauthoryear{{Jung} et~al.,}{{Jung}
  et~al.}{2022}]{2022MNRAS.515...22J}
{Jung} S.~L.,  et~al., 2022, \mn@doi [\mnras] {10.1093/mnras/stac1622}, \href
  {https://ui.adsabs.harvard.edu/abs/2022MNRAS.515...22J} {515, 22}

\bibitem[\protect\citeauthoryear{{Kang}, {Ryu}, {Cen}  \& {Ostriker}}{{Kang}
  et~al.}{2007}]{2007ApJ...669..729K}
{Kang} H.,  {Ryu} D.,  {Cen} R.,   {Ostriker} J.~P.,  2007, \mn@doi [\apj]
  {10.1086/521717}, \href
  {https://ui.adsabs.harvard.edu/abs/2007ApJ...669..729K} {669, 729}

\bibitem[\protect\citeauthoryear{{Katz} \& {White}}{{Katz} \&
  {White}}{1993}]{1993ApJ...412..455K}
{Katz} N.,  {White} S. D.~M.,  1993, \mn@doi [\apj] {10.1086/172935}, \href
  {https://ui.adsabs.harvard.edu/abs/1993ApJ...412..455K} {412, 455}

\bibitem[\protect\citeauthoryear{{Kennicutt}}{{Kennicutt}}{1998}]{1998ApJ...498..541K}
{Kennicutt} Robert~C. J.,  1998, \mn@doi [\apj] {10.1086/305588}, \href
  {https://ui.adsabs.harvard.edu/abs/1998ApJ...498..541K} {498, 541}

\bibitem[\protect\citeauthoryear{{Kormendy} \& {Ho}}{{Kormendy} \&
  {Ho}}{2013}]{2013ARA&A..51..511K}
{Kormendy} J.,  {Ho} L.~C.,  2013, \mn@doi [\araa]
  {10.1146/annurev-astro-082708-101811}, \href
  {https://ui.adsabs.harvard.edu/abs/2013ARA&A..51..511K} {51, 511}

\bibitem[\protect\citeauthoryear{{Kugel} \& {Borrow}}{{Kugel} \&
  {Borrow}}{2022}]{2022JOSS....7.4240K}
{Kugel} R.,  {Borrow} J.,  2022, \mn@doi [The Journal of Open Source Software]
  {10.21105/joss.04240}, \href
  {https://ui.adsabs.harvard.edu/abs/2022JOSS....7.4240K} {7, 4240}

\bibitem[\protect\citeauthoryear{Lam, Pitrou  \& Seibert}{Lam
  et~al.}{2015}]{lam2015numba}
Lam S.~K.,  Pitrou A.,   Seibert S.,  2015, in Proceedings of the Second
  Workshop on the LLVM Compiler Infrastructure in HPC. LLVM '15.
Association for Computing Machinery, New York, NY, USA,
  \mn@doi{10.1145/2833157.2833162}, \url
  {https://doi.org/10.1145/2833157.2833162}

\bibitem[\protect\citeauthoryear{{Le Brun}, {McCarthy}, {Schaye}  \&
  {Ponman}}{{Le Brun} et~al.}{2014}]{2014MNRAS.441.1270L}
{Le Brun} A. M.~C.,  {McCarthy} I.~G.,  {Schaye} J.,   {Ponman} T.~J.,  2014,
  \mn@doi [\mnras] {10.1093/mnras/stu608}, \href
  {https://ui.adsabs.harvard.edu/abs/2014MNRAS.441.1270L} {441, 1270}

\bibitem[\protect\citeauthoryear{{Li}, {Bryan}, {Ruszkowski}, {Voit}, {O'Shea}
  \& {Donahue}}{{Li} et~al.}{2015}]{2015ApJ...811...73L}
{Li} Y.,  {Bryan} G.~L.,  {Ruszkowski} M.,  {Voit} G.~M.,  {O'Shea} B.~W.,
  {Donahue} M.,  2015, \mn@doi [\apj] {10.1088/0004-637X/811/2/73}, \href
  {https://ui.adsabs.harvard.edu/abs/2015ApJ...811...73L} {811, 73}

\bibitem[\protect\citeauthoryear{{Li}, {Ruszkowski}  \& {Bryan}}{{Li}
  et~al.}{2017}]{2017ApJ...847..106L}
{Li} Y.,  {Ruszkowski} M.,   {Bryan} G.~L.,  2017, \mn@doi [\apj]
  {10.3847/1538-4357/aa88c1}, \href
  {https://ui.adsabs.harvard.edu/abs/2017ApJ...847..106L} {847, 106}

\bibitem[\protect\citeauthoryear{{Li} et~al.,}{{Li}
  et~al.}{2020}]{2020MNRAS.495.2930L}
{Li} Q.,  et~al., 2020, \mn@doi [\mnras] {10.1093/mnras/staa1385}, \href
  {https://ui.adsabs.harvard.edu/abs/2020MNRAS.495.2930L} {495, 2930}

\bibitem[\protect\citeauthoryear{{Lyutikov}}{{Lyutikov}}{2006}]{2006MNRAS.373...73L}
{Lyutikov} M.,  2006, \mn@doi [\mnras] {10.1111/j.1365-2966.2006.10835.x},
  \href {https://ui.adsabs.harvard.edu/abs/2006MNRAS.373...73L} {373, 73}

\bibitem[\protect\citeauthoryear{{Martizzi}, {Teyssier}  \& {Moore}}{{Martizzi}
  et~al.}{2012}]{2012MNRAS.420.2859M}
{Martizzi} D.,  {Teyssier} R.,   {Moore} B.,  2012, \mn@doi [\mnras]
  {10.1111/j.1365-2966.2011.19950.x}, \href
  {https://ui.adsabs.harvard.edu/abs/2012MNRAS.420.2859M} {420, 2859}

\bibitem[\protect\citeauthoryear{{Martizzi}, {Quataert}, {Faucher-Gigu{\`e}re}
  \& {Fielding}}{{Martizzi} et~al.}{2019}]{2019MNRAS.483.2465M}
{Martizzi} D.,  {Quataert} E.,  {Faucher-Gigu{\`e}re} C.-A.,   {Fielding} D.,
  2019, \mn@doi [\mnras] {10.1093/mnras/sty3273}, \href
  {https://ui.adsabs.harvard.edu/abs/2019MNRAS.483.2465M} {483, 2465}

\bibitem[\protect\citeauthoryear{{McAlpine} et~al.,}{{McAlpine}
  et~al.}{2016}]{2016A&C....15...72M}
{McAlpine} S.,  et~al., 2016, \mn@doi [Astronomy and Computing]
  {10.1016/j.ascom.2016.02.004}, \href
  {https://ui.adsabs.harvard.edu/abs/2016A&C....15...72M} {15, 72}

\bibitem[\protect\citeauthoryear{{McAlpine}, {Bower}, {Rosario}, {Crain},
  {Schaye}  \& {Theuns}}{{McAlpine} et~al.}{2018}]{2018MNRAS.481.3118M}
{McAlpine} S.,  {Bower} R.~G.,  {Rosario} D.~J.,  {Crain} R.~A.,  {Schaye} J.,
   {Theuns} T.,  2018, \mn@doi [\mnras] {10.1093/mnras/sty2489}, \href
  {https://ui.adsabs.harvard.edu/abs/2018MNRAS.481.3118M} {481, 3118}

\bibitem[\protect\citeauthoryear{{McCarthy} et~al.,}{{McCarthy}
  et~al.}{2007}]{2007MNRAS.376..497M}
{McCarthy} I.~G.,  et~al., 2007, \mn@doi [\mnras]
  {10.1111/j.1365-2966.2007.11465.x}, \href
  {https://ui.adsabs.harvard.edu/abs/2007MNRAS.376..497M} {376, 497}

\bibitem[\protect\citeauthoryear{{McCarthy} et~al.,}{{McCarthy}
  et~al.}{2010}]{2010MNRAS.406..822M}
{McCarthy} I.~G.,  et~al., 2010, \mn@doi [\mnras]
  {10.1111/j.1365-2966.2010.16750.x}, \href
  {https://ui.adsabs.harvard.edu/abs/2010MNRAS.406..822M} {406, 822}

\bibitem[\protect\citeauthoryear{{McCarthy}, {Schaye}, {Bower}, {Ponman},
  {Booth}, {Dalla Vecchia}  \& {Springel}}{{McCarthy}
  et~al.}{2011}]{2011MNRAS.412.1965M}
{McCarthy} I.~G.,  {Schaye} J.,  {Bower} R.~G.,  {Ponman} T.~J.,  {Booth}
  C.~M.,  {Dalla Vecchia} C.,   {Springel} V.,  2011, \mn@doi [\mnras]
  {10.1111/j.1365-2966.2010.18033.x}, \href
  {https://ui.adsabs.harvard.edu/abs/2011MNRAS.412.1965M} {412, 1965}

\bibitem[\protect\citeauthoryear{{McCarthy}, {Schaye}, {Bird}  \& {Le
  Brun}}{{McCarthy} et~al.}{2017}]{2017MNRAS.465.2936M}
{McCarthy} I.~G.,  {Schaye} J.,  {Bird} S.,   {Le Brun} A. M.~C.,  2017,
  \mn@doi [\mnras] {10.1093/mnras/stw2792}, \href
  {https://ui.adsabs.harvard.edu/abs/2017MNRAS.465.2936M} {465, 2936}

\bibitem[\protect\citeauthoryear{{McDonald} et~al.,}{{McDonald}
  et~al.}{2014}]{2014ApJ...794...67M}
{McDonald} M.,  et~al., 2014, \mn@doi [\apj] {10.1088/0004-637X/794/1/67},
  \href {https://ui.adsabs.harvard.edu/abs/2014ApJ...794...67M} {794, 67}

\bibitem[\protect\citeauthoryear{{Mitchell}, {McCarthy}, {Bower}, {Theuns}  \&
  {Crain}}{{Mitchell} et~al.}{2009}]{2009MNRAS.395..180M}
{Mitchell} N.~L.,  {McCarthy} I.~G.,  {Bower} R.~G.,  {Theuns} T.,   {Crain}
  R.~A.,  2009, \mn@doi [\mnras] {10.1111/j.1365-2966.2009.14550.x}, \href
  {https://ui.adsabs.harvard.edu/abs/2009MNRAS.395..180M} {395, 180}

\bibitem[\protect\citeauthoryear{{Muanwong}, {Thomas}, {Kay}  \&
  {Pearce}}{{Muanwong} et~al.}{2002}]{2002MNRAS.336..527M}
{Muanwong} O.,  {Thomas} P.~A.,  {Kay} S.~T.,   {Pearce} F.~R.,  2002, \mn@doi
  [\mnras] {10.1046/j.1365-8711.2002.05770.x}, \href
  {https://ui.adsabs.harvard.edu/abs/2002MNRAS.336..527M} {336, 527}

\bibitem[\protect\citeauthoryear{{Nobels}, {Schaye}, {Schaller}, {Bah{\'e}}  \&
  {Chaikin}}{{Nobels} et~al.}{2022}]{2022MNRAS.tmp.1955N}
{Nobels} F. S.~J.,  {Schaye} J.,  {Schaller} M.,  {Bah{\'e}} Y.~M.,   {Chaikin}
  E.,  2022, \mn@doi [\mnras] {10.1093/mnras/stac2061}, \href
  {https://ui.adsabs.harvard.edu/abs/2022MNRAS.515.4838N} {515, 4838}

\bibitem[\protect\citeauthoryear{{Oppenheimer} \& {Schaye}}{{Oppenheimer} \&
  {Schaye}}{2013}]{2013MNRAS.434.1043O}
{Oppenheimer} B.~D.,  {Schaye} J.,  2013, \mn@doi [\mnras]
  {10.1093/mnras/stt1043}, \href
  {https://ui.adsabs.harvard.edu/abs/2013MNRAS.434.1043O} {434, 1043}

\bibitem[\protect\citeauthoryear{{Oppenheimer}, {Babul}, {Bah{\'e}}, {Butsky}
  \& {McCarthy}}{{Oppenheimer} et~al.}{2021}]{2021Univ....7..209O}
{Oppenheimer} B.~D.,  {Babul} A.,  {Bah{\'e}} Y.,  {Butsky} I.~S.,   {McCarthy}
  I.~G.,  2021, \mn@doi [Universe] {10.3390/universe7070209}, \href
  {https://ui.adsabs.harvard.edu/abs/2021Univ....7..209O} {7, 209}

\bibitem[\protect\citeauthoryear{{Pearce}, {Thomas}, {Couchman}  \&
  {Edge}}{{Pearce} et~al.}{2000}]{2000MNRAS.317.1029P}
{Pearce} F.~R.,  {Thomas} P.~A.,  {Couchman} H.~M.~P.,   {Edge} A.~C.,  2000,
  \mn@doi [\mnras] {10.1046/j.1365-8711.2000.03773.x}, \href
  {https://ui.adsabs.harvard.edu/abs/2000MNRAS.317.1029P} {317, 1029}

\bibitem[\protect\citeauthoryear{{Pike}, {Kay}, {Newton}, {Thomas}  \&
  {Jenkins}}{{Pike} et~al.}{2014}]{2014MNRAS.445.1774P}
{Pike} S.~R.,  {Kay} S.~T.,  {Newton} R. D.~A.,  {Thomas} P.~A.,   {Jenkins}
  A.,  2014, \mn@doi [\mnras] {10.1093/mnras/stu1788}, \href
  {https://ui.adsabs.harvard.edu/abs/2014MNRAS.445.1774P} {445, 1774}

\bibitem[\protect\citeauthoryear{{Pillepich} et~al.,}{{Pillepich}
  et~al.}{2018}]{2018MNRAS.473.4077P}
{Pillepich} A.,  et~al., 2018, \mn@doi [\mnras] {10.1093/mnras/stx2656}, \href
  {https://ui.adsabs.harvard.edu/abs/2018MNRAS.473.4077P} {473, 4077}

\bibitem[\protect\citeauthoryear{{Planck Collaboration} et~al.,}{{Planck
  Collaboration} et~al.}{2020}]{planck.2018.cosmology}
{Planck Collaboration} et~al., 2020, \mn@doi [\aap]
  {10.1051/0004-6361/201833910}, \href
  {https://ui.adsabs.harvard.edu/abs/2020A&A...641A...6P} {641, A6}

\bibitem[\protect\citeauthoryear{{Planelles} et~al.,}{{Planelles}
  et~al.}{2021}]{2021MNRAS.507.5703P}
{Planelles} S.,  et~al., 2021, \mn@doi [\mnras] {10.1093/mnras/stab2436}, \href
  {https://ui.adsabs.harvard.edu/abs/2021MNRAS.507.5703P} {507, 5703}

\bibitem[\protect\citeauthoryear{{Ploeckinger} \& {Schaye}}{{Ploeckinger} \&
  {Schaye}}{2020}]{2020MNRAS.497.4857P}
{Ploeckinger} S.,  {Schaye} J.,  2020, \mn@doi [\mnras]
  {10.1093/mnras/staa2172}, \href
  {https://ui.adsabs.harvard.edu/abs/2020MNRAS.497.4857P} {497, 4857}

\bibitem[\protect\citeauthoryear{{Poole}, {Babul}, {McCarthy}, {Sanderson}  \&
  {Fardal}}{{Poole} et~al.}{2008}]{2008MNRAS.391.1163P}
{Poole} G.~B.,  {Babul} A.,  {McCarthy} I.~G.,  {Sanderson} A.~J.~R.,
  {Fardal} M.~A.,  2008, \mn@doi [\mnras] {10.1111/j.1365-2966.2008.14003.x},
  \href {https://ui.adsabs.harvard.edu/abs/2008MNRAS.391.1163P} {391, 1163}

\bibitem[\protect\citeauthoryear{{Prasad}, {Sharma}  \& {Babul}}{{Prasad}
  et~al.}{2015}]{2015ApJ...811..108P}
{Prasad} D.,  {Sharma} P.,   {Babul} A.,  2015, \mn@doi [\apj]
  {10.1088/0004-637X/811/2/108}, \href
  {https://ui.adsabs.harvard.edu/abs/2015ApJ...811..108P} {811, 108}

\bibitem[\protect\citeauthoryear{{Pratt} et~al.,}{{Pratt}
  et~al.}{2010}]{entropy_profiles_pratt2010}
{Pratt} G.~W.,  et~al., 2010, \mn@doi [\aap] {10.1051/0004-6361/200913309},
  \href {https://ui.adsabs.harvard.edu/abs/2010A&A...511A..85P} {511, A85}

\bibitem[\protect\citeauthoryear{{Price}}{{Price}}{2008}]{2008JCoPh.22710040P}
{Price} D.~J.,  2008, \mn@doi [Journal of Computational Physics]
  {10.1016/j.jcp.2008.08.011}, \href
  {https://ui.adsabs.harvard.edu/abs/2008JCoPh.22710040P} {227, 10040}

\bibitem[\protect\citeauthoryear{{Rasia} et~al.,}{{Rasia}
  et~al.}{2015}]{2015ApJ...813L..17R}
{Rasia} E.,  et~al., 2015, \mn@doi [\apjl] {10.1088/2041-8205/813/1/L17}, \href
  {https://ui.adsabs.harvard.edu/abs/2015ApJ...813L..17R} {813, L17}

\bibitem[\protect\citeauthoryear{{Robson} \& {Dav{\'e}}}{{Robson} \&
  {Dav{\'e}}}{2020}]{2020MNRAS.498.3061R}
{Robson} D.,  {Dav{\'e}} R.,  2020, \mn@doi [\mnras] {10.1093/mnras/staa2394},
  \href {https://ui.adsabs.harvard.edu/abs/2020MNRAS.498.3061R} {498, 3061}

\bibitem[\protect\citeauthoryear{{Rosas-Guevara} et~al.,}{{Rosas-Guevara}
  et~al.}{2015}]{2015MNRAS.454.1038R}
{Rosas-Guevara} Y.~M.,  et~al., 2015, \mn@doi [\mnras] {10.1093/mnras/stv2056},
  \href {https://ui.adsabs.harvard.edu/abs/2015MNRAS.454.1038R} {454, 1038}

\bibitem[\protect\citeauthoryear{{Ro{\v{s}}kar}, {Teyssier}, {Agertz},
  {Wetzstein}  \& {Moore}}{{Ro{\v{s}}kar} et~al.}{2014}]{2014MNRAS.444.2837R}
{Ro{\v{s}}kar} R.,  {Teyssier} R.,  {Agertz} O.,  {Wetzstein} M.,   {Moore} B.,
   2014, \mn@doi [\mnras] {10.1093/mnras/stu1548}, \href
  {https://ui.adsabs.harvard.edu/abs/2014MNRAS.444.2837R} {444, 2837}

\bibitem[\protect\citeauthoryear{{Schaller}, {Dalla Vecchia}, {Schaye},
  {Bower}, {Theuns}, {Crain}, {Furlong}  \& {McCarthy}}{{Schaller}
  et~al.}{2015}]{2015MNRAS.454.2277S}
{Schaller} M.,  {Dalla Vecchia} C.,  {Schaye} J.,  {Bower} R.~G.,  {Theuns} T.,
   {Crain} R.~A.,  {Furlong} M.,   {McCarthy} I.~G.,  2015, \mn@doi [\mnras]
  {10.1093/mnras/stv2169}, \href
  {https://ui.adsabs.harvard.edu/abs/2015MNRAS.454.2277S} {454, 2277}

\bibitem[\protect\citeauthoryear{Schaller, Gonnet, Chalk  \& Draper}{Schaller
  et~al.}{2016}]{2016pasc.conf....2S}
Schaller M.,  Gonnet P.,  Chalk A. B.~G.,   Draper P.~W.,  2016, in Proceedings
  of the Platform for Advanced Scientific Computing Conference. PASC '16.
Association for Computing Machinery, New York, NY, USA,
  \mn@doi{10.1145/2929908.2929916}, \url
  {https://doi.org/10.1145/2929908.2929916}

\bibitem[\protect\citeauthoryear{{Schaller}, {Gonnet}, {Draper}, {Chalk},
  {Bower}, {Willis}  \& {Hausammann}}{{Schaller}
  et~al.}{2018}]{schaller_2018_swift}
{Schaller} M.,  {Gonnet} P.,  {Draper} P.~W.,  {Chalk} A. B.~G.,  {Bower}
  R.~G.,  {Willis} J.,   {Hausammann} L.,  2018, {SWIFT: SPH With
  Inter-dependent Fine-grained Tasking} (\mn@eprint {ascl} {1805.020})

\bibitem[\protect\citeauthoryear{{Schaye}}{{Schaye}}{2004}]{2004ApJ...609..667S}
{Schaye} J.,  2004, \mn@doi [\apj] {10.1086/421232}, \href
  {https://ui.adsabs.harvard.edu/abs/2004ApJ...609..667S} {609, 667}

\bibitem[\protect\citeauthoryear{{Schaye} \& {Dalla Vecchia}}{{Schaye} \&
  {Dalla Vecchia}}{2008}]{2008MNRAS.383.1210S}
{Schaye} J.,  {Dalla Vecchia} C.,  2008, \mn@doi [\mnras]
  {10.1111/j.1365-2966.2007.12639.x}, \href
  {https://ui.adsabs.harvard.edu/abs/2008MNRAS.383.1210S} {383, 1210}

\bibitem[\protect\citeauthoryear{{Schaye} et~al.,}{{Schaye}
  et~al.}{2015}]{eagle.schaye.2015}
{Schaye} J.,  et~al., 2015, \mn@doi [\mnras] {10.1093/mnras/stu2058}, \href
  {https://ui.adsabs.harvard.edu/abs/2015MNRAS.446..521S} {446, 521}

\bibitem[\protect\citeauthoryear{{Schmidt}}{{Schmidt}}{1959}]{1959ApJ...129..243S}
{Schmidt} M.,  1959, \mn@doi [\apj] {10.1086/146614}, \href
  {https://ui.adsabs.harvard.edu/abs/1959ApJ...129..243S} {129, 243}

\bibitem[\protect\citeauthoryear{{Sembolini} et~al.,}{{Sembolini}
  et~al.}{2016a}]{2016MNRAS.457.4063S}
{Sembolini} F.,  et~al., 2016a, \mn@doi [\mnras] {10.1093/mnras/stw250}, \href
  {https://ui.adsabs.harvard.edu/abs/2016MNRAS.457.4063S} {457, 4063}

\bibitem[\protect\citeauthoryear{{Sembolini} et~al.,}{{Sembolini}
  et~al.}{2016b}]{2016MNRAS.459.2973S}
{Sembolini} F.,  et~al., 2016b, \mn@doi [\mnras] {10.1093/mnras/stw800}, \href
  {https://ui.adsabs.harvard.edu/abs/2016MNRAS.459.2973S} {459, 2973}

\bibitem[\protect\citeauthoryear{{Shakura} \& {Sunyaev}}{{Shakura} \&
  {Sunyaev}}{1973}]{1973A&A....24..337S}
{Shakura} N.~I.,  {Sunyaev} R.~A.,  1973, \aap, \href
  {https://ui.adsabs.harvard.edu/abs/1973A&A....24..337S} {24, 337}

\bibitem[\protect\citeauthoryear{{Sun}, {Voit}, {Donahue}, {Jones}, {Forman}
  \& {Vikhlinin}}{{Sun} et~al.}{2009}]{entropy_profiles_sun2009}
{Sun} M.,  {Voit} G.~M.,  {Donahue} M.,  {Jones} C.,  {Forman} W.,
  {Vikhlinin} A.,  2009, \mn@doi [\apj] {10.1088/0004-637X/693/2/1142}, \href
  {https://ui.adsabs.harvard.edu/abs/2009ApJ...693.1142S} {693, 1142}

\bibitem[\protect\citeauthoryear{{Talbot}, {Bourne}  \& {Sijacki}}{{Talbot}
  et~al.}{2021}]{2021MNRAS.504.3619T}
{Talbot} R.~Y.,  {Bourne} M.~A.,   {Sijacki} D.,  2021, \mn@doi [\mnras]
  {10.1093/mnras/stab804}, \href
  {https://ui.adsabs.harvard.edu/abs/2021MNRAS.504.3619T} {504, 3619}

\bibitem[\protect\citeauthoryear{{Timmerman}, {van Weeren}, {McDonald},
  {Ignesti}, {McNamara}, {Hlavacek-Larrondo}  \& {R{\"o}ttgering}}{{Timmerman}
  et~al.}{2021}]{2021A&A...646A..38T}
{Timmerman} R.,  {van Weeren} R.~J.,  {McDonald} M.,  {Ignesti} A.,  {McNamara}
  B.~R.,  {Hlavacek-Larrondo} J.,   {R{\"o}ttgering} H.~J.~A.,  2021, \mn@doi
  [\aap] {10.1051/0004-6361/202039075}, \href
  {https://ui.adsabs.harvard.edu/abs/2021A&A...646A..38T} {646, A38}

\bibitem[\protect\citeauthoryear{{Tormen}, {Bouchet}  \& {White}}{{Tormen}
  et~al.}{1997}]{1997MNRAS.286..865T}
{Tormen} G.,  {Bouchet} F.~R.,   {White} S. D.~M.,  1997, \mn@doi [\mnras]
  {10.1093/mnras/286.4.865}, \href
  {https://ui.adsabs.harvard.edu/abs/1997MNRAS.286..865T} {286, 865}

\bibitem[\protect\citeauthoryear{{Towler}, {Kay}  \& {Altamura}}{{Towler}
  et~al.}{2022}]{2022arXiv221101239T}
{Towler} I.,  {Kay} S.,   {Altamura} E.,  2022, arXiv e-prints, \href
  {https://ui.adsabs.harvard.edu/abs/2022arXiv221101239T} {p. arXiv:2211.01239}

\bibitem[\protect\citeauthoryear{{Tozzi} \& {Norman}}{{Tozzi} \&
  {Norman}}{2001}]{2001ApJ...546...63T}
{Tozzi} P.,  {Norman} C.,  2001, \mn@doi [\apj] {10.1086/318237}, \href
  {https://ui.adsabs.harvard.edu/abs/2001ApJ...546...63T} {546, 63}

\bibitem[\protect\citeauthoryear{{Tremmel} et~al.,}{{Tremmel}
  et~al.}{2019}]{2019MNRAS.483.3336T}
{Tremmel} M.,  et~al., 2019, \mn@doi [\mnras] {10.1093/mnras/sty3336}, \href
  {https://ui.adsabs.harvard.edu/abs/2019MNRAS.483.3336T} {483, 3336}

\bibitem[\protect\citeauthoryear{{Van Rossum} \& {Drake Jr}}{{Van Rossum} \&
  {Drake Jr}}{1995}]{van1995python}
{Van Rossum} G.,  {Drake Jr} F.~L.,  1995, Python tutorial.
~ Vol. 620, Centrum voor Wiskunde en Informatica, Amsterdam, The Netherlands

\bibitem[\protect\citeauthoryear{{Vikhlinin}, {Kravtsov}, {Forman}, {Jones},
  {Markevitch}, {Murray}  \& {Van Speybroeck}}{{Vikhlinin}
  et~al.}{2006}]{vikhlinin_2006_profile_slope}
{Vikhlinin} A.,  {Kravtsov} A.,  {Forman} W.,  {Jones} C.,  {Markevitch} M.,
  {Murray} S.~S.,   {Van Speybroeck} L.,  2006, \mn@doi [\apj]
  {10.1086/500288}, \href
  {https://ui.adsabs.harvard.edu/abs/2006ApJ...640..691V} {640, 691}

\bibitem[\protect\citeauthoryear{Virtanen et~al.,}{Virtanen
  et~al.}{2020}]{virtanen2020scipy}
Virtanen P.,  et~al., 2020, Nature methods, 17, 261

\bibitem[\protect\citeauthoryear{{Voit}, {Bryan}, {Balogh}  \& {Bower}}{{Voit}
  et~al.}{2002}]{2002ApJ...576..601V}
{Voit} G.~M.,  {Bryan} G.~L.,  {Balogh} M.~L.,   {Bower} R.~G.,  2002, \mn@doi
  [\apj] {10.1086/341864}, \href
  {https://ui.adsabs.harvard.edu/abs/2002ApJ...576..601V} {576, 601}

\bibitem[\protect\citeauthoryear{{Voit}, {Kay}  \& {Bryan}}{{Voit}
  et~al.}{2005}]{vkb_2005}
{Voit} G.~M.,  {Kay} S.~T.,   {Bryan} G.~L.,  2005, \mn@doi [\mnras]
  {10.1111/j.1365-2966.2005.09621.x}, \href
  {https://ui.adsabs.harvard.edu/abs/2005MNRAS.364..909V} {364, 909}

\bibitem[\protect\citeauthoryear{{Weinberger}, {Ehlert}, {Pfrommer}, {Pakmor}
  \& {Springel}}{{Weinberger} et~al.}{2017}]{2017MNRAS.470.4530W}
{Weinberger} R.,  {Ehlert} K.,  {Pfrommer} C.,  {Pakmor} R.,   {Springel} V.,
  2017, \mn@doi [\mnras] {10.1093/mnras/stx1409}, \href
  {https://ui.adsabs.harvard.edu/abs/2017MNRAS.470.4530W} {470, 4530}

\bibitem[\protect\citeauthoryear{{Wetzel}, {Tinker}  \& {Conroy}}{{Wetzel}
  et~al.}{2012}]{2012MNRAS.424..232W}
{Wetzel} A.~R.,  {Tinker} J.~L.,   {Conroy} C.,  2012, \mn@doi [\mnras]
  {10.1111/j.1365-2966.2012.21188.x}, \href
  {https://ui.adsabs.harvard.edu/abs/2012MNRAS.424..232W} {424, 232}

\bibitem[\protect\citeauthoryear{{Wiersma}, {Schaye}, {Theuns}, {Dalla Vecchia}
   \& {Tornatore}}{{Wiersma} et~al.}{2009}]{2009MNRAS.399..574W}
{Wiersma} R. P.~C.,  {Schaye} J.,  {Theuns} T.,  {Dalla Vecchia} C.,
  {Tornatore} L.,  2009, \mn@doi [\mnras] {10.1111/j.1365-2966.2009.15331.x},
  \href {https://ui.adsabs.harvard.edu/abs/2009MNRAS.399..574W} {399, 574}

\bibitem[\protect\citeauthoryear{{Yang} \& {Reynolds}}{{Yang} \&
  {Reynolds}}{2016}]{2016ApJ...829...90Y}
{Yang} H. Y.~K.,  {Reynolds} C.~S.,  2016, \mn@doi [\apj]
  {10.3847/0004-637X/829/2/90}, \href
  {https://ui.adsabs.harvard.edu/abs/2016ApJ...829...90Y} {829, 90}

\makeatother
\end{thebibliography}

\appendix

\section{Reduced SN heating temperature}
\label{appx:sn-heating-temperature}

\begin{figure*}
	\includegraphics[width=2\columnwidth]{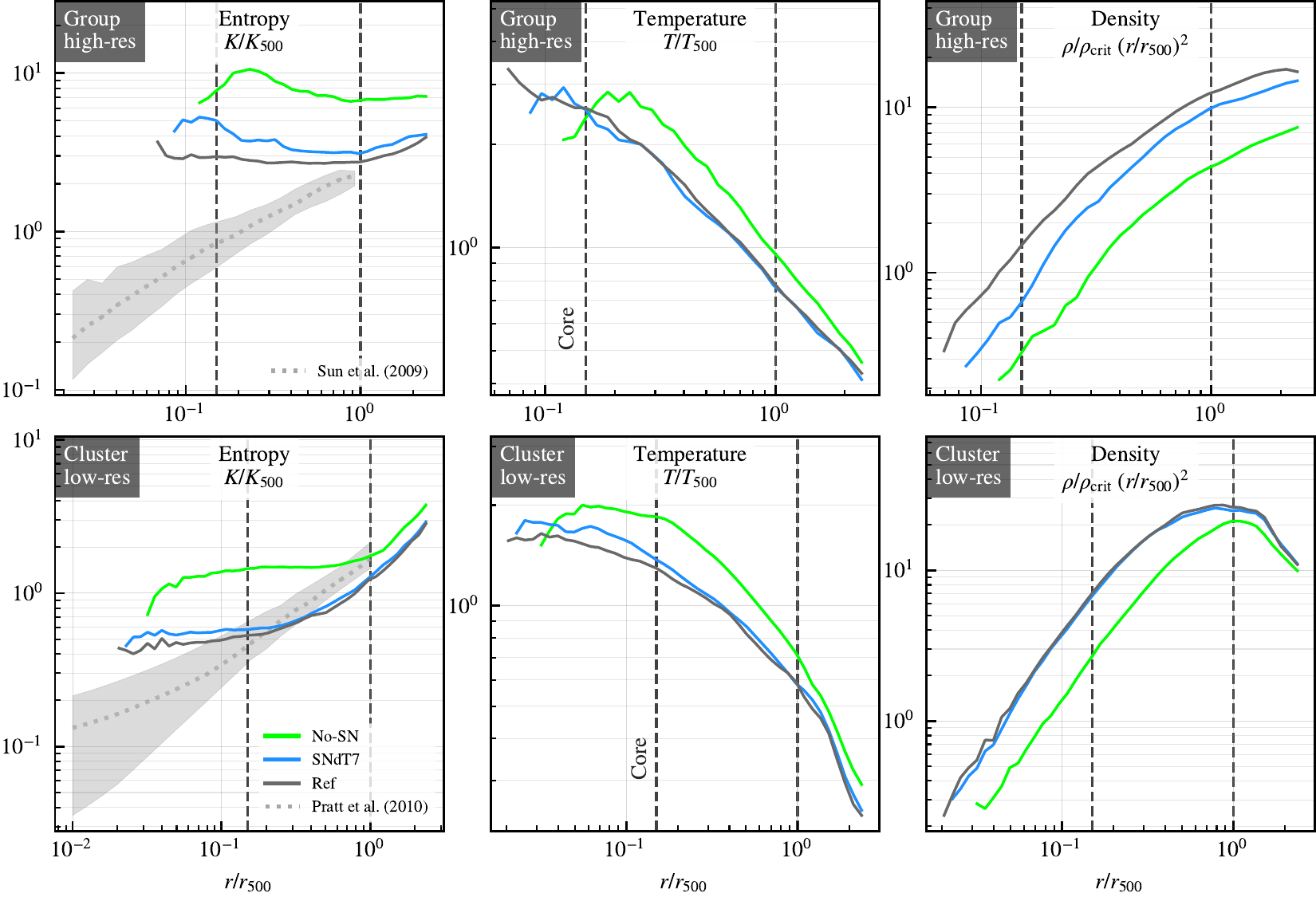}
    \caption{{As in Fig. \ref{fig:profiles_nofeedback}, but including runs with reduced heating temperature in SN feedback (blue). Here, we only show the profiles for the group at high-res (top) and the cluster at low-res (bottom).}}
    \label{fig:sndt_variations}
\end{figure*}

{To further examine the onset of cooling losses leading to anomalous star-formation in models with no SN feedback, we have produced additional simulations with reduced SN heating temperature, $\Delta T_{\rm SN} = 10^{7}~{\rm K}$. In Fig. \ref{fig:sndt_variations}, we compare this model, labelled as SNdT7, with Ref ($\Delta T_{\rm SN} = 10^{7.5}~{\rm K}$) and No-SN ($\Delta T_{\rm SN} = 0~{\rm K}$) for the group at high resolution (same as EAGLE) and the cluster at low resolution (8 times lower than EAGLE) at $z=0$. We found that the group was more susceptible to a decrease in $\Delta T_{\rm SN}$: the density of the gas drops at all radii, especially in the core. This effect results in higher entropy, despite the mass-weighted temperature profile being consistent with Ref. On the other hand, the SNdT7 cluster shows profiles similar to Ref. The temperature profile being slightly higher in the core means that the entropy is higher, but the magnitude of this effect is far smaller than disabling SN feedback completely (No-SN).\\
For the group at high-res, the Ref model produces a star fraction (in $r_{500}$) $f_\star = 0.028$, the SNdT7 variation returns $f_\star = 0.046$ and the No-SN solution gives $f_\star = 0.084$. Similarly, the low-res cluster gives $f_\star = 0.019$ for Ref, $f_\star = 0.023$ for SNdT7 and $f_\star = 0.054$ for No-SN. In both objects, we measure an increasing star fraction associated with a decreasing hot gas fraction. These results are consistent with the predictions of \cite{2012MNRAS.426..140D} and other simulations of galaxy-sized halos which found a larger stellar mass in weak SN feedback regimes \citep[e.g.][]{2014MNRAS.444.2837R}. Here, we show that cooling losses are more prevalent in the group than in the cluster and the overall effect is to raise the entropy profile.}

\bsp	
\label{lastpage}
\end{document}